\documentclass[10pt]{article}
\usepackage[a4paper, left=1in, right=1in, top=1in, bottom=1in]{geometry}
\usepackage{graphicx} 
\usepackage{empheq}
\usepackage{amsthm}
\usepackage{amsmath}
\usepackage{amsfonts}
\usepackage{amssymb}
\usepackage{tikz-cd}
\usepackage{harpoon}
\usepackage{subcaption}
\usepackage{float}
\usepackage{colortbl}
\usepackage{physics}
\usepackage{subcaption}
\usepackage{hyperref}
\usepackage{indentfirst}
\usepackage{pdfpages}
\usepackage{authblk}

\usepackage[backend=biber, style=numeric, sorting=none]{biblatex}
\hypersetup{colorlinks,linkcolor={blue},citecolor={blue},urlcolor={red}}

\title{Self-supporting Structure of Bird Nests}
\author[1]{Yuming Chen}
\author[1]{Yifang Xu}
\author[1]{Yiran Yao}
\author[2]{Sihui Wang\thanks{Corresponding author: wangsihui@nju.edu.cn}}
\affil[1]{Basis International School Nanjing, 210000 Nanjing, China}
\affil[2]{School of Physics, Nanjing University, 210093 Nanjing, China}
\date{\today}

\newcommand{\keywords}[1]{\textbf{Keywords:} #1}
\DeclareBibliographyDriver{article}{%
  \usebibmacro{bibindex}%
  \usebibmacro{begentry}%
  \usebibmacro{author/translator+others}%
  \setunit{\labelnamepunct}\newblock
  \usebibmacro{title}%
  \newunit
  \printfield{journaltitle}%
  \newunit
  \printfield{volume}%
  \setunit{\addspace}%
  \printfield{number}%
  \setunit{\addcomma\space}%
  \printfield{pages}%
  \setunit{\addspace}%
  \usebibmacro{date}%
  \newunit\newblock
  \iftoggle{bbx:doi}
    {\printfield{doi}}
    {}%
  \newunit\newblock
  \usebibmacro{addendum+pubstate}%
  \setunit{\bibpagerefpunct}\newblock
  \usebibmacro{pageref}%
  \usebibmacro{finentry}}

\begin{document}
\maketitle
\begin{abstract}

The mystery behind the bird nest's construction is not well understood. Our study focuses on the stability of a self-supporting nest-like structure. Firstly, we derived a stable/unstable phase boundary for the structure at the fixed coefficient of friction with varying geometrical parameters through force analysis. Structures with a lower height and greater friction coefficient between rods are more stable. The theoretical phase boundary matched the experiment results well.

Then we investigate the nest structure's stability under applied weight. Static structures with lower height and more rods (five>four>three) are more stable. Our theory also predicts a transition from plastic phase to elastic phase. These theoretical predictions are all confirmed by experiment. In the experiment, we also find that wet rod structures are more stable than dry ones. The structures can support up to 100 times of it's weight. 

Finally, we test the nest structure's stability under vibration. When there are no weights applied, we are able to identify the appropriate geometric configuration that can withstand the greatest vibration ($1g$ of vibration acceleration and vibration energy up to $4\times10^{-4}$ times of a rod's maximum potential energy). The critical vibration energy and acceleration depend on the applied weights. They are increased by two and one order of magnitude respectively under proper weight. We also make potential energy analysis to explain the stability of the structure.

\end{abstract}

\newpage
\tableofcontents
\newpage
\section{Introduction}
\begin{figure}[H]
    \centering
    \begin{minipage}{0.5\textwidth}
        \centering
        \includegraphics[width=0.8\linewidth]{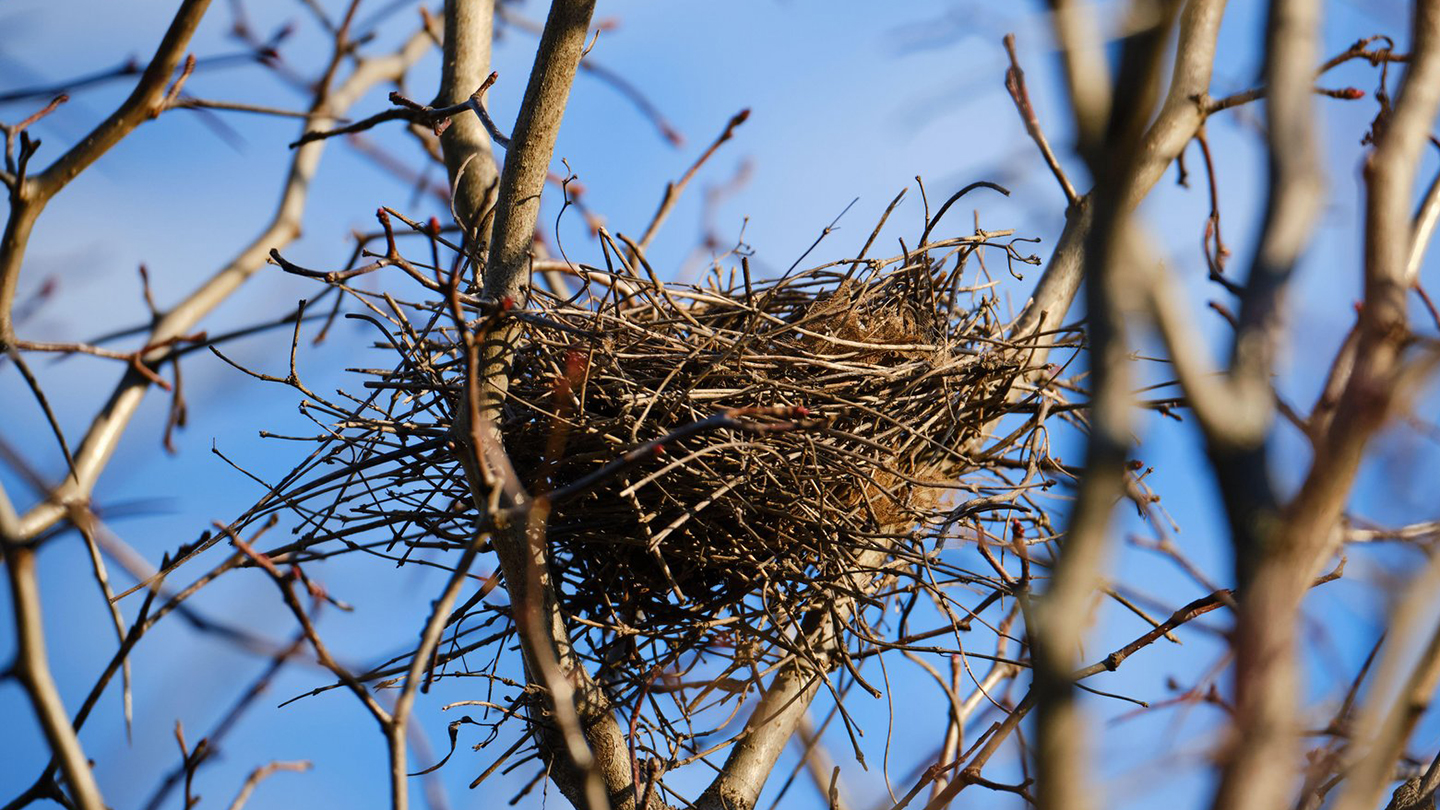}
        \caption{Bird Nest (from Internet)}
        \label{fig:natural bird nest}
    \end{minipage}%
    \begin{minipage}{0.5\textwidth}
        \centering
        \includegraphics[width=0.8\linewidth]{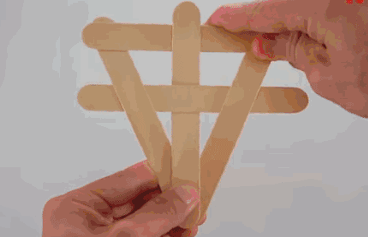}
        \caption{Nest is not a Stick Bomb}
        \label{fig:bomb1}
    \end{minipage}
\end{figure}

Birds choose simple elements from then environment and synthesize them into their nests (figure \ref{fig:natural bird nest}). However, the mystery behind the bird nest's construction is still not well understood \cite{weiner_2020_mechanics}. Reference \cite{roberts_2020_why}
simplifies a nest to random packing of sticks (slender grains) and hypothesizes that ``the nest state" results from the ``jamming" of its elements \cite{liu2010jamming} that prevent them from falling apart. Inspired by these works, we will study the mechanics of a nest structure made of bamboo sticks and try to answer how and why birds can build a stable structure by merely packing.

\begin{figure}[H]
    \centering
    \begin{minipage}{0.5\textwidth}
        \centering
        \includegraphics[width=0.8\linewidth]{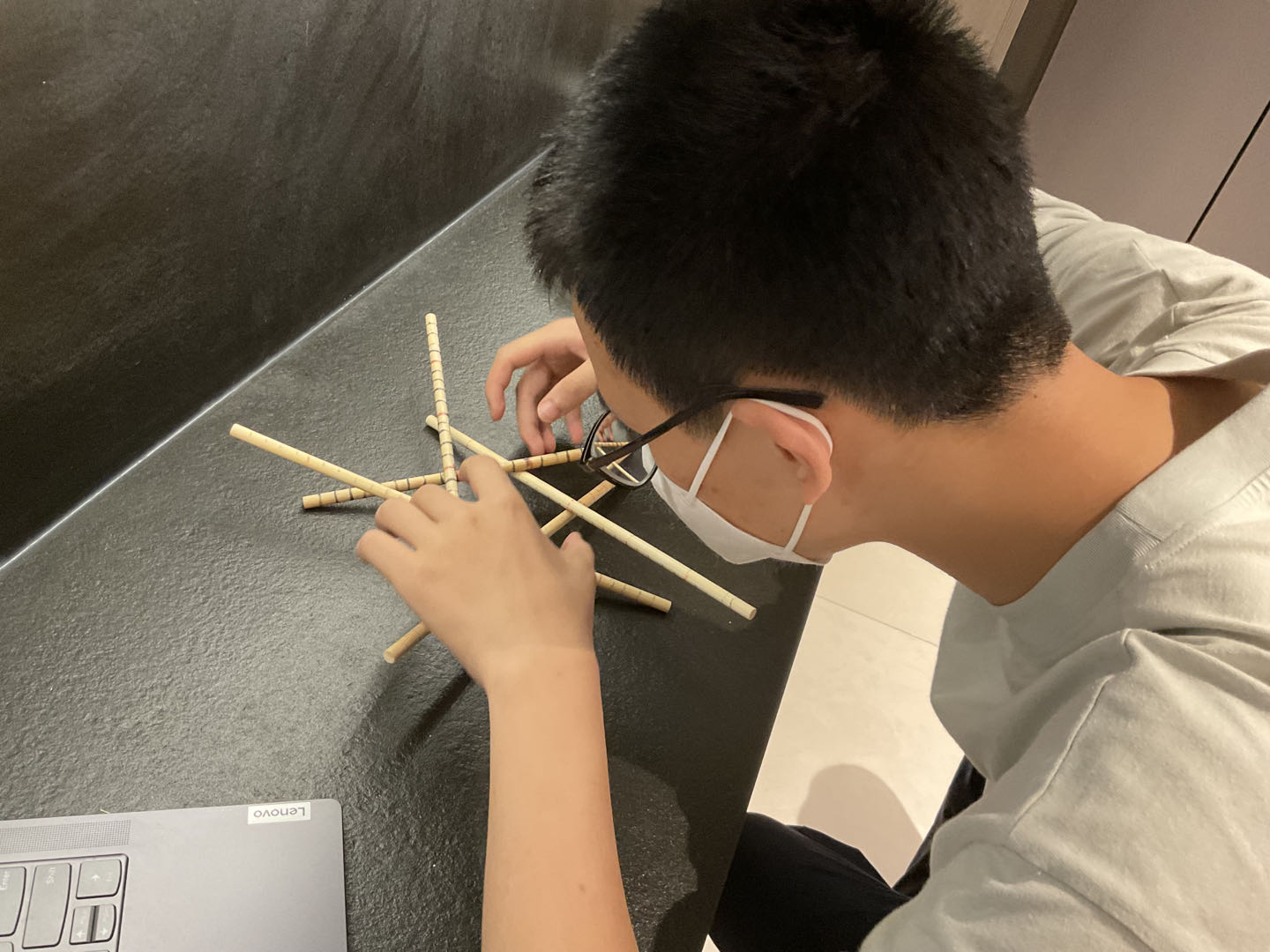}
        \caption{Build the Structure}
        \label{fig:build}
    \end{minipage}%
    \begin{minipage}{0.5\textwidth}
        \centering
        \includegraphics[width=0.8\linewidth]{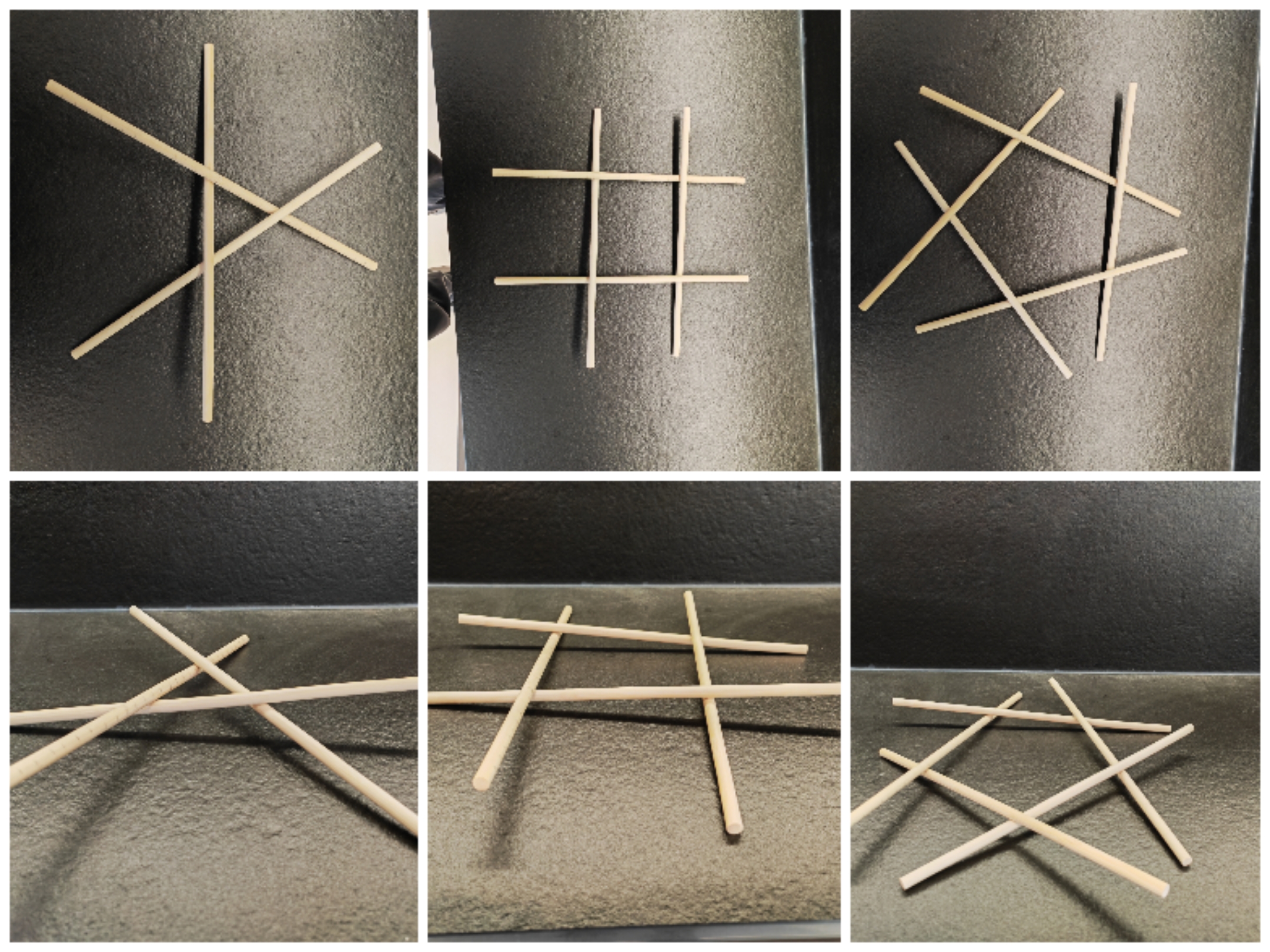}
        \caption{Rod Structures}
        \label{fig:rod_structures}
    \end{minipage}
\end{figure}

We will not regard a nest as a stick bomb \cite{icebomb2020} (explodes and releases the elastic potential energy when falls to the ground), as shown in figure \ref{fig:bomb1}, with stored elastic potential energy to keep itself stable. In a stick bomb, it is the external force that causes the elastic deformation which produces the normal forces and produces the friction to hold the whole structure. In this aspect, we suppose that the normal forces inside a nest arise to resist its own gravity and balance the structure. 
Neither will we use a container to hold the random sticks together as done in \cite{yashrajbhosale_2022_micromechanical}. Like a real bird does, we try to obtain a stable structure that can stand by itself by merely packing. Then, we find such a self-supporting structure with only a few sticks (see figure \ref{fig:build} and \ref{fig:rod_structures}),  where the normal force inside a nest arises to resist its own gravity. To build a practical bird nest, our structure may act as the base of the nest that supports the weight of a bunch of filling sticks and the birds and eggs.

Our study focuses on the stability of this self-supporting nest structure. First of all, we derive the stability conditions and obtain a phase diagram of a free nest structure by force analysis and verify the phase diagram by experiment. Then we test the nest structure's stability under applied weight. Finally, we test the nest structure's stability under vibration and applied weight. 
Ultimately, this paper aims to explore new possibilities of the potential application of the nest structure in architecture, packaging, and other fields of industry. By analyzing its properties, we aim to pave the way for more sustainable and resilient materials.

\newpage
\section{Static Stability of Free Bird Nest Structure}

Some structures are able to stand steadily whereas others collapse or slide down to a stable level. To derive the physical conditions that allow rods to stand stable, we analyze the forces and torques in this system. First, we theoretically analyze the forces in the system and mathematically derive the conditions. Then we will verify the conditions through experiments. 

\subsection{Theory}
\subsubsection{Variables}
\begin{figure}[H]
  \begin{minipage}{0.5\textwidth}
    \small
    \caption{Variable List}
    \begin{tabular}{ll}
        Symbol & Description \\
        \hline
        $L$ & Length of the Rod \\
        $d_1, d_2, d_3$ & Length of lower, middle, and upper segment \\
        $m$ & Mass of the rod \\
        $t$ & Thickness (diameter) of rod \\
        $n$ & Number of rods \\
        $\theta$ & Angle the between rods and the ground \\
        $\alpha$ & Interior Angle of a regular n-gon \\
        $\mu$ & Friction between rods \\
        $\mu'$ & Friction between rods and table \\
        $h$ & Height of rods \\
    \end{tabular}
  \end{minipage}%
  \begin{minipage}{0.5\textwidth}
    \centering
    \includegraphics[width=0.9\textwidth]{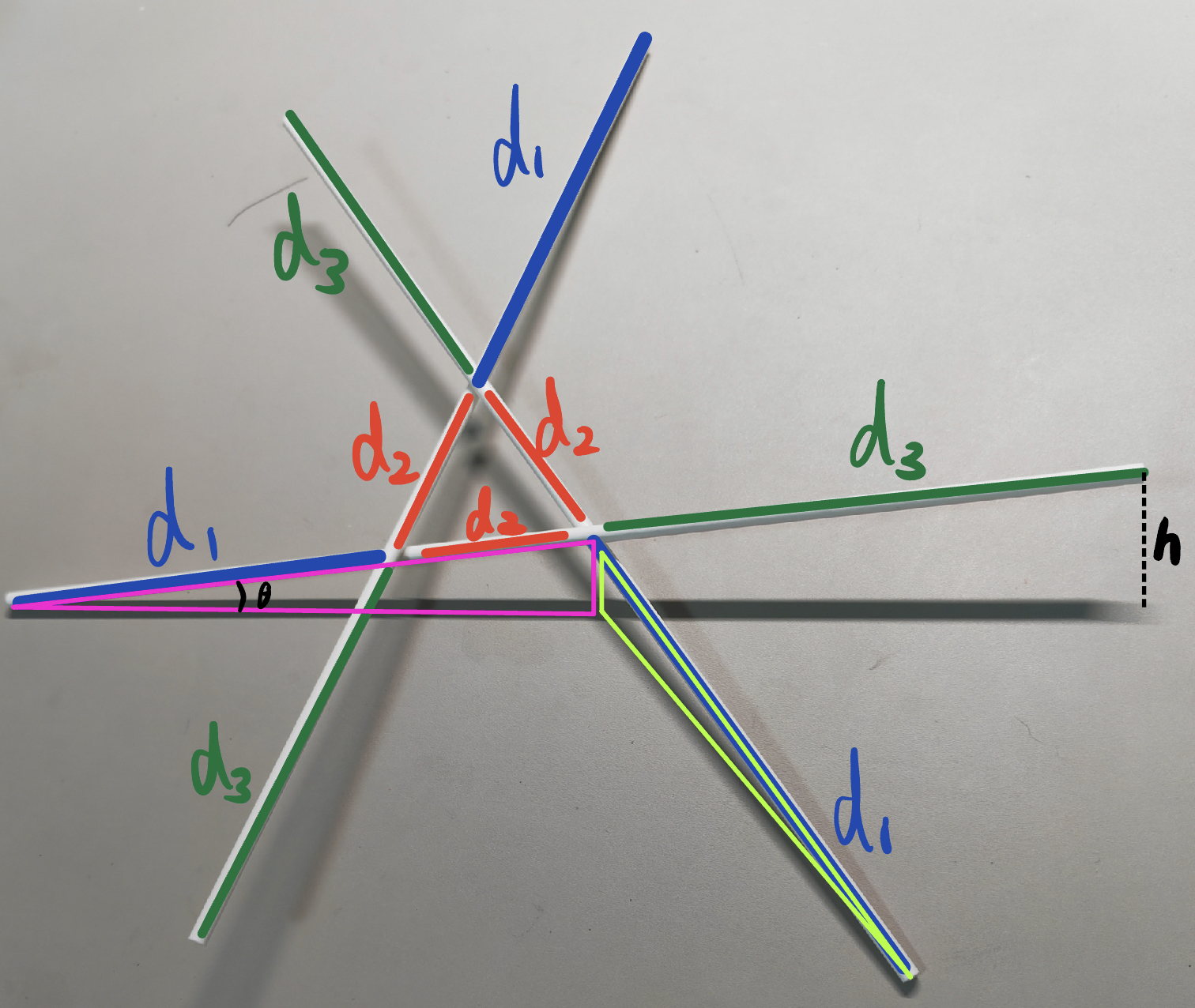}
    \caption{Visual Representation of Variables}
    \label{Visual Representation of Variables}
  \end{minipage}
\end{figure}

\subsubsection{Angle with Ground}
In figure \ref{Visual Representation of Variables}, the angle $\theta$ each rod makes with the ground can be determined geometrically. The height of the pink triangle on the left is $(d_1+d_2)\sin\theta$, and the height of the green triangle on the right is $d_1\sin\theta$. The two heights differ by the diameter of the rod, $t$. 
\begin{align}
    d_1\sin\theta+d_2\sin\theta &= d_1\sin\theta+t, \\
    \sin\theta &= \frac{t}{d_2}.
\end{align}
Thus, the height of a structure $h$ is 
\begin{equation}
    h=L\sin\theta=L\frac{t}{d2}.
\end{equation}

\subsubsection{Force and Torque Analysis}
There are 7 forces in total on the rod: 3 pairs of normal forces and friction forces, along with gravity. If the structure is in equilibrium, forces and torques are balanced for each rod. To simplify, we only investigate rod structures that are symmetrical.
\begin{figure}[H]
  \centering
  \includegraphics[width=0.5\textwidth]{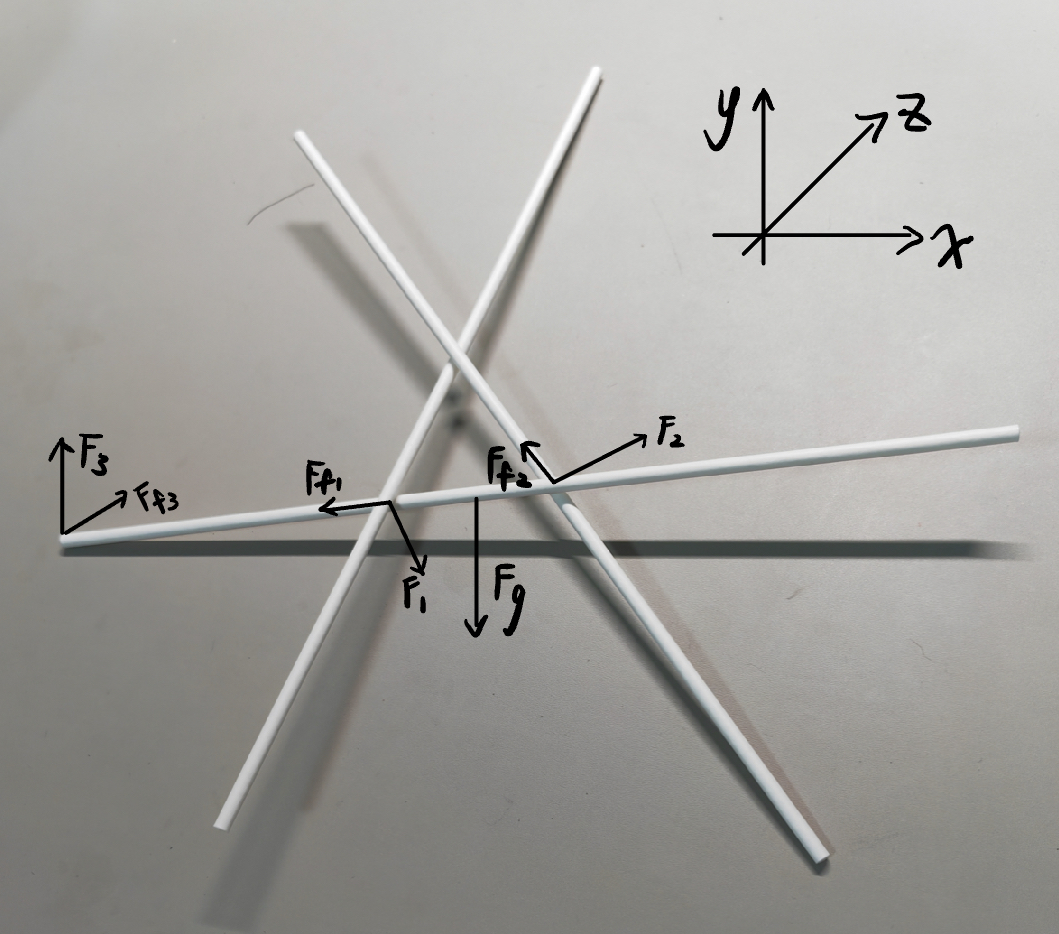}
  \caption{Free Body Diagram}
  \label{Free Body Diagram}
\end{figure}

The weights of the rods are distributed evenly across the three contact points with the ground. Thus, $F_3=mg$ and $F_{f3}\leq \mu'mg$. As the structure is symmetric, $F_1=F_2$ and $F_{f1}=F_{f2}$.\\

We establish a 3-D coordinate system. The x-z plane is the table. $+x$ is from left to right, and $+z$ is from closer to further. $y$ is the height from the ground. We then represent the forces as vectors. 
\begin{align}\label{force_vector_free}
&\begin{array}{llcl}
\vec{F}_3 & = & \begin{bmatrix} 0 & mg & 0 \end{bmatrix}, \\
\vec{F}_{f3} & = & \begin{bmatrix} F_{f3x} & 0 & F_{f3z} \end{bmatrix}, \\
\vec{F}_g & = & \begin{bmatrix} 0 & -mg & 0 \end{bmatrix},\\
\vec{F}_1 & = & \begin{bmatrix} -F_1\sin\theta\cos\alpha & -F_1\cos\theta & -F_1\sin\theta\sin\alpha \end{bmatrix}, \\
\vec{F}_{f1} & = & \begin{bmatrix} -F_{f1}\cos\theta & -F_{f1}\sin\theta & 0 \end{bmatrix},\\
\vec{F}_2 & = & \begin{bmatrix} F_2\sin\theta\cos\alpha & F_2\cos\theta & -F_2\sin\theta\sin\alpha \end{bmatrix},\\
\vec{F}_{f2} & = & \begin{bmatrix} F_{f2}\cos\theta\cos\alpha & F_{f2}\sin\theta & -F_{f2}\cos\theta\sin\alpha \end{bmatrix}.
\end{array}
\end{align}

Take the contact point to the ground as the center of rotation. The radius of each force can be represented as the following vectors: 
\begin{align}\label{radius_vector_free}
&\begin{array}{lcl}
\vec{R}_{F3} & = & \vec{R}_{Ff_3} =\begin{bmatrix} 0 & 0 & 0 \end{bmatrix},\\
\vec{R}_{F1} & = & \vec{R}_{Ff_1}=\begin{bmatrix} d_1\cos\theta & d_1\sin\theta & 0 \end{bmatrix},\\
\vec{R}_{Ff2} & = & \vec{R}_{Ff2}=\begin{bmatrix} (d_1+d_2)\cos\theta & (d_1+d_2)\sin\theta & 0 \end{bmatrix},\\
\vec{R}_G & = & \begin{bmatrix} \frac{L}{2}\cos\theta & \frac{L}{2}\sin\theta & 0 \end{bmatrix}.
\end{array}
\end{align}

Forces and torques need to be balanced out for the structure to stay in equilibrium. By $\Sigma \vec{F}=\vec{0}$ and $\Sigma \vec{R}\times \vec{F}=\vec{0}$, we can set up the following system of equation. 
\begin{small}
\begin{empheq}[left={\empheqlbrace}]{align}
&- F_{1} \sin(\theta) \cos(\alpha) + F_{2} \sin(\theta) \cos(\alpha) - F_{f1} \cos(\theta) - F_{f2} \cos(\alpha) \cos(\theta) + F_{f3x} = 0, \\
&- F_{1} \cos(\theta) + F_{2} \cos(\theta) - F_{f1} \sin(\theta) + F_{f2} \sin(\theta) = 0, \\
&- F_{1} \sin(\alpha) \sin(\theta) - F_{2} \sin(\alpha) \sin(\theta) + F_{f2} \sin(\alpha) \cos(\theta) + F_{f3z} = 0, \\
&F_{1} d_{1} \sin(\alpha) \sin^2(\theta) + F_{2} (d_{1} + d_{2}) \sin(\alpha) \sin^2(\theta) - F_{f2} (d_{1} + d_{2}) \sin(\alpha) \sin(\theta) \cos(\theta) = 0, \\
&- F_{1} d_{1} \sin(\alpha) \sin(\theta) \cos(\theta) - F_{2} (d_{1} + d_{2}) \sin(\alpha) \sin(\theta) \cos(\theta) + F_{f2} (d_{1} + d_{2}) \sin(\alpha) \cos^2(\theta) = 0, \\
\begin{split}
&- F_{1} d_{1} \sin^2(\theta) \cos(\alpha) + F_{1} d_{1} \cos^2(\theta) + F_{2} (d_{1} + d_{2}) \sin^2(\theta) \cos(\alpha) - F_{2} (d_{1} + d_{2}) \cos^2(\theta) \\
&- F_{f2} (d_{1} + d_{2}) \sin(\theta) \cos(\alpha) \cos(\theta) - F_{f2} (d_{1} + d_{2}) \sin(\theta) \cos(\theta) + \frac{L g m \cos(\theta)}{2} = 0.
\end{split}
\end{empheq}
\end{small}

The first three equations are derived from force equilibrium in the $x$, $y$, and $z$ direction respectively. The latter three equations are derived from torque equilibrium in the $x$, $y$, and $z$ directions respectively. With $F_1=F_2=F_N$ and $F_{f1}=F_{f2}=F_f$, we can then simplify the set of equations to the following: 
\begin{small}
\begin{empheq}[left={\empheqlbrace}]{align}
&F_{f3x}-F_f\cos\theta(1+\cos\alpha) = 0 \label{eq:forcex},\\
&F_{f3z}-2F_N\sin\theta\sin\alpha+F_f\cos\theta\sin\alpha = 0\label{eq:forcez},\\
&F_N\sin^2\theta\sin\alpha(2d_1+d_2)-F_f(d_1+d_2)\sin\theta\cos\theta\sin\alpha = 0 \label{eq:torquex},\\
&-F_N\sin\theta\cos\theta\sin\alpha(2d_1+d_2)+F_f(d_1+d_2)\cos^2\theta\sin\alpha = 0\label{eq:torquey},\\
\begin{split}
&F_Nd_1(\cos^2\theta-\sin^2\theta\cos\alpha) + F_N(d_1+d_2)(\sin^2\theta\cos\alpha-\cos^2\theta) \\
&- F_f(d_1+d_2)\sin\theta\cos\theta(1+\cos\alpha)+\frac{Lmg\cos\theta}{2} = 0.
\end{split}\label{eq:torquez}
\end{empheq}
\end{small}

We can derive the condition for $\mu$ by adding equation \eqref{eq:torquex} and equation \eqref{eq:torquey}.
\begin{equation}\label{eq:FNFfrelation}
    F_f=F_N\frac{2d_1+d_2}{d_1+d_2}\tan\theta.
\end{equation}

By $F_f\leq\mu F_N$, 
\begin{equation}\label{eq:mucondition}
    \mu \geq\frac{2d_1+d_2}{d_1+d_2}\frac{t}{\sqrt{d_2^2-t^2}}.
\end{equation}

Substitute  equation \eqref{eq:FNFfrelation} into equation \eqref{eq:torquez}, 
\begin{equation}\label{eq:deriveF_N}
    F_N=\frac{Lmg\cos\theta}{4d_1\sin^2\theta(1+\cos\alpha)+2d_2}.
\end{equation}

Substitute equation \eqref{eq:deriveF_N} into equation \eqref{eq:FNFfrelation},
\begin{equation}\label{eq:Ff_final}
    F_f=\frac{Lmg\sin\theta(2d_1+d_2)}{(4d_1\sin^2\theta(1+\cos\alpha)+2d_2)(d_1+d_2)}.
\end{equation}

Substitute equation \eqref{eq:FNFfrelation} into equation \eqref{eq:forcex} and equation \eqref{eq:forcez},
\begin{equation}\label{eq:Ff3x}
    F_{f3x}=F_N\frac{(2d_1+d_2)\sin\theta(1+\cos\alpha)}{d_1+d_2},
\end{equation}
\begin{equation}\label{eq:Ff3z}
    F_{f3z}=F_N\frac{d_2\sin\theta\sin\alpha}{d_1+d_2}.
\end{equation}

Combining the components of $F_{f3}$, we can derive the condition,
\begin{equation}\label{eq:mu'condition}
    \mu'^2 \geq \frac{L^2 (d_2^2 \sin^2 \alpha + (2d_1 + d_2)^2 (1 + \cos \alpha)^2) \sin^2 \theta \cos^2 \theta}{4 (d_1 + d_2)^2 (2d_1 (1 + \cos \alpha) \sin^2 \theta + d_2)^2}
\end{equation}

Equation \eqref{eq:mucondition} is the condition for static friction coefficient between rods $\mu$, and equation \eqref{eq:mu'condition} is the condition for static friction coefficient between rods and the table $\mu'$. When $\mu'$ is big enough, the condition for $\mu$ would be the limiting factor, and the stability of the structure will depend on $d_1$ and $d_2$ only.
\subsubsection{Energy Analysis}
We attempt to analyze the rods structure's behavior through potential energy level. Consider the following structure. The rod on the left is supported by the rod on the right. Their point of contact with the ground is fixed, so what distance of $x$ gives the maximum energy?
\begin{figure}[H]
    \centering
    \includegraphics[width=0.5\textwidth]{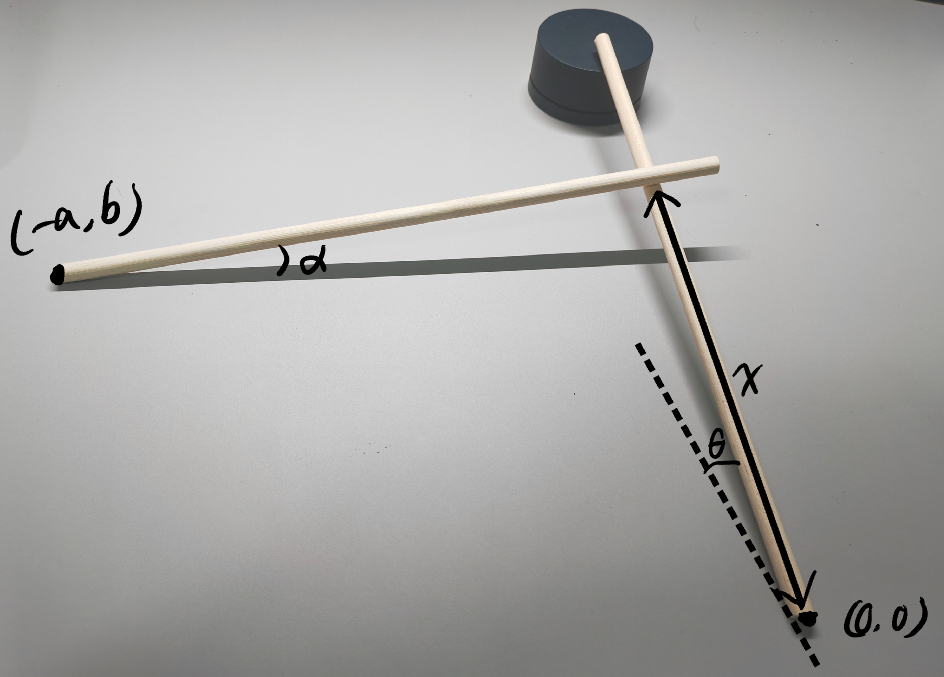}
    \caption{Two Rods Structure and Variables}
    \label{fig:two_rods_energy}
\end{figure}
\begin{equation}
    \tan\alpha=\frac{x\sin\theta}{\sqrt{a^2+(x\cos\theta-b)^2}}.
\end{equation}

The potential energy of the rod A
\begin{equation}
    E\propto y_{o_a}=\frac{L}{2}\sin\alpha=\frac{L}{2}\frac{x\sin\theta}{\sqrt{a^2+b^2+x^2-2b\cos\theta x}}.
\end{equation}

We solve for extremes by taking the derivative
\begin{equation}
    \frac{dy_{o_a}}{dx}=\frac{{L g m (a^2 + b^2 - b x \cos \theta) \sin \theta}}{{2 (a^2 + b^2 - 2 b x \cos \theta + x^2)^{\frac{3}{2}}}}.
\end{equation}

One local extrema exists at
\begin{equation}
    x=\frac{a^2+b^2}{b\cos\theta}.
\end{equation}

\begin{figure}[H]
    \centering
    \includegraphics[width=0.8\textwidth]{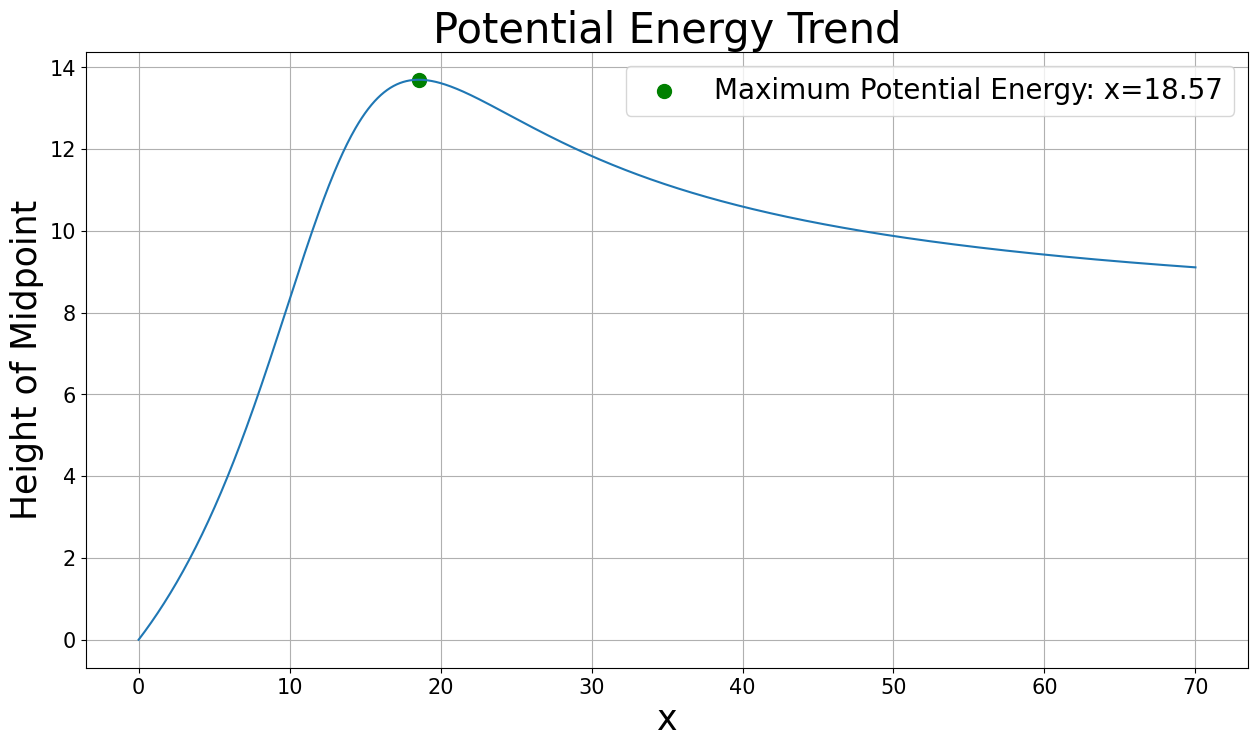}
    \caption{Potential Energy Trend of Rod A, $L=30$, $a=5$, $b=15$, $\theta=\frac{\pi}{6}$}
    \label{fig:potential energy}
\end{figure}

Figure \ref{fig:potential energy} shows the potential energy of rod A as a function of $x$, the distance from the intersection to the landing point of rod B. There is one global maxima. The system's potential energy will drop rapidly as $x$ decreases, and will drop quickly as $x$ increases.
\subsection{Experiments}
\subsubsection{Measurement of Friction Coefficient}
To measure the coefficient between rods, we first tape two rods to a flat cardboard, creating a track on which another rod can slide down as we incline the board. 
\begin{figure}[H]
    \centering
    \begin{minipage}{0.45\textwidth}
        \centering
        \includegraphics[width=0.9\linewidth]{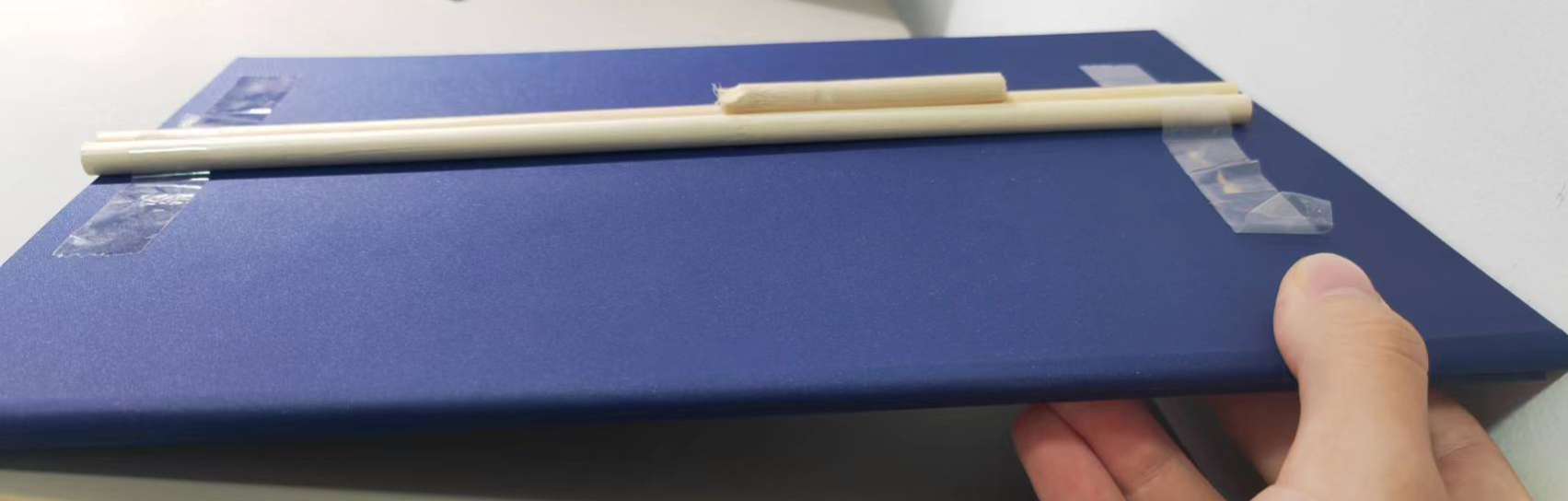}
        \caption{Friction Coefficient Measurement}
        \label{fig:friction_coefficient}
    \end{minipage}\hfill
    \begin{minipage}{0.45\textwidth}
        \centering
        \includegraphics[width=0.9\linewidth]{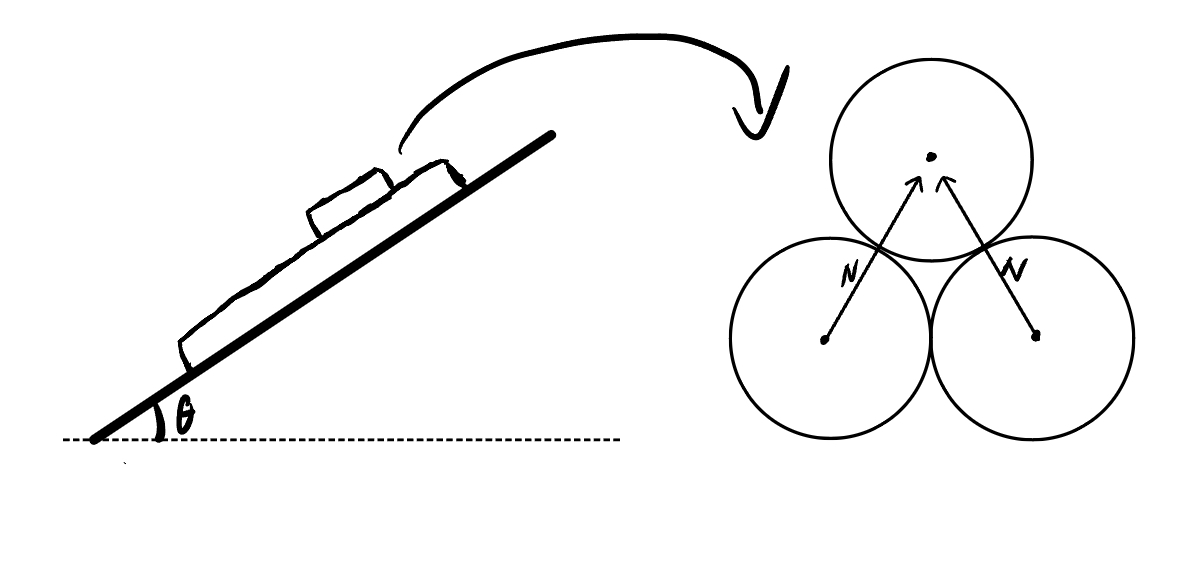}
        \caption{Free Body Diagram}
        \label{fig:fbd_friction}
    \end{minipage}
\end{figure}

The sum of friction forces is $2N\mu$ where $N$ is the normal force from one bottom rod and is equal to a component of gravity along the plane. Thus, we can calculate $\mu$ by the threshold incline angle $\theta$ using the below equations

\begin{equation}
    \begin{aligned}
        2N\mu&=mg\sin\theta,\\
        2N\cos 30&=mg\cos\theta,\\
        \mu&=\frac{\sqrt{3}}{2}\tan\theta.
    \end{aligned}
\end{equation}

\subsubsection{Experiment Results}
We test our theoretical prediction with structures of different materials, lengths, diameters, and number of rods. We vary the combination of $d_1$ and $d_2$ and check whether the structure can self-support. We then compare the experiment results to theoretical prediction. Under sufficient $\mu'$, we focus only on the condition for $\mu$. Structures with $d_1$ and $d_2$ above the curve are stable whereas those below is unstable.
\begin{figure}[H]
  \centering
  \begin{subfigure}[b]{0.45\textwidth}
    \includegraphics[width=\textwidth]{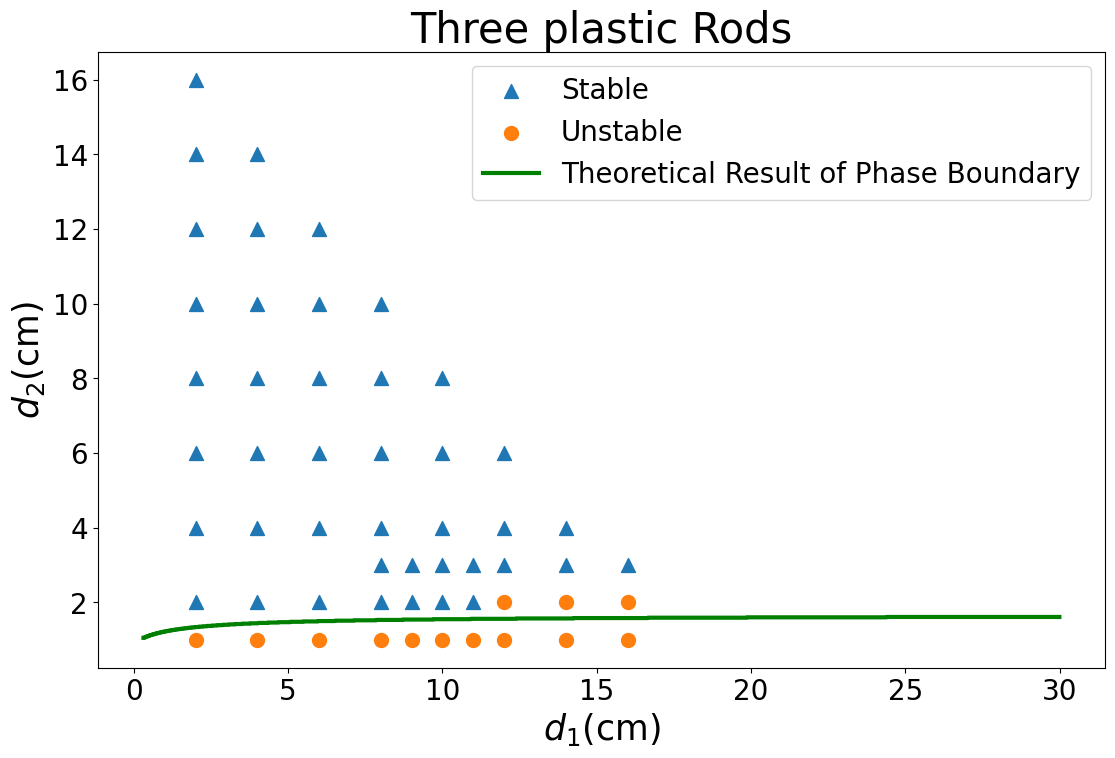}
    \caption{Case 1: Three Plastic Rods}
    \label{fig:3plastic}
  \end{subfigure}
  \hfill
  \begin{subfigure}[b]{0.45\textwidth}
    \includegraphics[width=\textwidth]{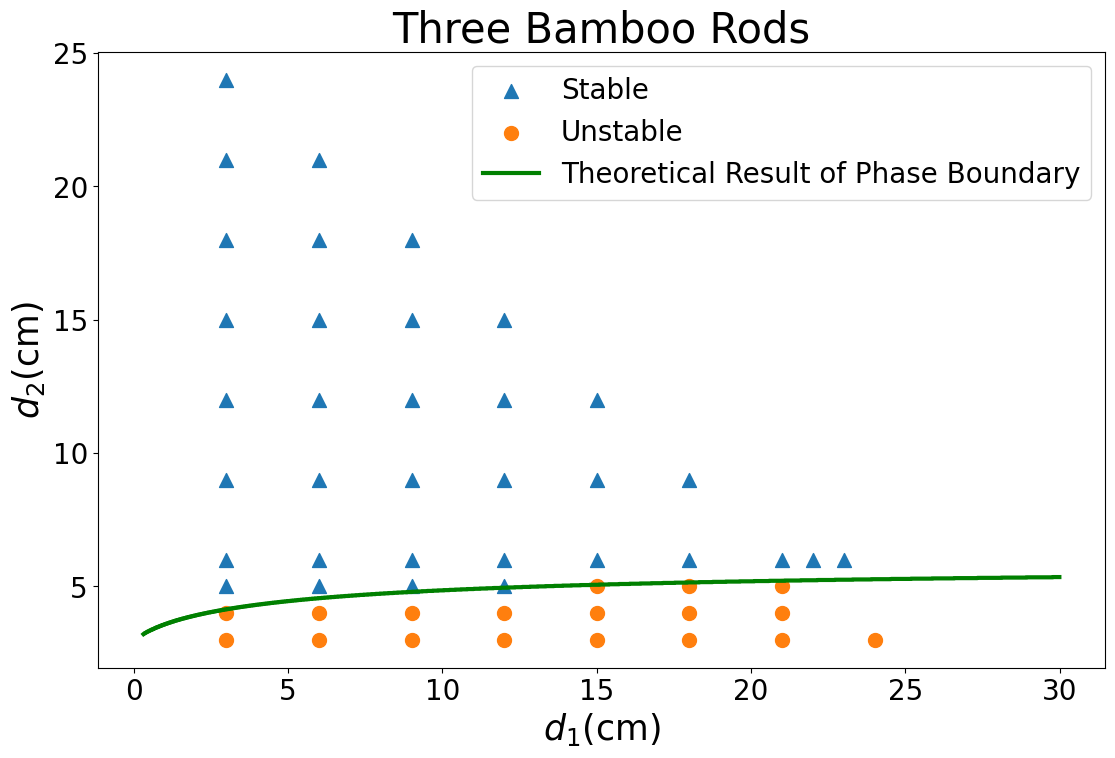}
    \caption{Case 2: Three Bamboo Rods}
    \label{fig:3bamboo}
  \end{subfigure}

  \vspace{10pt} 

  \begin{subfigure}[b]{0.45\textwidth}
    \centering
    \includegraphics[width=\textwidth]{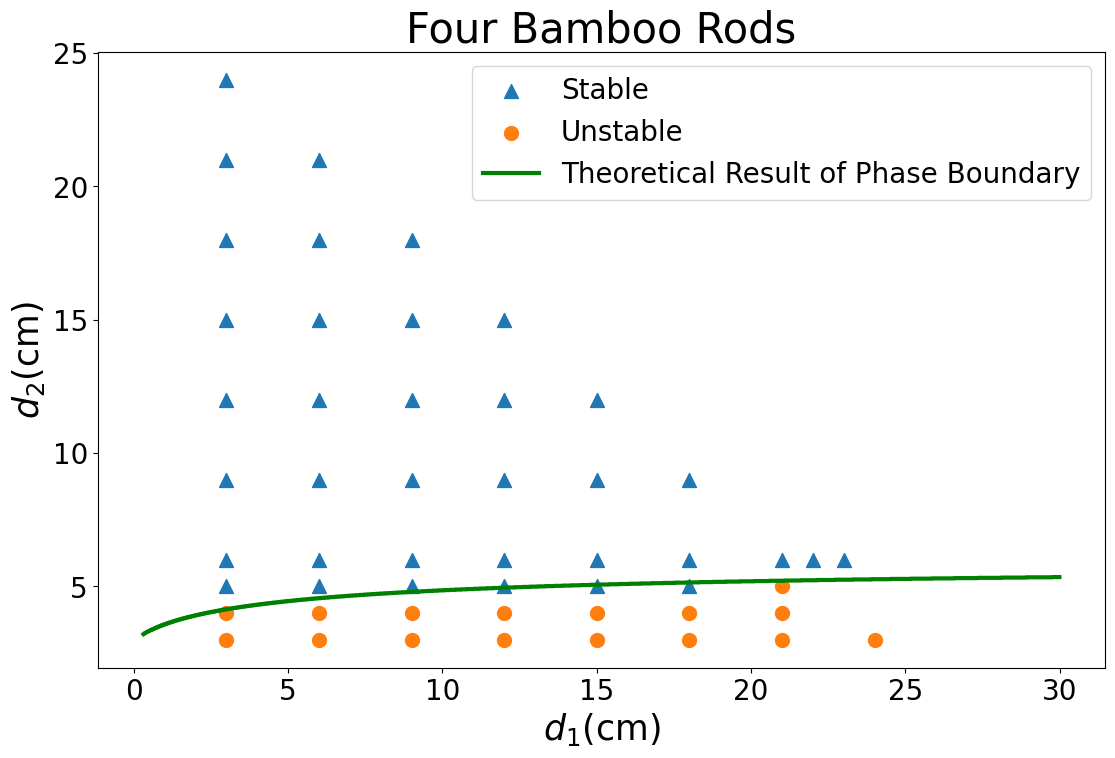}
    \caption{Case 3: Four Bamboo Rods}
    \label{fig:4bamboo}
  \end{subfigure}
  \hfill
  \begin{minipage}[b]{0.45\textwidth}
    \centering
    \renewcommand{\arraystretch}{2.5}
    \begin{tabular}{|p{0.15\textwidth}|p{0.17\textwidth}|p{0.17\textwidth}|p{0.17\textwidth}|p{0.07\textwidth}|p{0.07\textwidth}|}
      \hline
      Test Case & Number of Rods & Material & Diameter (cm) & L (cm) & $\mu$ \\
      \hline
      Case 1 & 3 & Plastic & 0.30 & 20 & 0.36 \\
      Case 2 & 3 & Bamboo & 0.80 & 30 & 0.27 \\
      Case 3 & 4 & Bamboo & 0.80 & 30 & 0.27 \\
      \hline
    \end{tabular}
    \captionof{table}{Summary of Test Cases}
    \label{tab:test_cases}
  \end{minipage}
  \caption{Stability Diagram for Theoretical and Experiment Results }
  \label{fig:experiments_and_cases}
\end{figure}

The theoretical stability boundary is mainly determined by $\mu$ and $d_2$, see equation \eqref{eq:mucondition},  and is independent of mass and number of rods---the experiment result of three and four rods is almost the same.

Unstable experimental data may fall beyond the curve. The reason is that the stable structures near the boundary is rather weak and can easily be disturbed by very small disturbance.

In conclusion, the theoretical and experimental results matched. Theory better predicted the stability of bamboo rods than plastic rods. This is likely because the thin plastic rods deformed in shape, which affected the friction coefficient.

\newpage
\section{Static Stability of Bird Nest Structure under Applied Weight}
\subsection{Theory}

A stable structure should maintain its shape when holding weights. The objective of this section is to investigate the relationship between applied weight and structural deformation. We aim to quantify this ability by applying incremental loads on the top of rod structures and subsequently measuring the corresponding decrease in height. 

\begin{table}[h]
    \centering
    \begin{tabular}{|l|p{0.5\textwidth}|}
        \hline
        Variable & Description \\
        \hline
        \(d_2\) & The length of the middle segment of the rods. \\
        \(M\)   & Mass of the applied load. \\
        \(h\)   & Height of the structures.\\
        \(n\)   & Number of rods. \\
        \hline
    \end{tabular}
\end{table}

We introduce the new force from applied weight besides the force vectors in equation \eqref{force_vector_free}. $F_3$ changes alongside, restoring balance of forces.
\begin{equation}
    \begin{aligned}
        &F_a=\begin{bmatrix}
            0 & -\frac{Mg}{n} & 0
        \end{bmatrix},\\
        &F_3'=\begin{bmatrix}
            0 & mg+\frac{Mg}{n} & 0
        \end{bmatrix}.
    \end{aligned}
\end{equation}

We need to resolve the balance of torque. The radius of applied force is
\begin{equation}
    R_a=\begin{bmatrix}
        L\cos\theta & L\sin\theta & 0
    \end{bmatrix}.
\end{equation}

The new force doesn't affect condition for $\mu$ (Equation \eqref{eq:mucondition}), but would alter condition for $\mu'$. 
\begin{equation}
    F_N'=\frac{L(M/n+m/2)g\cos\theta}{2d_1\sin^2\theta(1+\cos\alpha)+d_2}
    \label{eq:new_fn},
\end{equation}
\begin{equation}
     \mu'^2 \geq \frac{L^2 (\frac{M/n+m/2}{M/n+m})^2 (d_2^2 \sin^2 \alpha + (2d_1 + d_2)^2 (1 + \cos \alpha)^2) \sin^2 \theta \cos^2 \theta}{(d_1 + d_2)^2 (2d_1 (1 + \cos \alpha) \sin^2 \theta + d_2)^2}
    \label{eq:new_mu}
\end{equation}

\subsection{Experiment Procedure}
\paragraph{Equipments}Same bamboo rods as the first experiment (cylindrical shape, length 30cm, diameter 8mm, dry and soaked overnight); Fifteen $\times$ 200g weights; an Acrylic Board.
\paragraph{Friction Coefficient with the Table}
We apply force along the rod and measure the minimum angle that the rod can make with the ground. \begin{figure}[H]
    \centering
    \begin{minipage}{0.5\textwidth}
        \centering
        \includegraphics[width=0.5\linewidth]{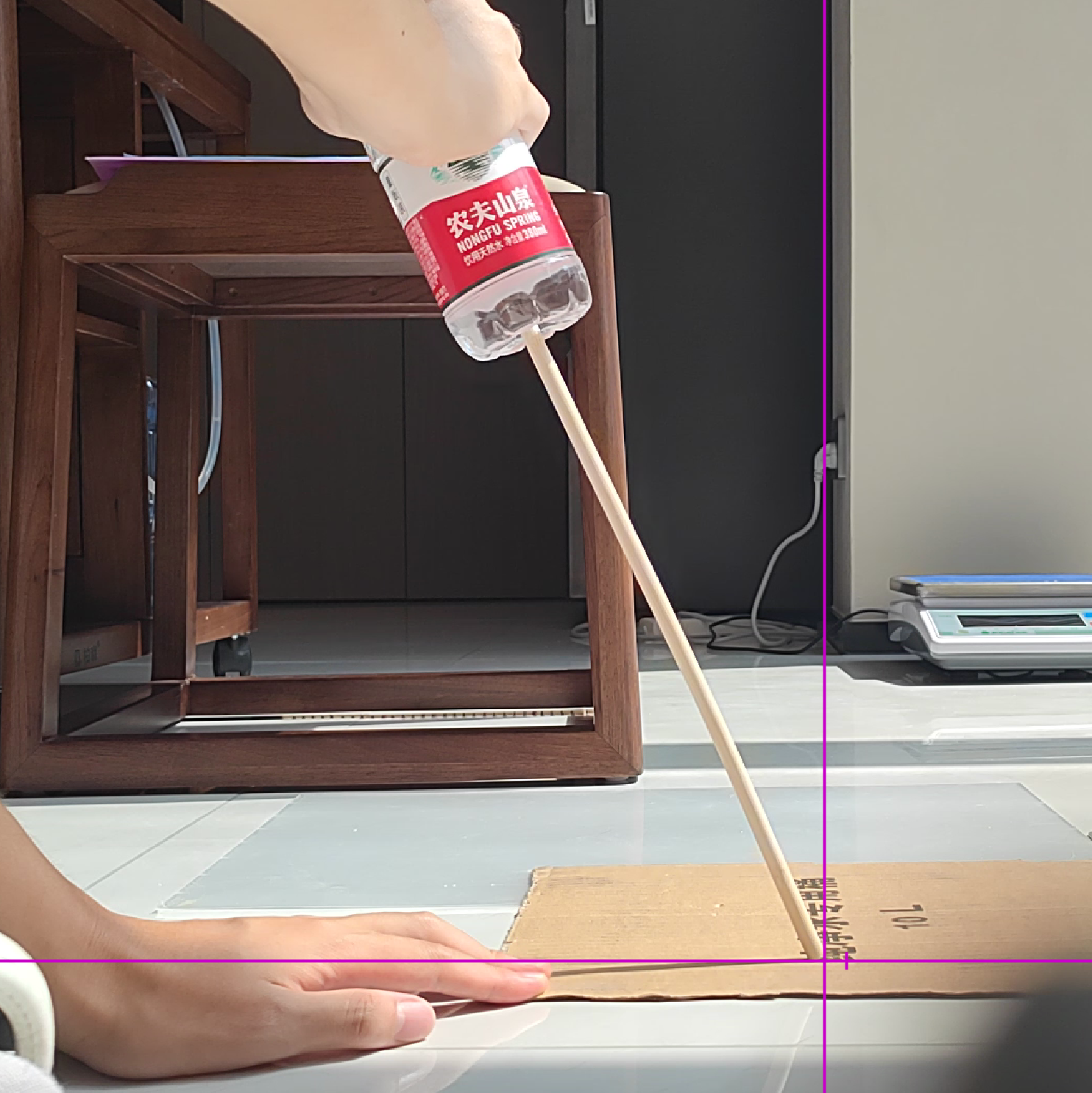}
        \caption{Friction Coefficient Measurement}
        \label{fig:mu'_measure}
    \end{minipage}%
    \begin{minipage}{0.5\textwidth}
        \centering
        \includegraphics[width=0.5\linewidth]{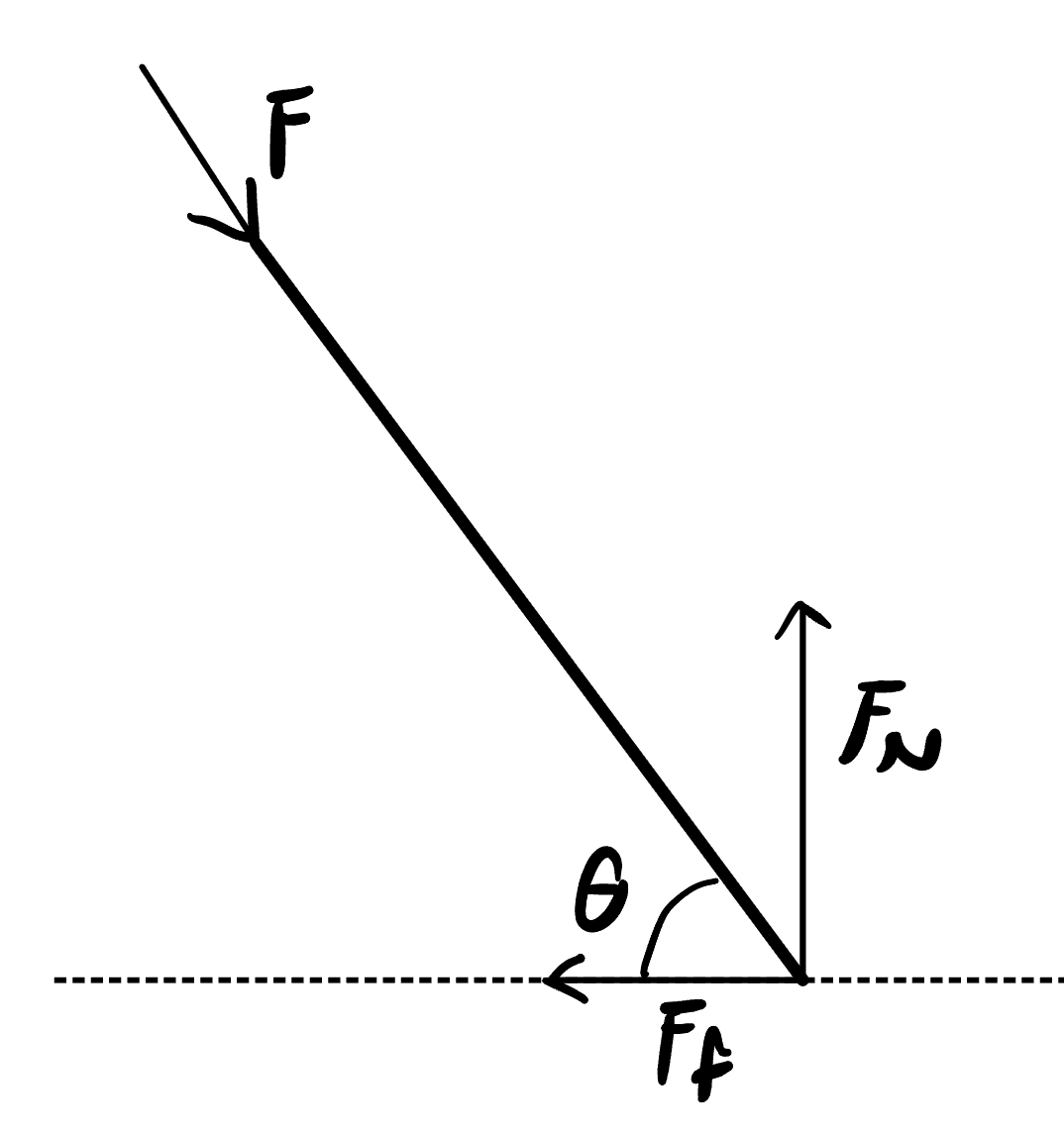}
        \caption{Free Body Diagram of Friction Measurement}
        \label{fig:fbd_friction}
    \end{minipage}
\end{figure}
We can derive $\mu'$ from the angle with ground.
\begin{equation}\label{eq:mu'_cal}
    \mu'=\frac{F_f}{F_N}=\frac{F \cos\theta}{F\sin\theta}=\cot\theta.
\end{equation}
\paragraph{Young's Modulus} We use the method introduced in \cite{song2016bending} to calculate the Young's Modules of dry and wet rods. We first place the rod on two thin blades and hang a weight load at the middle of the rod. Young's modules can be calculated as
\begin{equation}
    Y=\frac{\Delta mgl^3}{12\pi r^4 \Delta z}.
\end{equation}

\begin{figure}[H]
    \centering
    \includegraphics[width=0.5\textwidth]{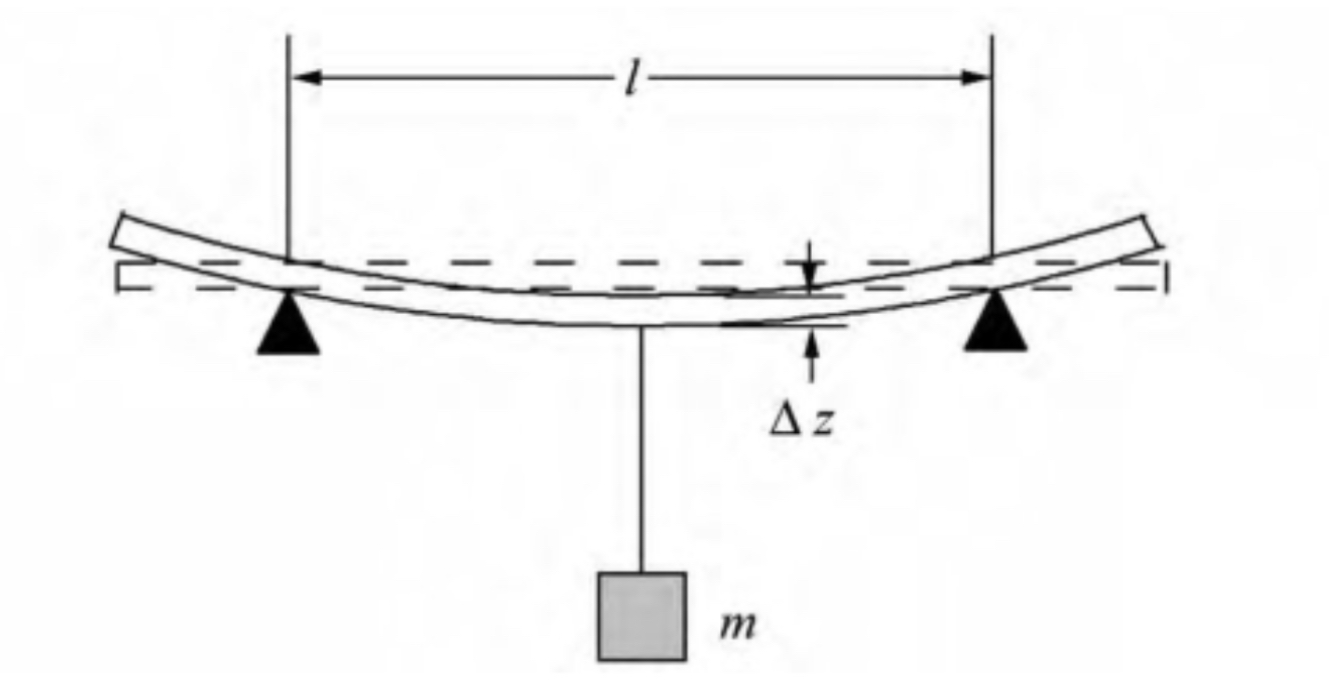}
    \caption{Young's Modules Measurement}
    \label{fig:young's modules}
\end{figure}
The parameters of the rods we use are shown in the below table. 

\begin{table}[H]
    \centering
    \begin{tabular}{|c|c|c|c|c|c|c|}
    \hline
    Type of Rod & \( r \) (cm) & \( \Delta m \) (kg) & \(m\) (g)& \( l \) (cm) & \( \Delta z \) (cm) & \( Y \) (\( \text{Pa} \)) \\
    \hline
    Wet Rods & 0.80 & 2.6 & 9.0 & 30.0 & 1.2 & \( 5.7 \times 10^9 \) \\
    \hline
    Dry Rods & 0.80 & 2.6 & 9.75 & 30.0 & 1.0 & \( 7.2 \times 10^9 \) \\
    \hline
    \end{tabular}
\end{table}

For three-rods, four-rods, and five-rods structures, we set $d_1=10$cm and vary $d_2$ from 5cm (structures with $d_2<5$ cannot stand on its own) to 13cm, increasing 2 centimeters at a time. We also soak rods into water overnight to observe the effect of moisture. 

After we construct the structure, we place an acrylic board onto the top, creating a platform that equally distribute weight to the tips of the rods. 

Then, we place 200g weight onto the acrylic platform one at a time while recording the new height. 
\begin{figure}[H]
    \centering
    \includegraphics[width=0.4\textwidth]{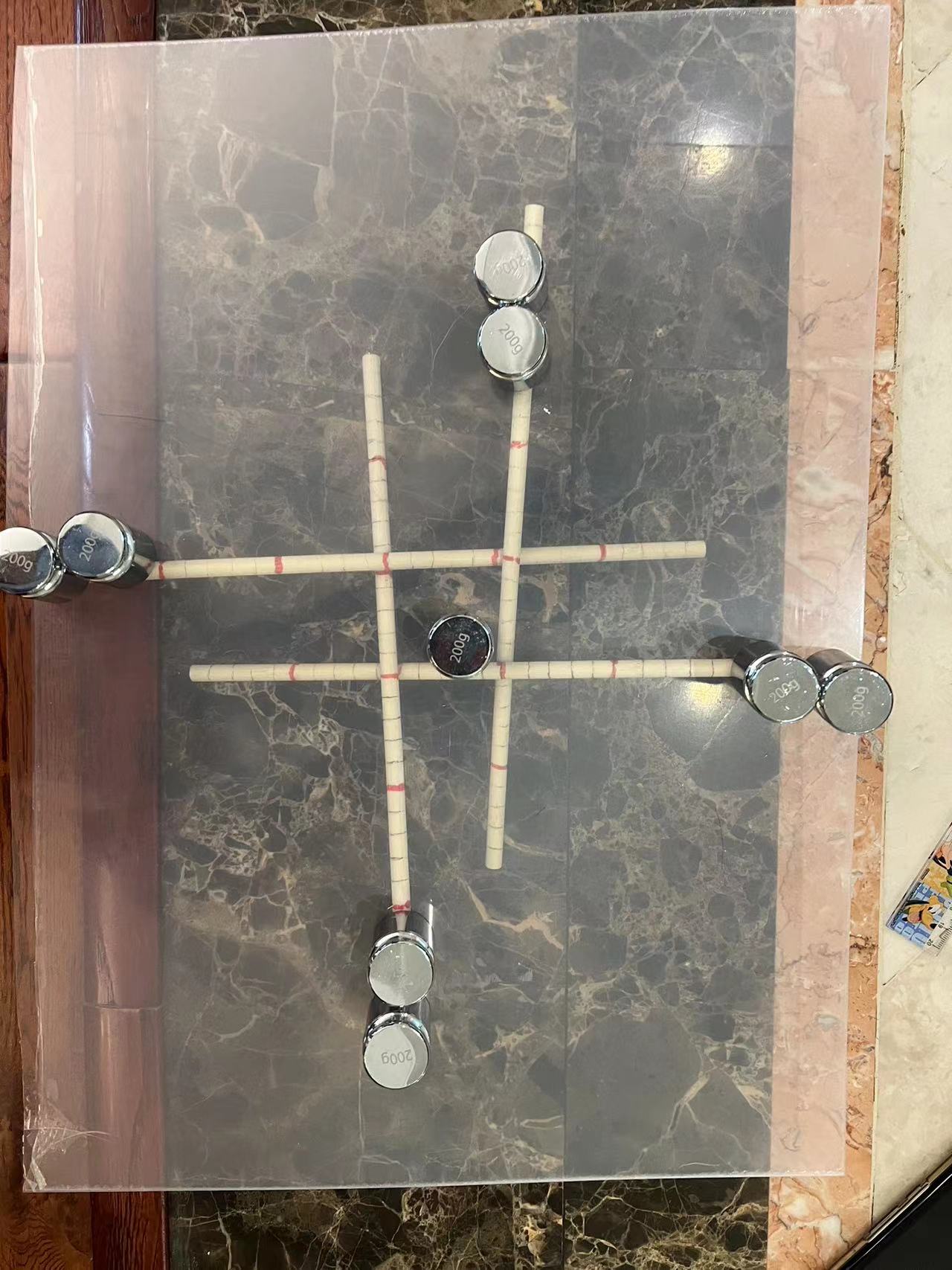}
    \caption{Applying Weight on Acrylic Platform}
    \label{fig:platform}
\end{figure}

\subsection{Experiment Results and Analysis}
The three figures below are experiment results for n=3, 4 and 5, obtained on June 4th 2023. 
\begin{figure}[H]
    \centering
    \begin{subfigure}{0.8\textwidth}
        \centering
        \includegraphics[width=\textwidth]{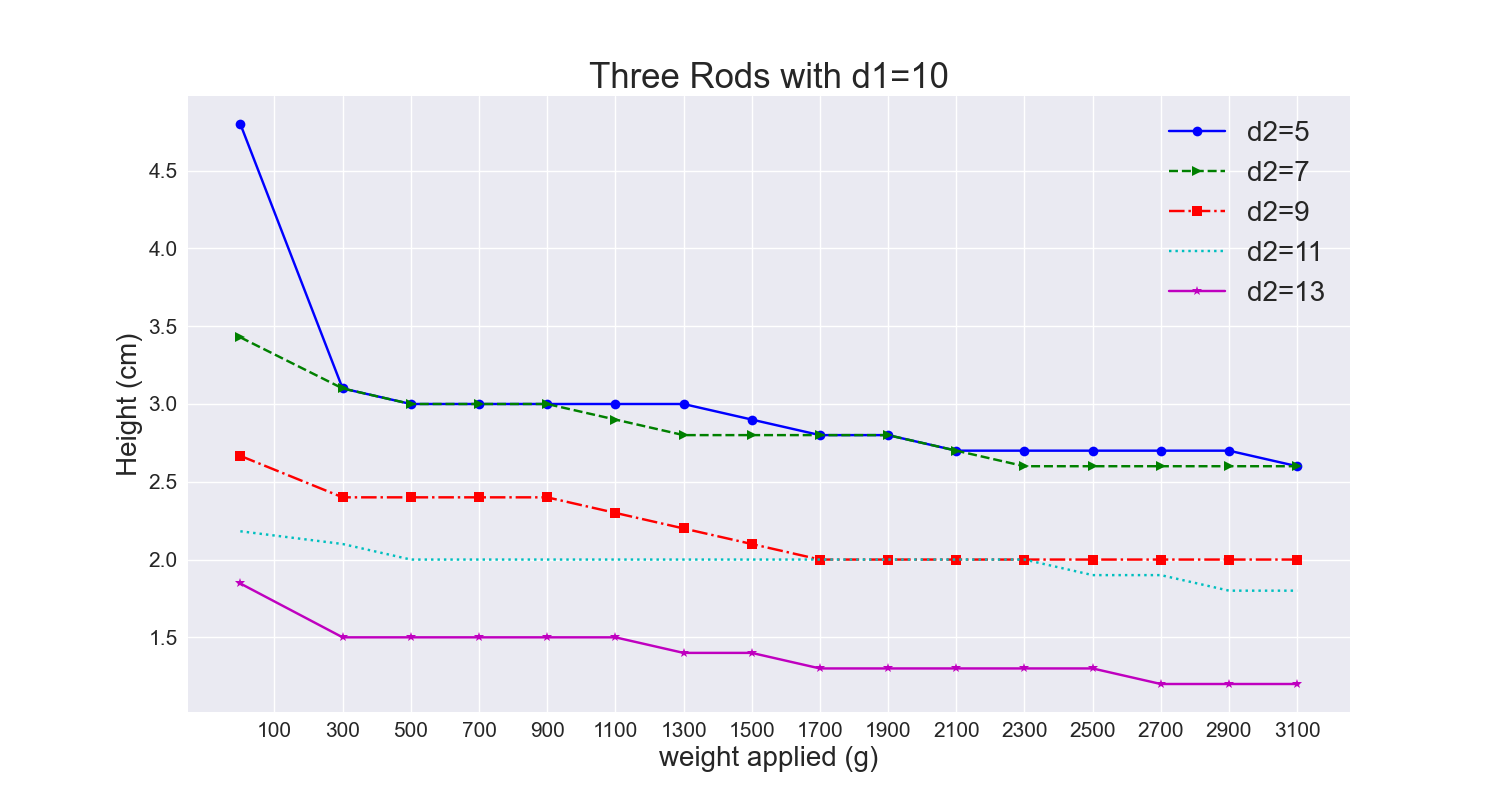}
        \caption{Three Dry Rods,experiment date June 4th}
        \label{fig:three}
    \end{subfigure}
    
    \begin{subfigure}{0.8\textwidth}
        \centering
        \includegraphics[width=\textwidth]{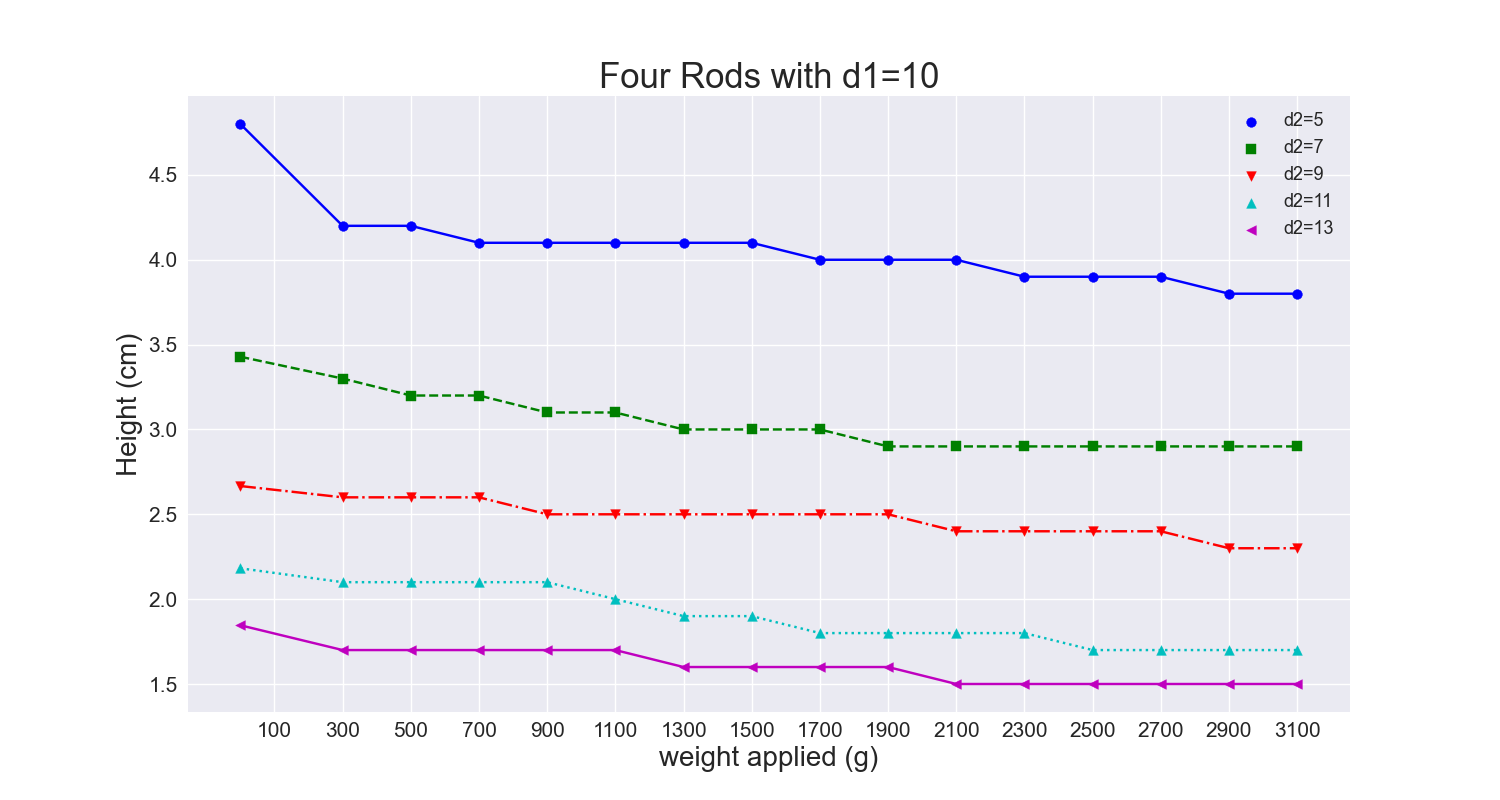}
        \caption{Four Dry Rods, experiment date June 4th}
        \label{fig:four}
    \end{subfigure}
    
    \begin{subfigure}{0.8\textwidth}
        \centering
        \includegraphics[width=\textwidth]{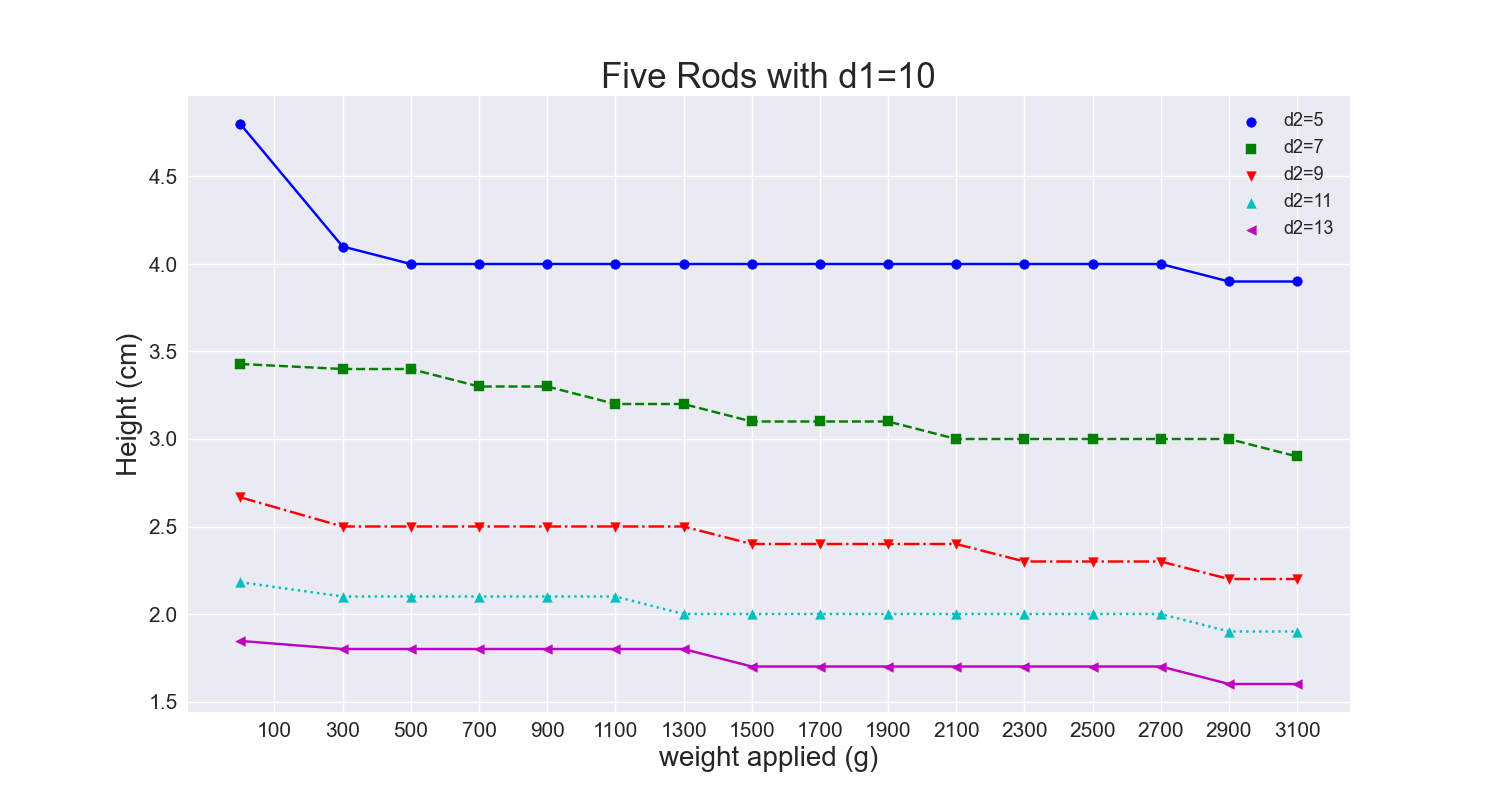}
        \caption{Five Dry Rods, experiment date June 4th}
        \label{fig:five}
    \end{subfigure}
    
    \caption{Height Change of Three, Four, and Five Rods Structures under Weight}
    \label{fig:pv_main}
\end{figure}
All the structures can withstand up to 60-100 times of their own weights, see figure \ref{fig:pv_main}. The structure drops at a rapid rate with the initial few weights. Then, it reaches a steady state in which the height decreases much slower with very small slope. Due to the limit of the measurement accuracy, the curve exhibits a decreasing behavior in a staircase manner.

For three rods structures, the curve with $d_2=5$ drops so fast initially that it collapses to the curve with $d_2=7$. Similarly, the curve with $d_2=9$ also collapses to the curve with $d_2=11$. This is not the case for four and five rods system, such that there are no crossings between curves and the rank by initial height is same as the rank by final height.  Therefore, the four and five rods systems are more stable and rigid than the 3 rod one.
\begin{figure}[H]
    \centering
    \begin{subfigure}{0.5\textwidth}
        \includegraphics[width=\textwidth]{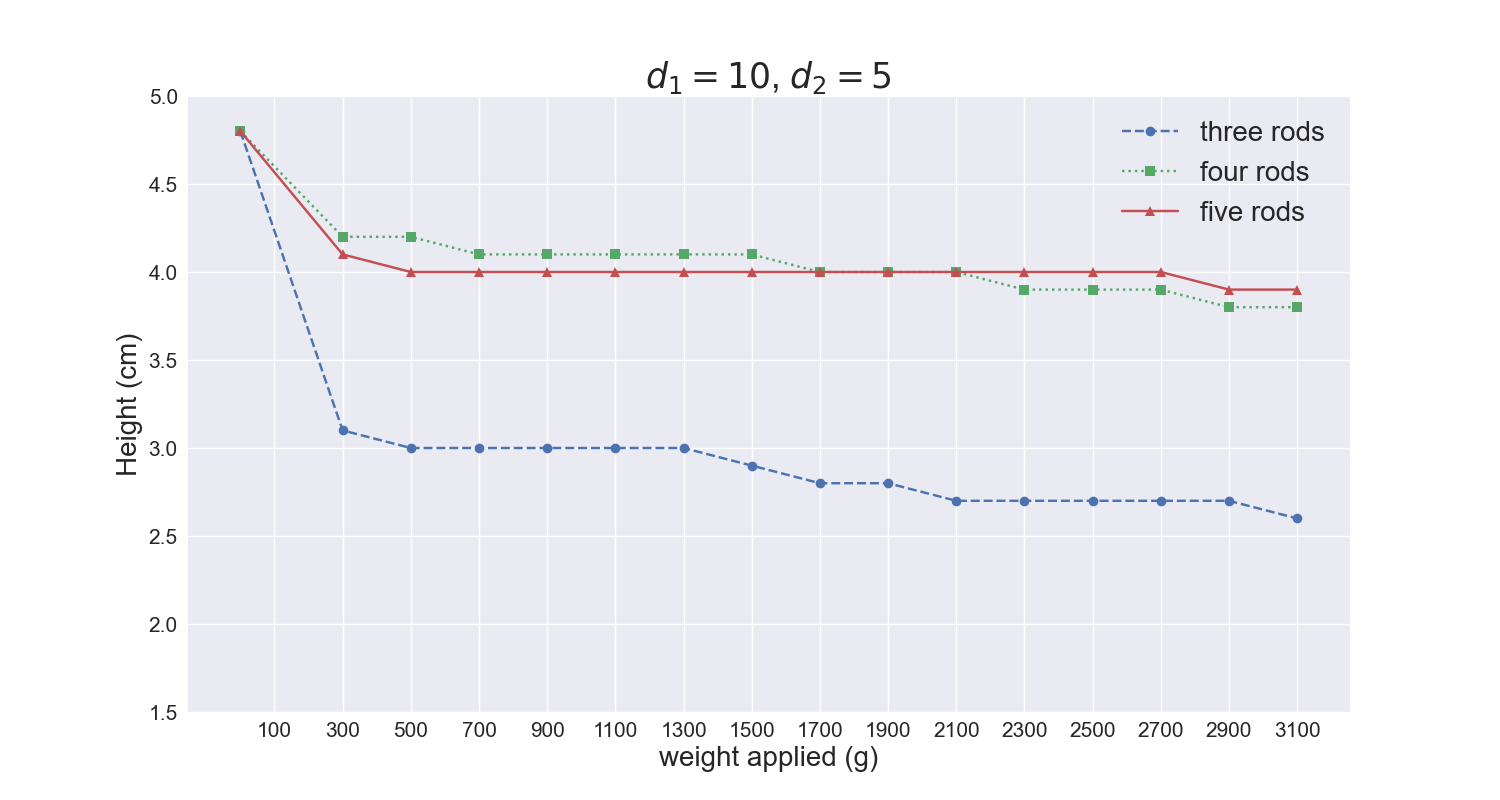}
        \caption{\(d_2=5\)}
        \label{fig:five}
    \end{subfigure}%
    \begin{subfigure}{0.5\textwidth}
        \includegraphics[width=\textwidth]{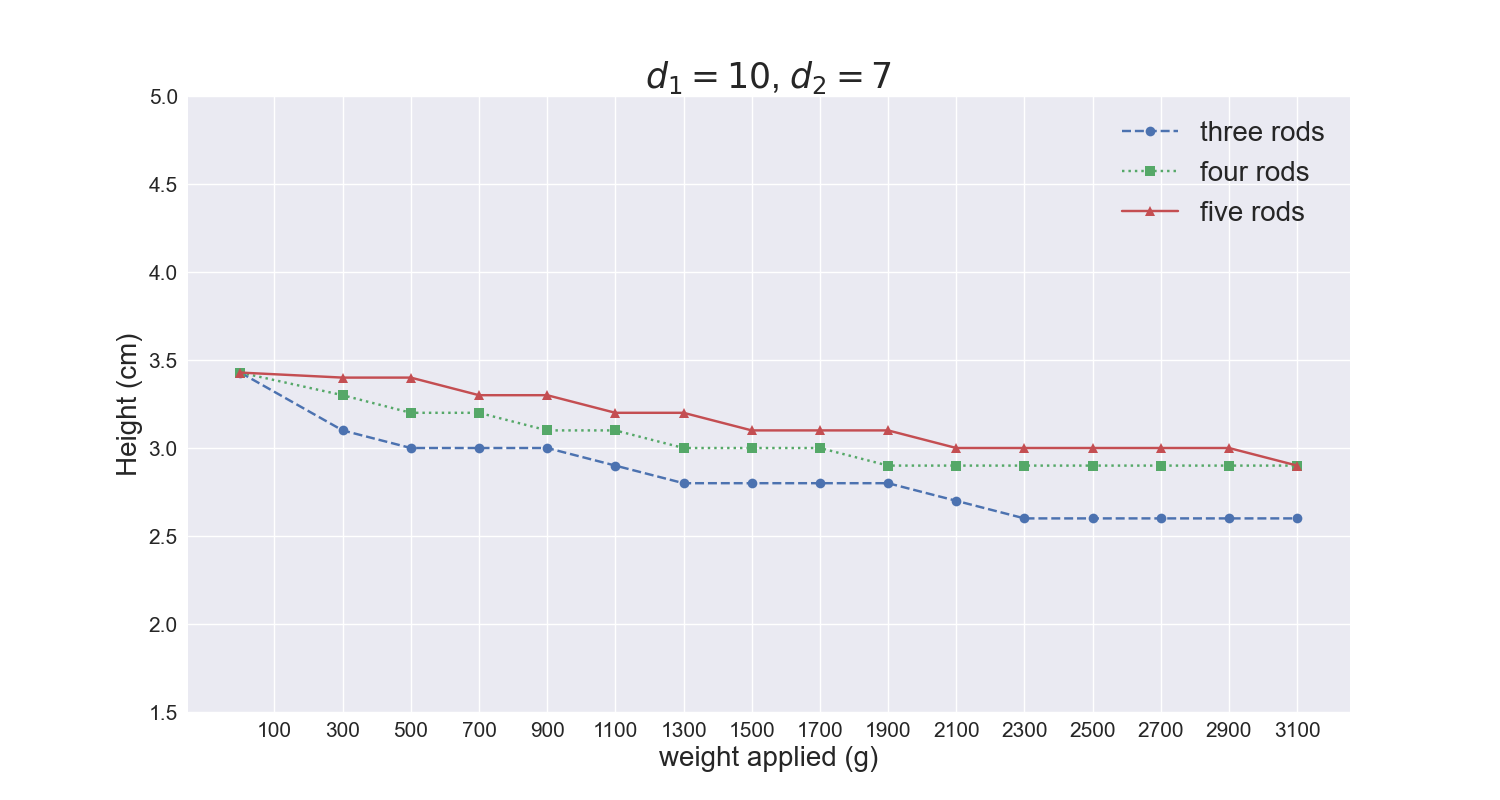}
        \caption{\(d_2=7\)}
        \label{fig:seven}
    \end{subfigure}
    \begin{subfigure}{0.5\textwidth}
        \includegraphics[width=\textwidth]{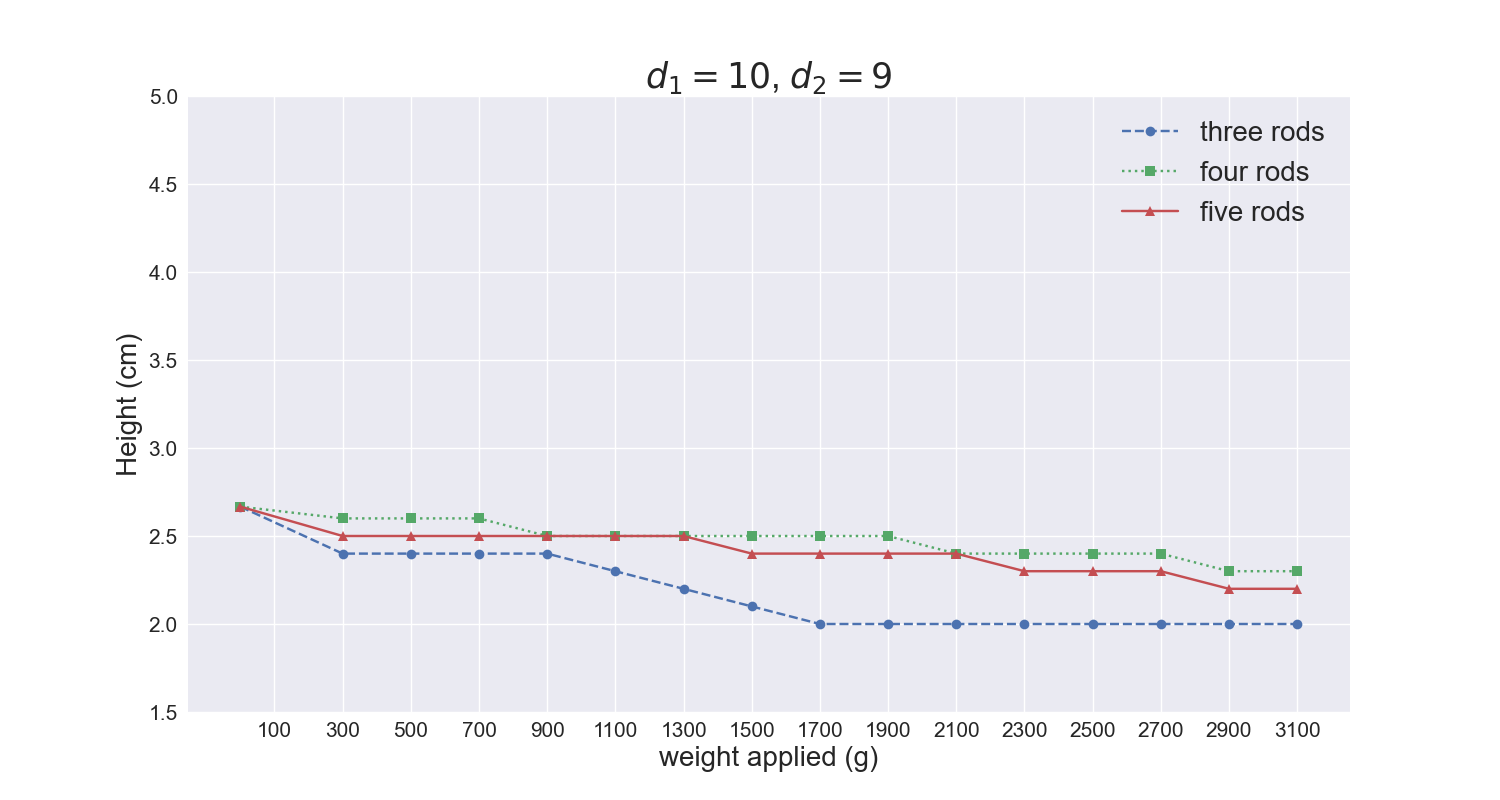}
        \caption{\(d_2=9\)}
        \label{fig:nine}
    \end{subfigure}%
    \begin{subfigure}{0.5\textwidth}
        \includegraphics[width=\textwidth]{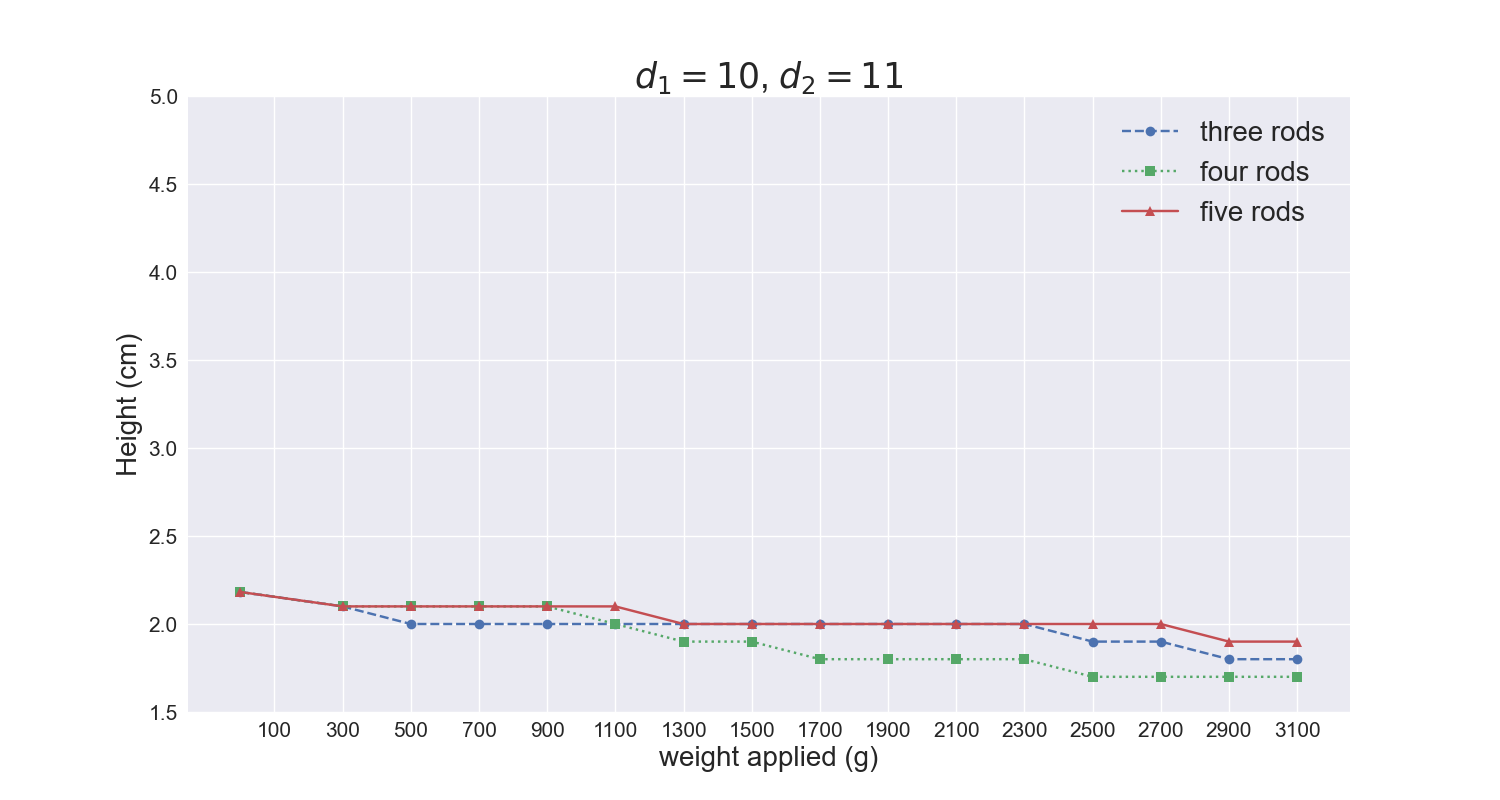}
        \caption{\(d_2=11\)}
        \label{fig:eleven}
    \end{subfigure}
    \begin{subfigure}{0.5\textwidth}
        \centering
        \includegraphics[width=\textwidth]{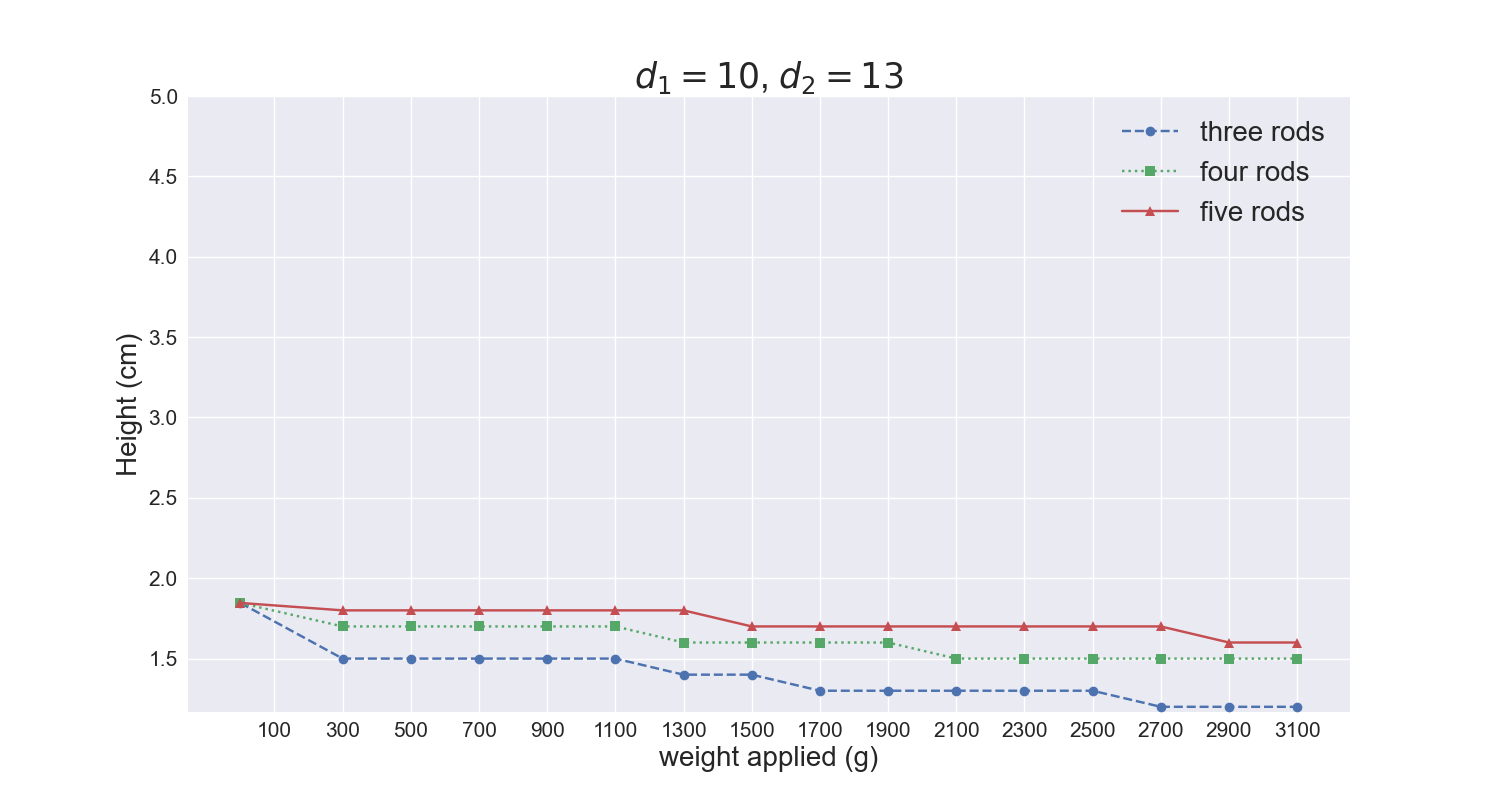}
        \caption{\(d_2=13\)}
        \label{fig:thirteen}
    \end{subfigure}
    \caption{Height of Three Dry Rods Structures for Different \(d_2\) Values}
    \label{fig:height_arange_by_d2}
\end{figure}
In figure \ref{fig:height_arange_by_d2}, three, four, and five rods structures with the same $d_2$ are plotted together. Observing figure \ref{fig:five}, the three rods structure with $d_2=5$ drops significantly in height, whereas 4 and 5 rods structures nearly retain at the same plateau much higher than 3 rods structures. For bigger values of $d_2$, the difference between three, four, and five rods is reduced significantly. 

\begin{figure}[H]
    \centering
    \includegraphics[width=0.8\textwidth]{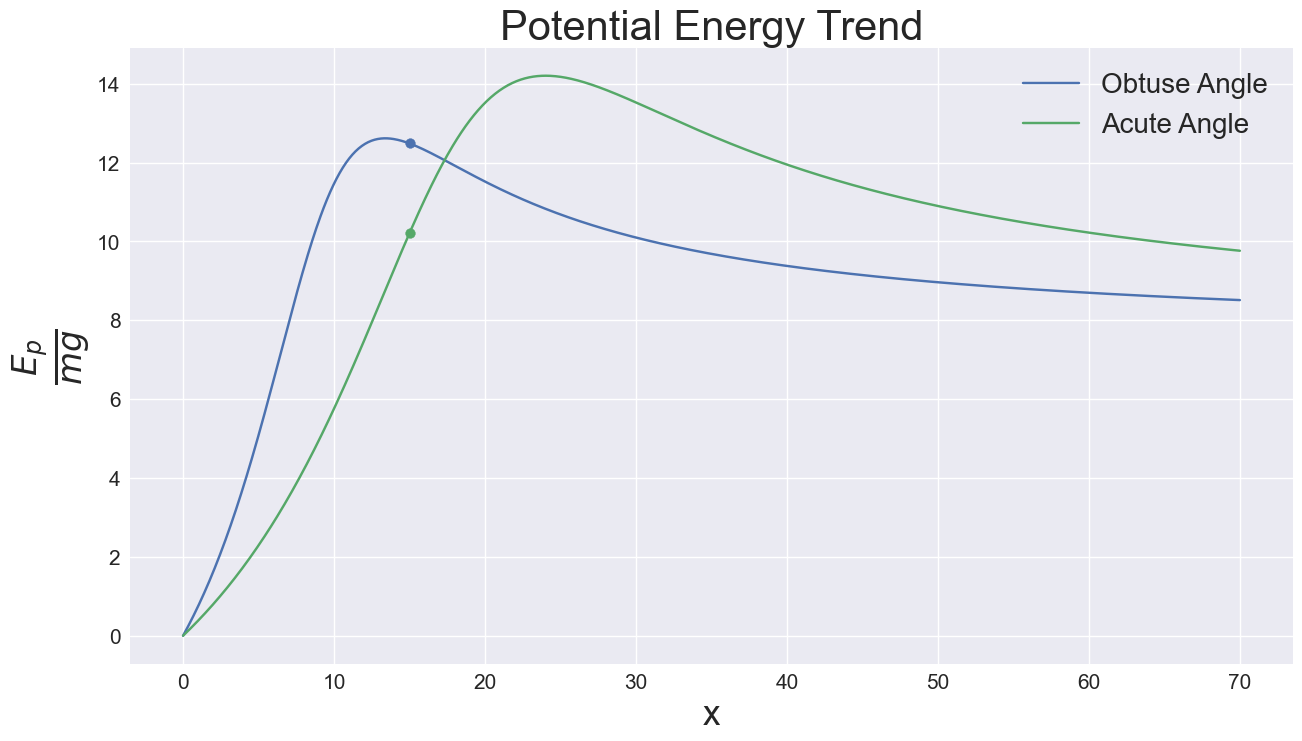}
    \caption{Potential Energy Trend for Acute and Obtuse Angle}
    \label{fig:angle_energy}
\end{figure}
\begin{figure}[H]
    \centering
    \begin{minipage}{0.45\textwidth}
        \centering
        \includegraphics[width=0.8\linewidth]{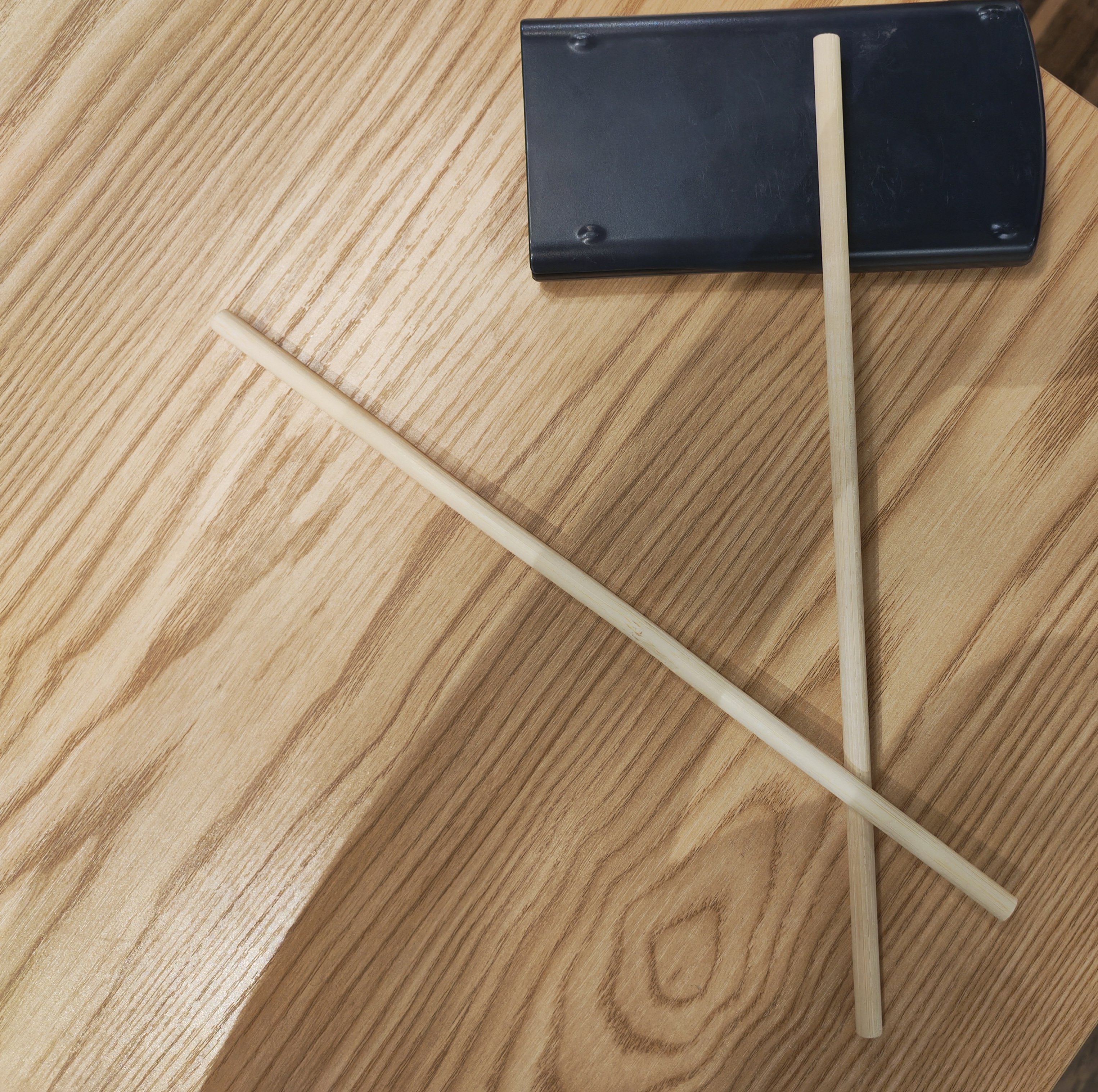}
        \caption{Structure with Acute Angle}
        \label{fig:acute}
    \end{minipage}\hfill
    \begin{minipage}{0.45\textwidth}
        \centering
        \includegraphics[width=0.8\linewidth]{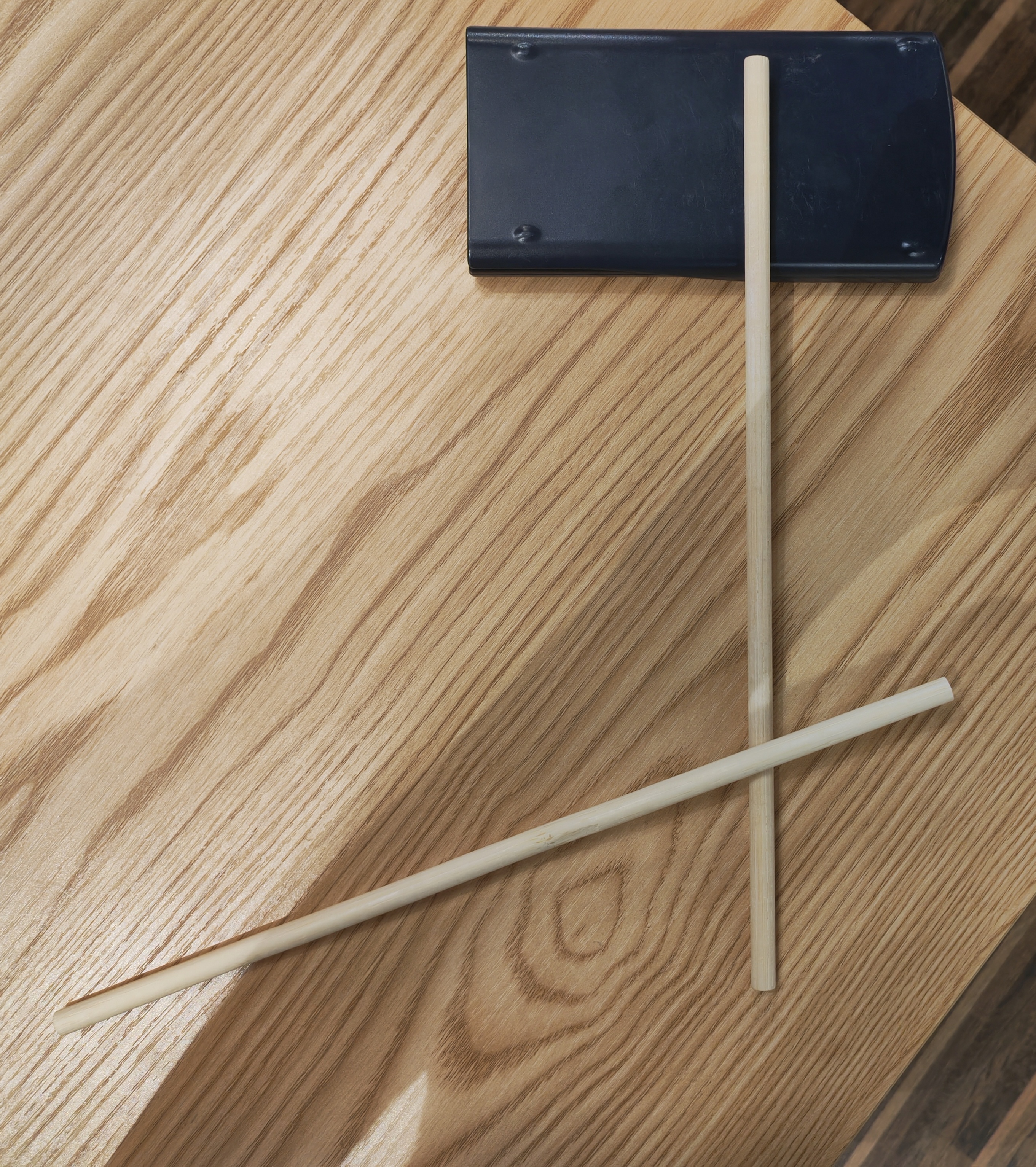}
        \caption{Structure with Obtuse Angle}
        \label{fig:obtuse}
    \end{minipage}
\end{figure}
One difference between the stability behavior of three and five rods' structure may be attributed to the potential energy trend. The two neighboring rods of five rods structures form an obtuse angle (figure \ref{fig:obtuse}), whereas three rods structures form acute angle (figure \ref{fig:acute}). The potential energy curves for these two cases are plotted in figure \ref{fig:angle_energy} as functions of $x$ (the distance on the lower rod between the contacting point of the rods and the end it touches the ground, see figure \ref{fig:two_rods_energy}). A larger $x$ denotes a higher contacting point. Two points are indicated in figure \ref{fig:angle_energy} for $x=15$. The point in case of obtuse angle is on the right side of a peak , showing that it has to go over an energy barrier before sliding down; in the case of acute angle, however, the potential energy is on a steep slope on the left side of a peak, meaning that it is much easier to slide down. Thus, five rods structures are more stable since the barrier of the energy peak prevents it from sliding down.

\begin{figure}[H]
    \centering
    \begin{subfigure}[b]{0.48\textwidth}
        \includegraphics[width=\textwidth]{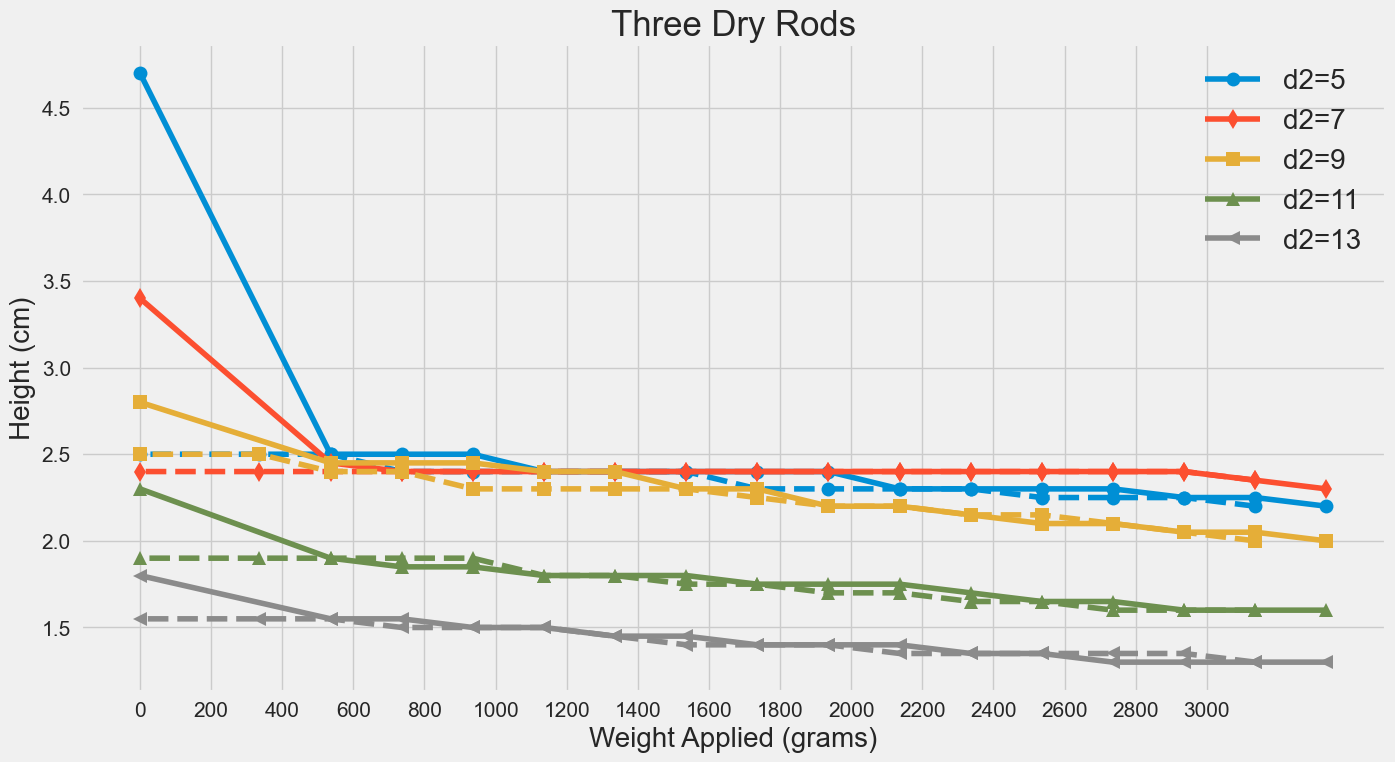}
        \caption{Three Dry Rods, experiment date August 25th}
        \label{fig:three_dry}
    \end{subfigure}
    \begin{subfigure}[b]{0.48\textwidth}
        \includegraphics[width=\textwidth]{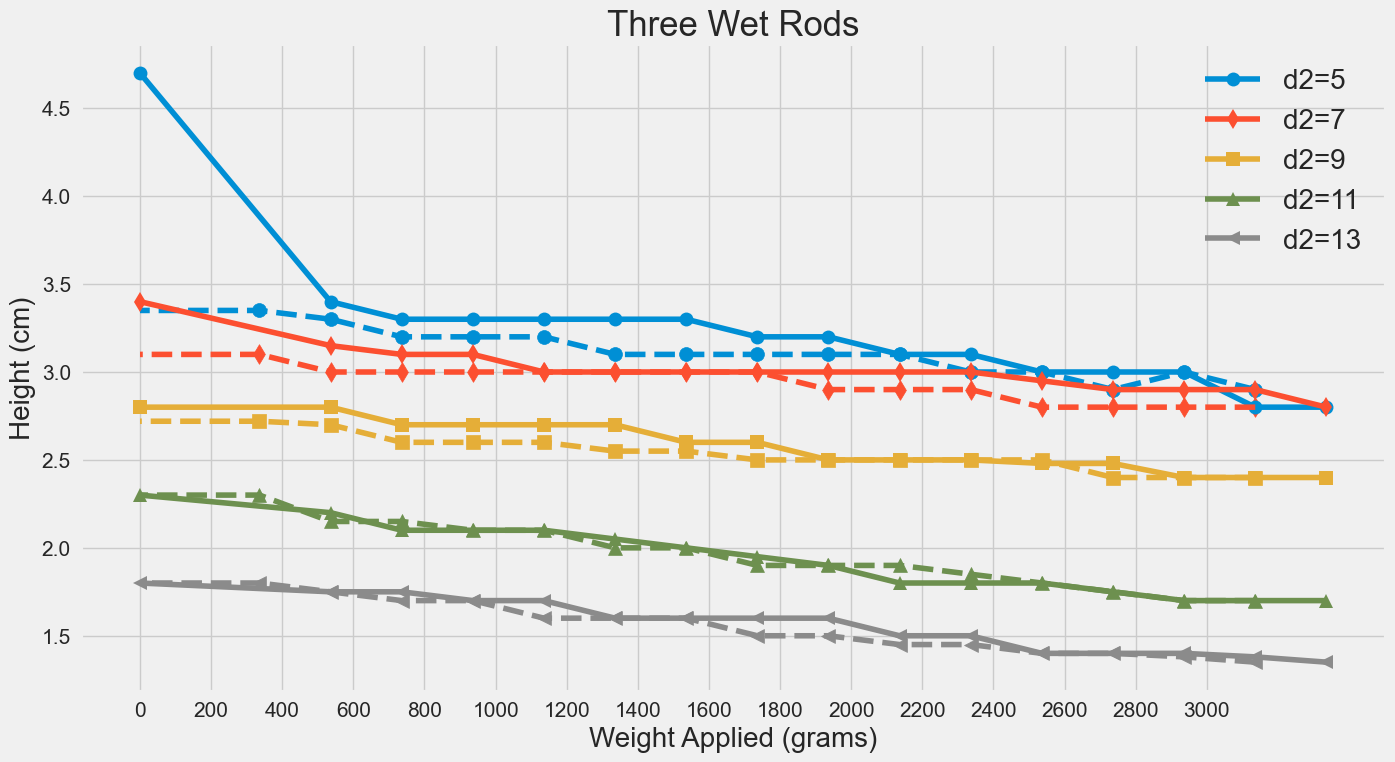}
        \caption{Three Wet Rods, experiment date August 25th}
        \label{fig:three_wet}
    \end{subfigure}
    
    \begin{subfigure}[b]{0.48\textwidth}
        \includegraphics[width=\textwidth]{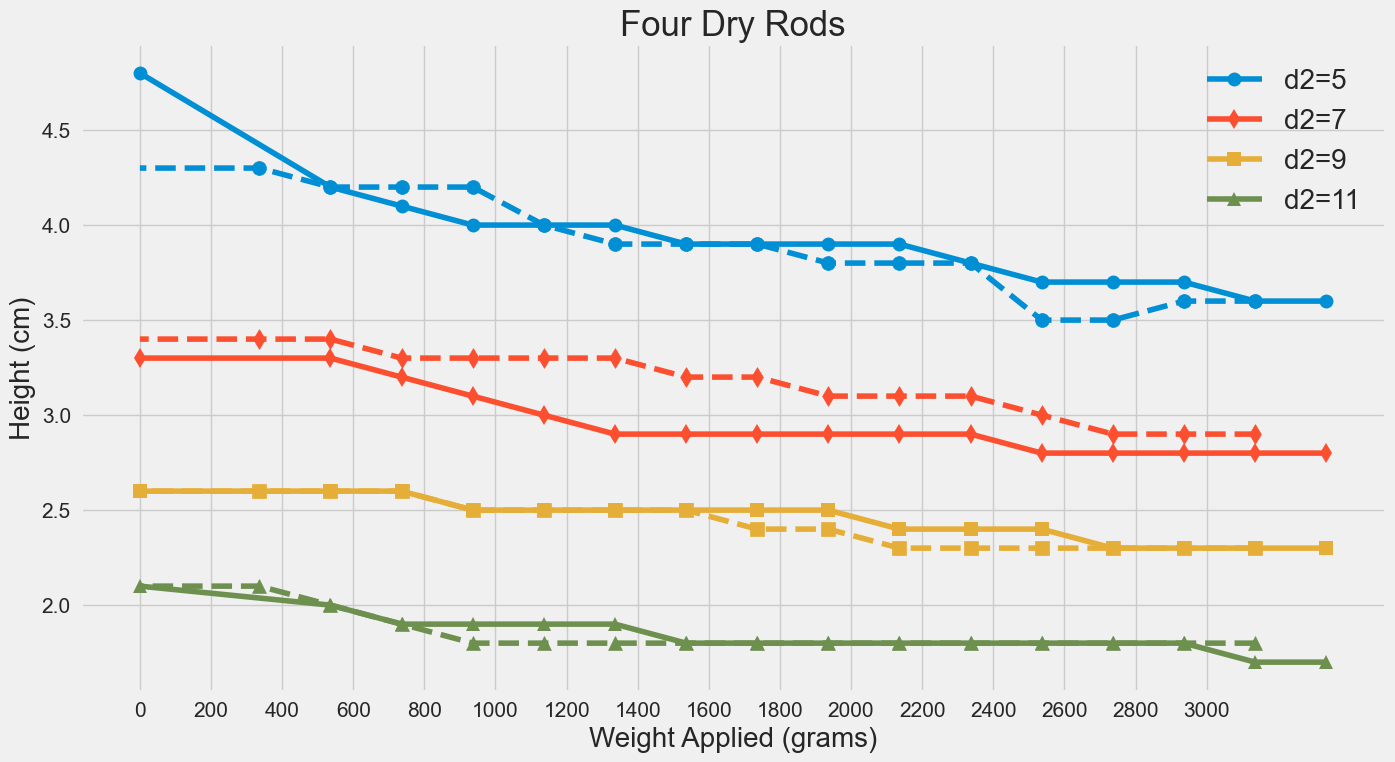}
        \caption{Four Dry Rods, experiment date August 1st}
        \label{fig:four_dry}
    \end{subfigure}
    \begin{subfigure}[b]{0.48\textwidth}
        \includegraphics[width=\textwidth]{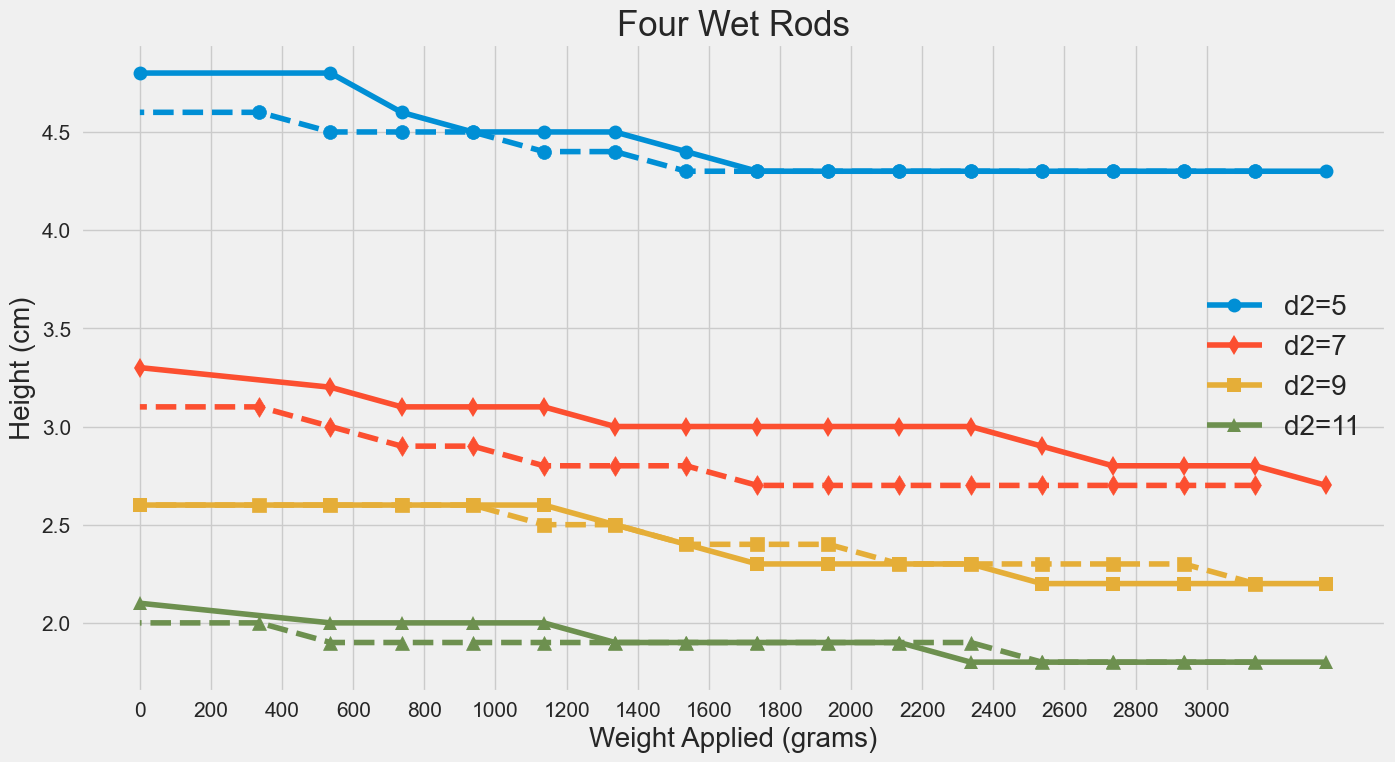}
        \caption{Four Wet Rods, experiment date August 1st}
        \label{fig:four_wet}
    \end{subfigure}
    \caption{Three/Four Dry/Wet Rod under Weight, the curve with lighter color represent "reformation"}
    \label{fig:hysteresis}
\end{figure}

In the height-weight figure of three rods (figure \ref{fig:pv_main}), multiple metastable states exist. The curve of $d_2=5$ jumps to $d_2=7$ curve when the first weight is applied. The $d_2=9$ curve merges with $d_2=11$ under higher weight. We presume that the nest will not be able to retain its original height after removing the weights, so we conducted hysteresis experiment as shown in figure \ref{fig:hysteresis}. This experiment was performed two months after the first one. The changing weather and environmental conditions may affect the results. This time, we also made comparison experiment between dry and wet rods. Their friction coefficients and Young’s Modulus were measured and listed in table \ref{table:1}.

\begin{table}[H]
\centering
\begin{tabular}{|c|c|c|}
\hline
 & Dry Rods & Wet Rods \\ 
 \hline
 Static Friction Coefficient & 0.202 & 0.485 \\  
 \hline
 Young's Modulus (\(\times 10^9\) Pa) & 7.15 & 5.69 \\   
 \hline
\end{tabular}
\caption{Physical properties of dry and wet rods}
\label{table:1}
\end{table} 

After all fifteen weights are applied, we then remove them one by one to observe the hysteresis loop of the different rod structures. Solid (dotted) curves are the results when applying (removing) weight. In dry rod experiment (figure \ref{fig:three_dry}), the results are mostly similar to figure \ref{fig:three} but different in terms that, this time, the three rod curves for $d_2=5,7,9$ first merge and then separate afterwards. The four rod curves never intersect. In the wet rod curves for $d_2=5,7$ merge slowly and other curves do not intersect with each other.

Now we look at the hysteresis properties. As we have expected, in figure \ref{fig:three_dry} the curves for d2=5,7 jump to d2=9 curve, their initial heights are not retained after removing the weights. The wet rod experiment for d2=5 is the same case, see figure \ref{fig:three_wet}.

\begin{figure}[H]
    \centering
    \includegraphics[width=0.8  \textwidth]{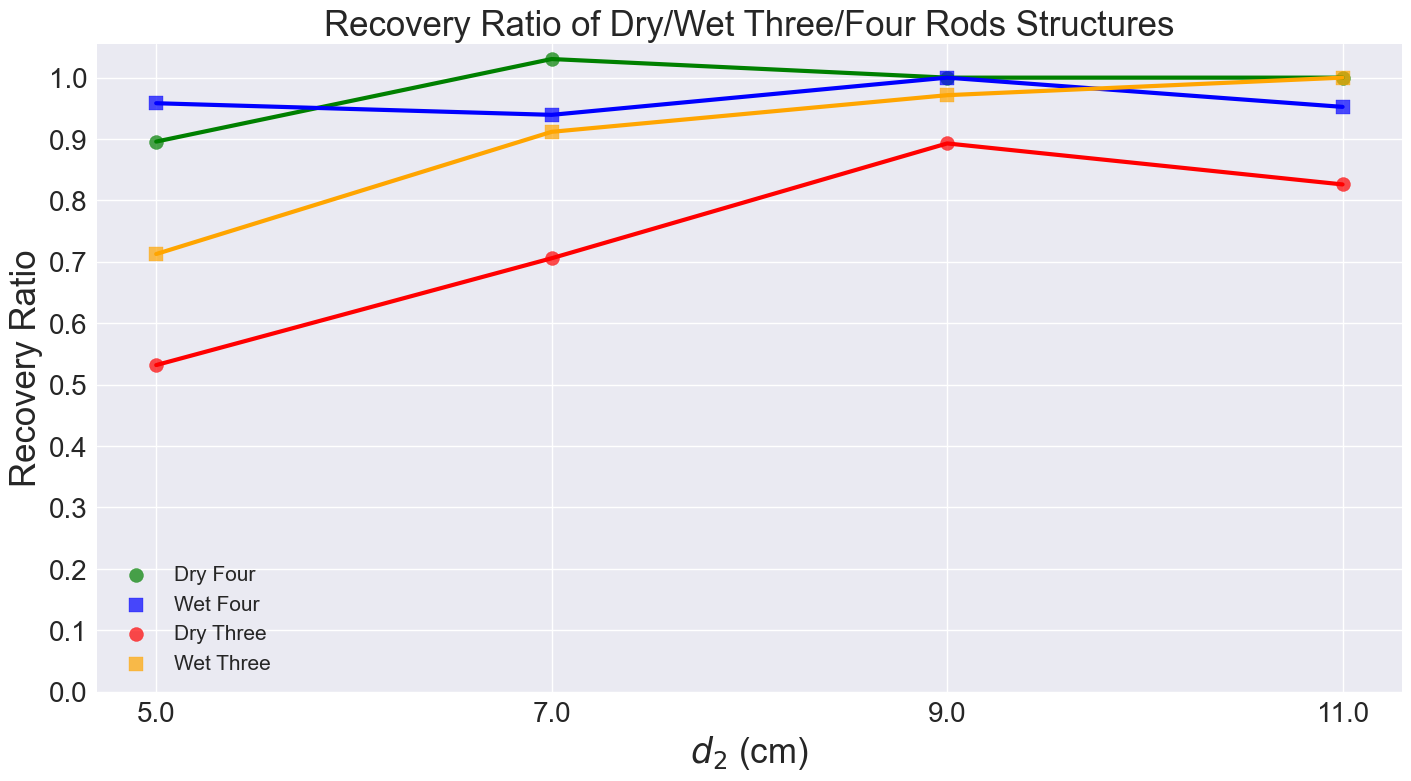}
    \caption{Recovery Ratio of Wet/Dry Three/Four Rods structures}
    \label{fig:rec_ratio}
\end{figure}
In figure \ref{fig:four_dry}, there also exist relatively small hysteresis loops. In some cases, the recovering height might ever be above the original one. Yet, we still regard the structures as an elastic phase approximately. If we define elastic phase as a recovery ratio greater than 90\%, then most four rods structures are elastic, whereas the three wet rods structures are elastic when $d_2\geq7$, see figure \ref{fig:rec_ratio}. For the structures not being able to recover to their initial heights, the points that deviate from linearity will be regarded as plastic phase. Thus, applying weight can transfer the structures from plastic phase to elastic phase. 

\begin{figure}[H]
    \centering
    \includegraphics[width=0.75\textwidth]{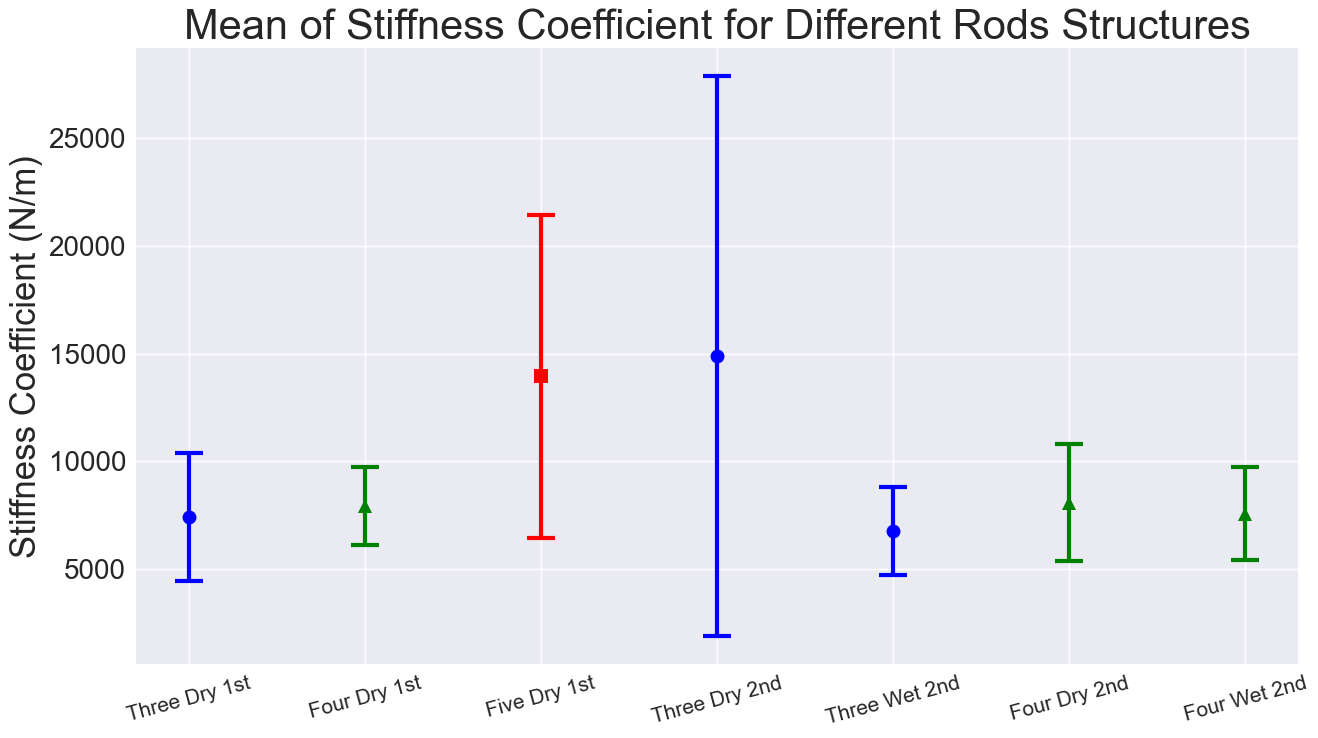}
    \caption{Stiffness Coefficient for Dry/Wet Three/Four Rods}
    \label{fig:stiffness}
\end{figure}
We then define the stiffness coefficient as $k=\frac{\Delta W}{\Delta h}$. The data are obtained by taking the average of the slopes over different $d_2$ from figures \ref{fig:pv_main} and \ref{fig:hysteresis}. Only the linear ranges are used. As shown in figure \ref{fig:stiffness}, five rods structures are significantly stiffer than four and three rods structures. The wet structures are less stiff than dry structures because of a lower young's modulus. 

\subsection{Comparison to Theory}
According to theory, applying weight does not change the condition for $\mu$ (equation \ref{eq:mucondition}) but changed the condition for $\mu'$ (equation \ref{eq:new_mu}). If friction coefficient with the table cannot support the structure, it will slide down and $d_2$ will increase. 
\begin{figure}[H]
    \centering
    \includegraphics[width=0.8  \textwidth]{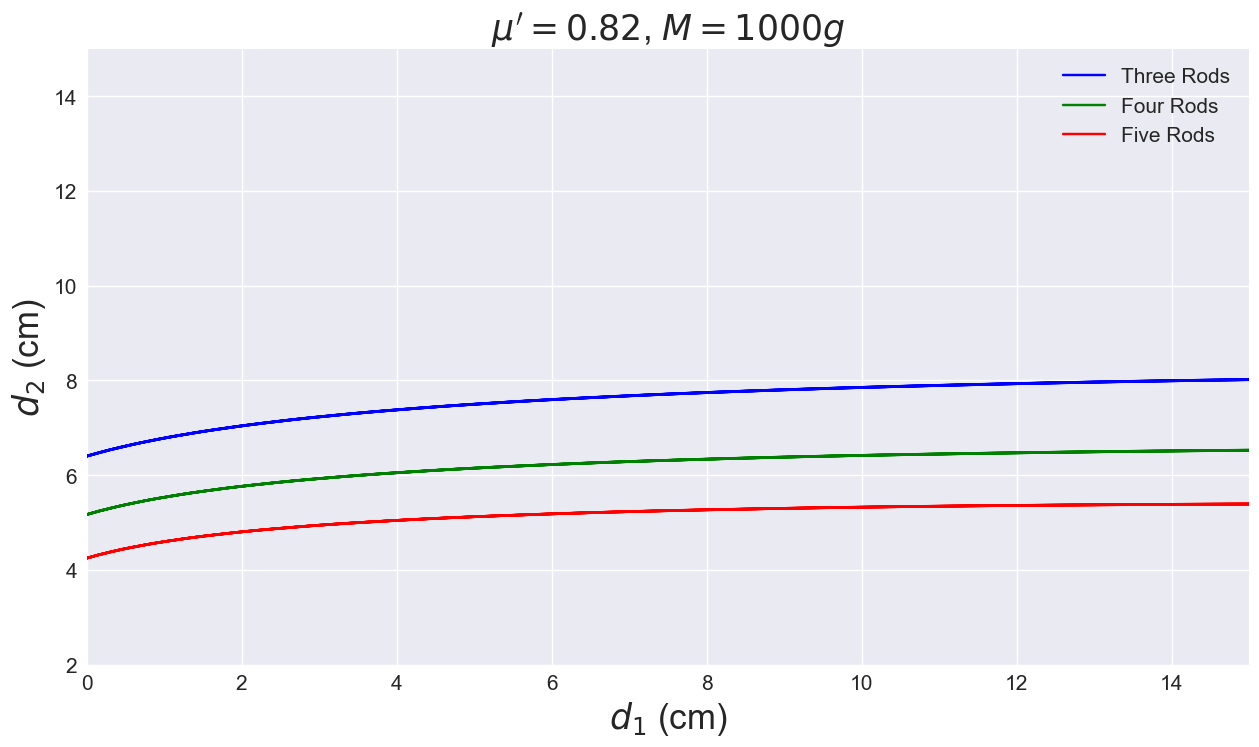}
    \caption{Theoretical Phase Boundary between Stable and Unstable States}
    \label{fig:mu_345}
\end{figure}

Figure \ref{fig:mu_345} is the theoretical phase boundaries between stable and unstable structures for three, four and five rods structures when $\mu'=0.8$2, applied weight $M=1000g$. The area above each curve is the stable phase and beneath each is the unstable phase. Both the theory (figure \ref{fig:mu'_345}) and the experiment (figure \ref{fig:height_arange_by_d2}) prove five rods structures can remain stable under a lower $d_2$. It is not just because of a lower weight per rod, but also because number of rods affect the parameter $\alpha$ (interior angle of a regular polygon), and a bigger $\alpha$ would yield a bigger stability area. 

The experiment confirms that the structures will decline rapidly at the beginning and slowly later (figure \ref{fig:pv_main}). Also, the wet rods structures are more stable because of a higher recovery ratio than dry rods (figure \ref{fig:rec_ratio}) and a lower standard deviation of stiffness coefficient (figure \ref{fig:stiffness}) than dry rods. These two trends can be explained by the condition for minimum  $\mu'$ which is proportional to  factor $\frac{M/n+m/2}{M/n+m}$  (equation \ref{eq:new_mu}). The factor ranges from $\frac{1}{2}$ to 1. When $M$ is close to zero, the ratio increases rapidly as $M$ increases, whereas the term would approach 1 as $M$ is sufficiently big.  For an insufficient value of $\mu'$ , this explains the behavior of the initial rapid decline and later slower decline. In addition, a bigger $m$ (heavier rods) would yield a lower phase boundary, which explains the relatively higher stability of the wet rods. 


\newpage
\section{Stability of Bird Nest Structure Under Vibration}

\textbf{Thermodynamic analogy}

For better understanding how changes in vibration and weight affect the structure's rigidity, stability, and other behavior, we try to describe the system's behavior in terms of thermodynamics. If the applied weights are regarded as ``pressure", and the vibration energy regarded as ``temperature", then transition from solid states (plastic phase and elastic phase) to collapsed state (liquid phase) happens  as ``temperature" and ``pressure" changes. 

Now we introduce the dimensionless temperature: the ratio between kinetic energy and maximum potential energy of the rods structures,
\begin{equation}
    \beta=\frac{\frac{1}{2}m\omega^2A^2}{\frac{1}{2}mgL}=\frac{\omega^2A^2}{gL}
    \label{eq:beta}
\end{equation}
where $\omega=2\pi f$, and $f$ is the frequency of vibration, A is the tested amplitude of the test bench, g is gravitational acceleration, L is the length of the rods. Another parameter is the ratio between acceleration due to vibration and gravitational acceleration,
\begin{equation}
    \gamma=\frac{\omega^2A}{g}
\end{equation}

In this following experiments, we describe the stability of the nest under vibration in terms of $\beta$ and $\gamma$. The critical ``temperature" $\beta$ represents the threshold beyond which the structure loses stability and undergoes a dramatic shift in state,e.g. transition between stable states (plastic and elastic) or from stable states to collapsed state (``liquid").

\subsection{Vibration Experiment 1} 

 {\large \textbf{Experiment setup and procedure}} \newline 

\begin{table}[h]
    \centering
    \begin{tabular}{|l|p{0.5\textwidth}|}
        \hline
        Number  & Equipments and Parameters\\
        \hline
        a   &  Test Bench\\
        b  & vibrometer (Precision to 1 $\mu$m) \\
        c &  Electronic scale (Precision to 0.1 g)\\
        d &  15*200g weights \\
        \hline
    \end{tabular}
\end{table}

\begin{figure}[H]
    \centering
    \begin{subfigure}{0.45\textwidth}
        \includegraphics[width=\textwidth]{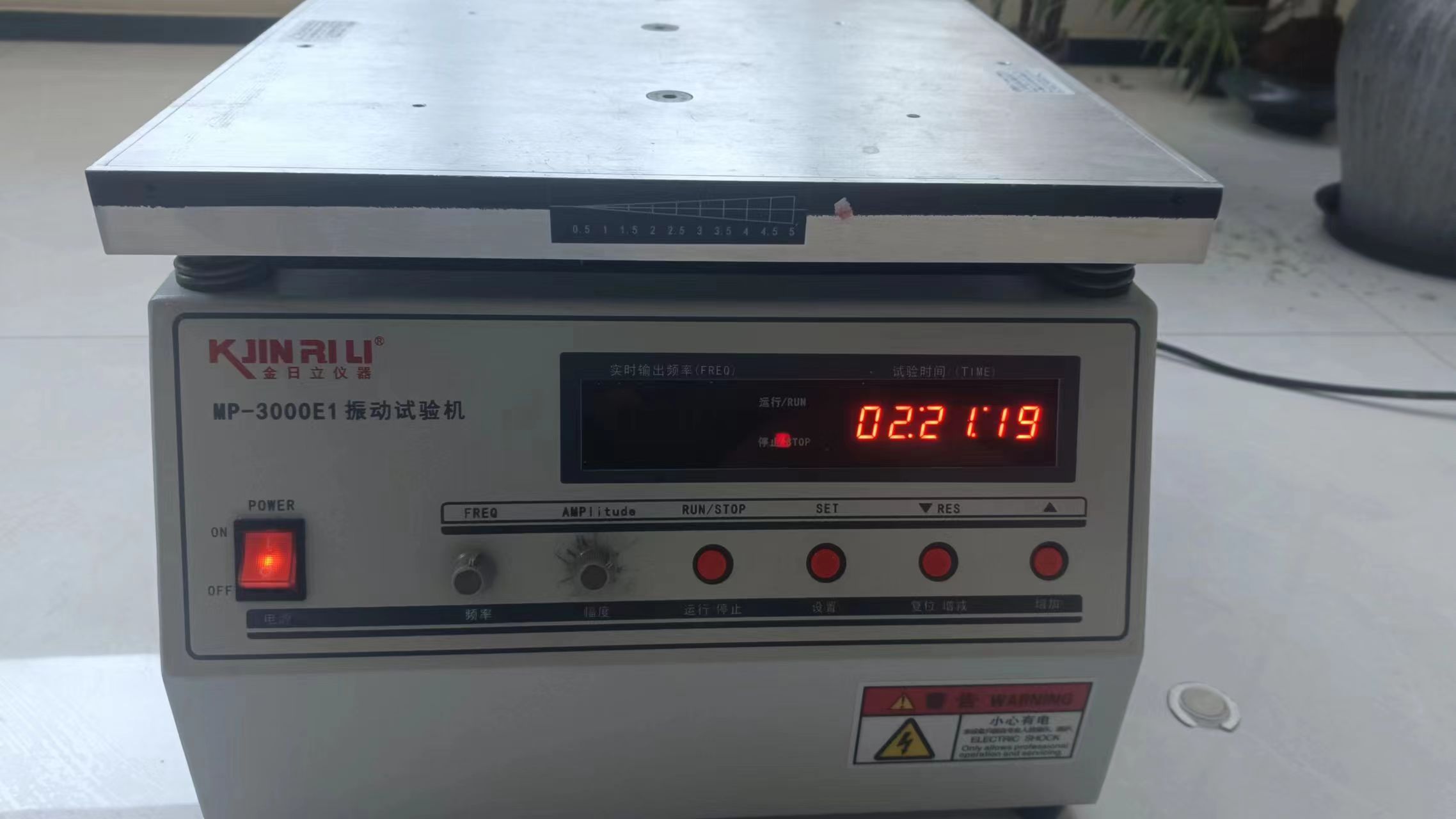}
        \caption{Test Bench}
        \label{fig:Test bench}
    \end{subfigure}
    \hfill
    \begin{subfigure}{0.45\textwidth}
        \includegraphics[width=\textwidth]{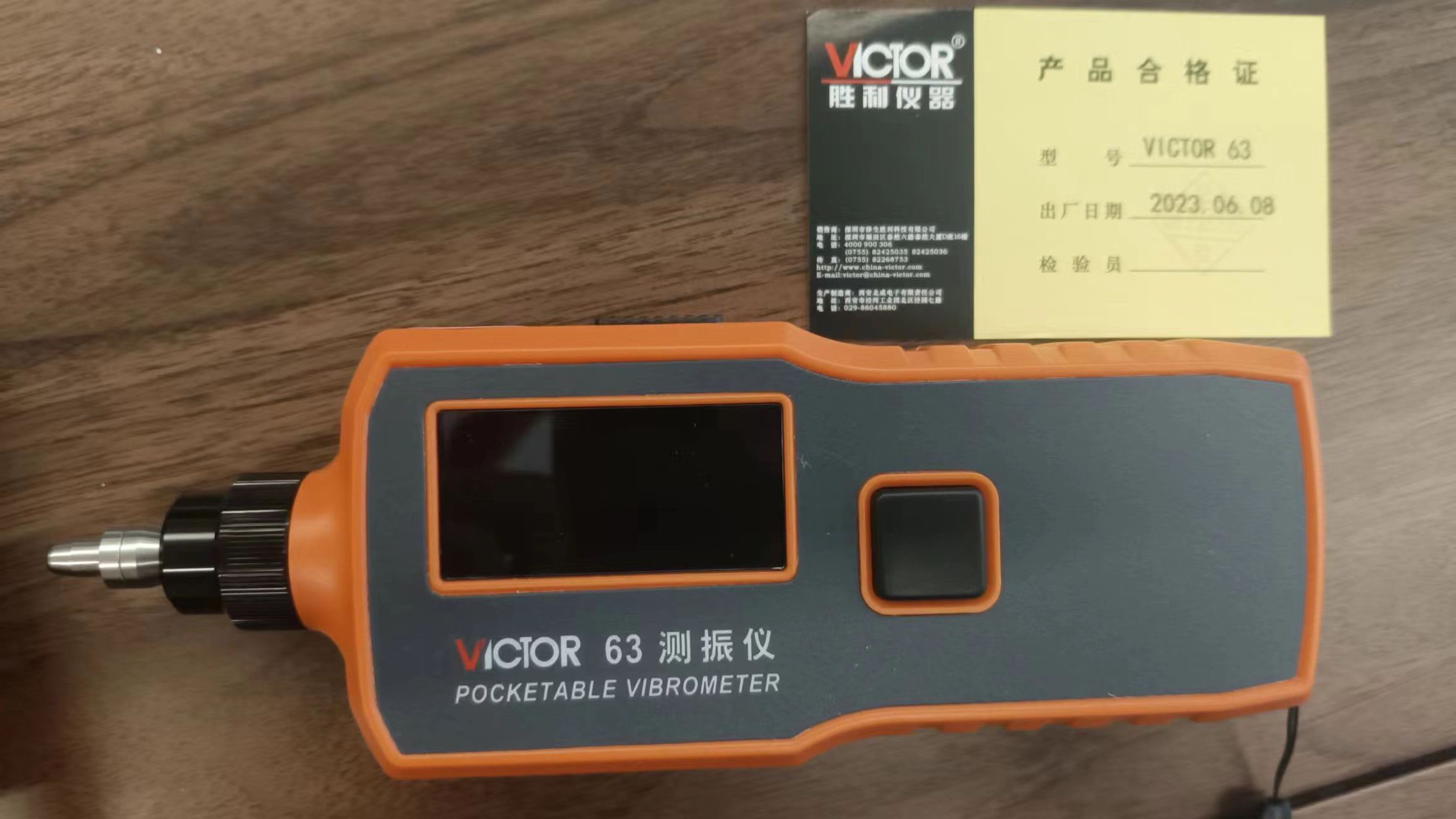}
        \caption{Vibrometer}
        \label{fig:vibrometer}
    \end{subfigure}
    
    \begin{subfigure}{0.45\textwidth}
        \includegraphics[width=\textwidth]{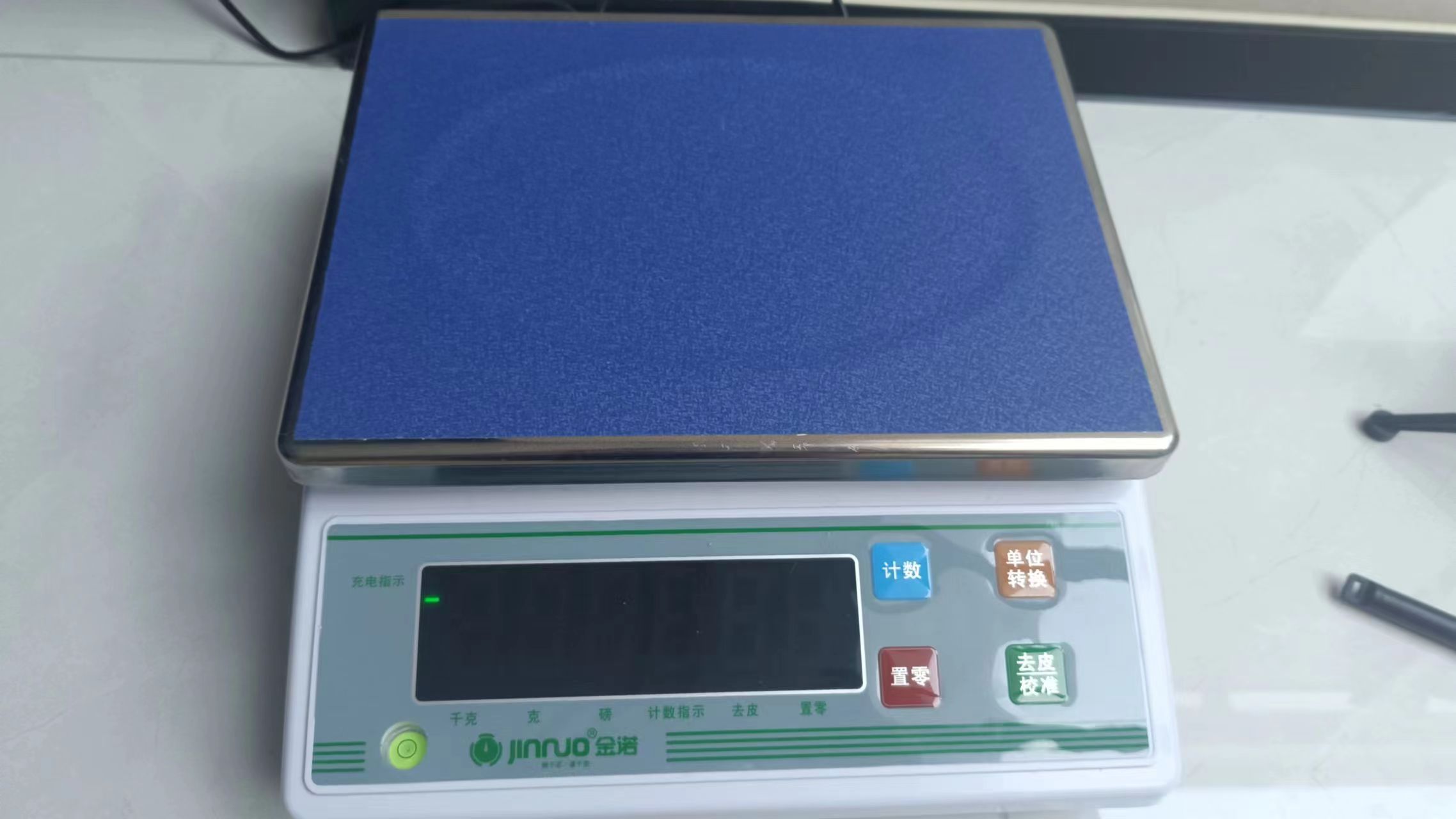}
        \caption{Electronic Scale}
        \label{fig:Electronic Scale}
    \end{subfigure}
    \hfill
    \begin{subfigure}{0.45\textwidth}
        \includegraphics[width=\textwidth]{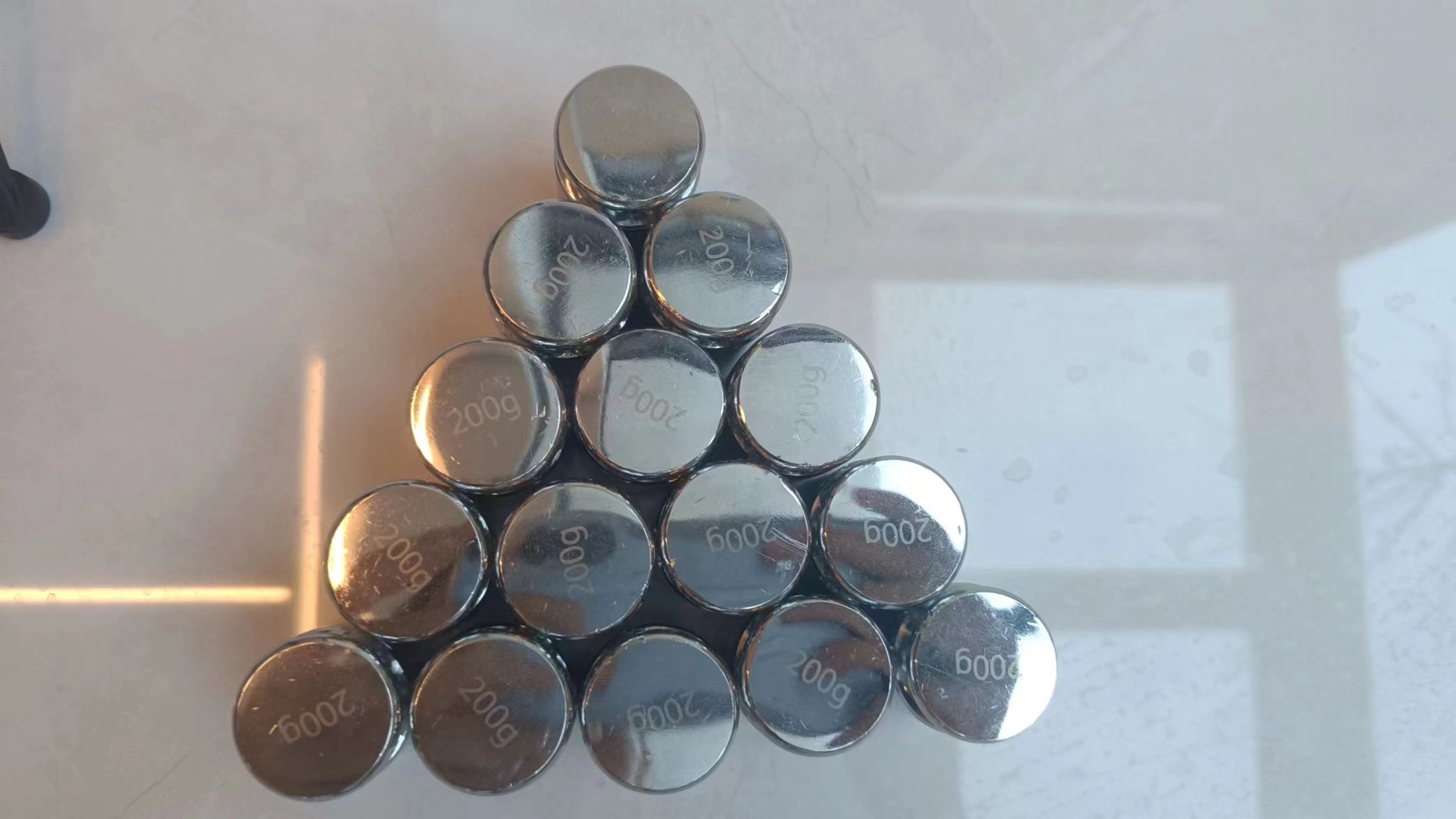}
        \caption{Weights}
        \label{fig:Weights}
    \end{subfigure}
    \caption{Equipments}
    \label{fig:grid_of_images}
\end{figure}

 In this experiment, the vibration frequency of the test bench is set as 18.4Hz. Three and four wet rods structures are tested. Make $d_1$ fixed at 10cm and vary $d_2$ from 5cm to 17cm, increasing 2cm at a time. Put the vibrometer onto the test bench to measure the amplitude and $\gamma$. Turn on the test bench and gradually increase the amplitude and stop when the structure begin to slide or move. Thus, we can determine the relationship between $d_2$, A and $\gamma$. Similarly, make $d_2$ fixed at 9cm and vary $d_1$ from 1cm to 13cm, increasing 2cm at a time. Find the critical amplitude A and $\gamma$ for $d_2$. \newline \newline {\large \textbf{Results and Analysis}}

\begin{figure}[H]
    \centering
    \begin{subfigure}[b]{0.45\textwidth}
        \centering
        \includegraphics[width=\textwidth]{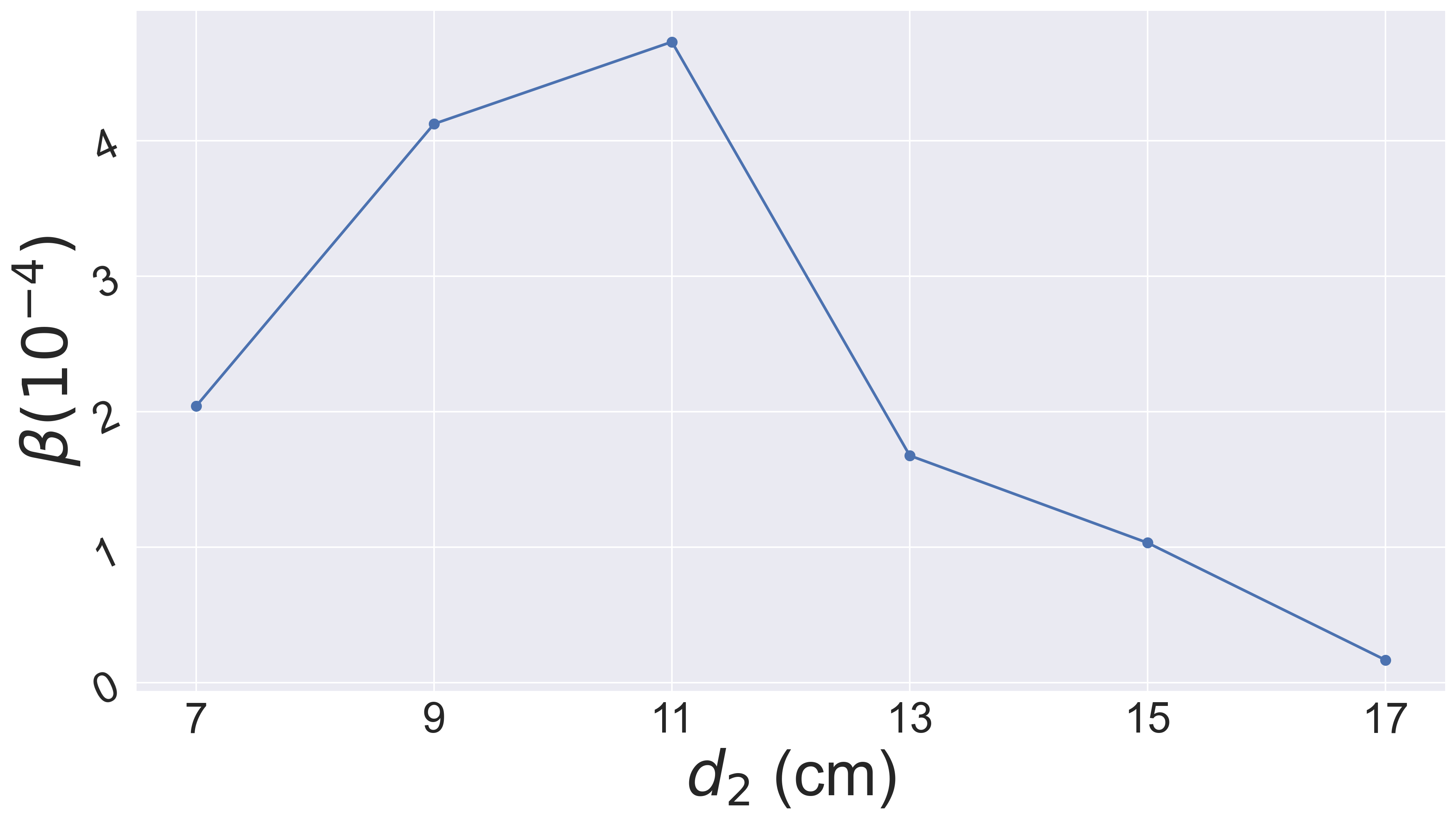}
        \caption{Critical ``temperature"}
        \label{fig:wet_d2_3_b}
    \end{subfigure}
    \hspace{0.05\textwidth} 
    \begin{subfigure}[b]{0.45\textwidth}
        \centering
        \includegraphics[width=\textwidth]{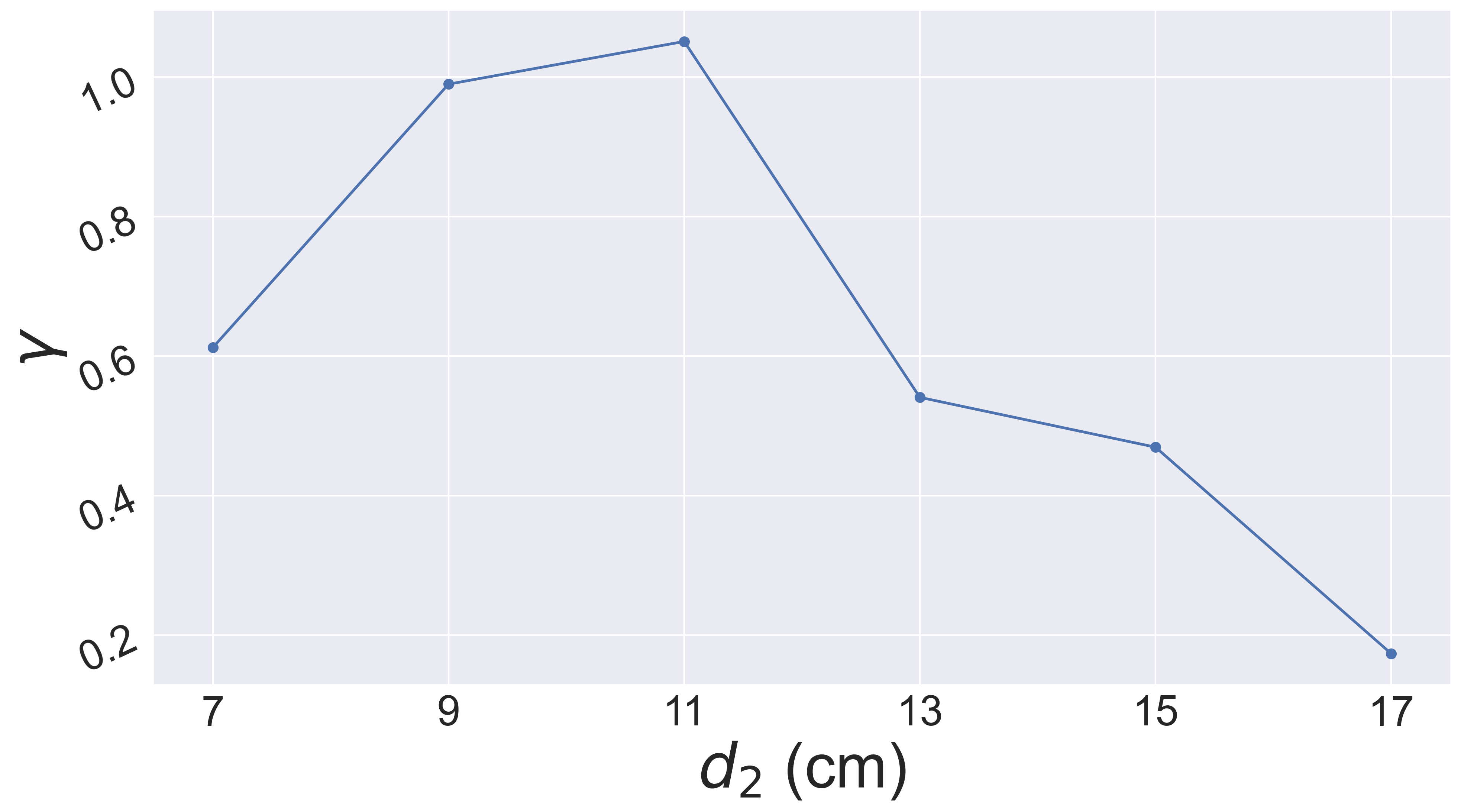}
        \caption{Critical acceleration}
        \label{fig:wet_d2_3_a}
    \end{subfigure}
    \caption{Critical ``temperature"  and acceleration, $d_1$=10cm with 3 rods}
    \label{fig:critical_vibration_energy 3 d2}
\end{figure}

\begin{figure}[H]
    \centering
    \begin{subfigure}[b]{0.45\textwidth}
        \centering
        \includegraphics[width=\textwidth]{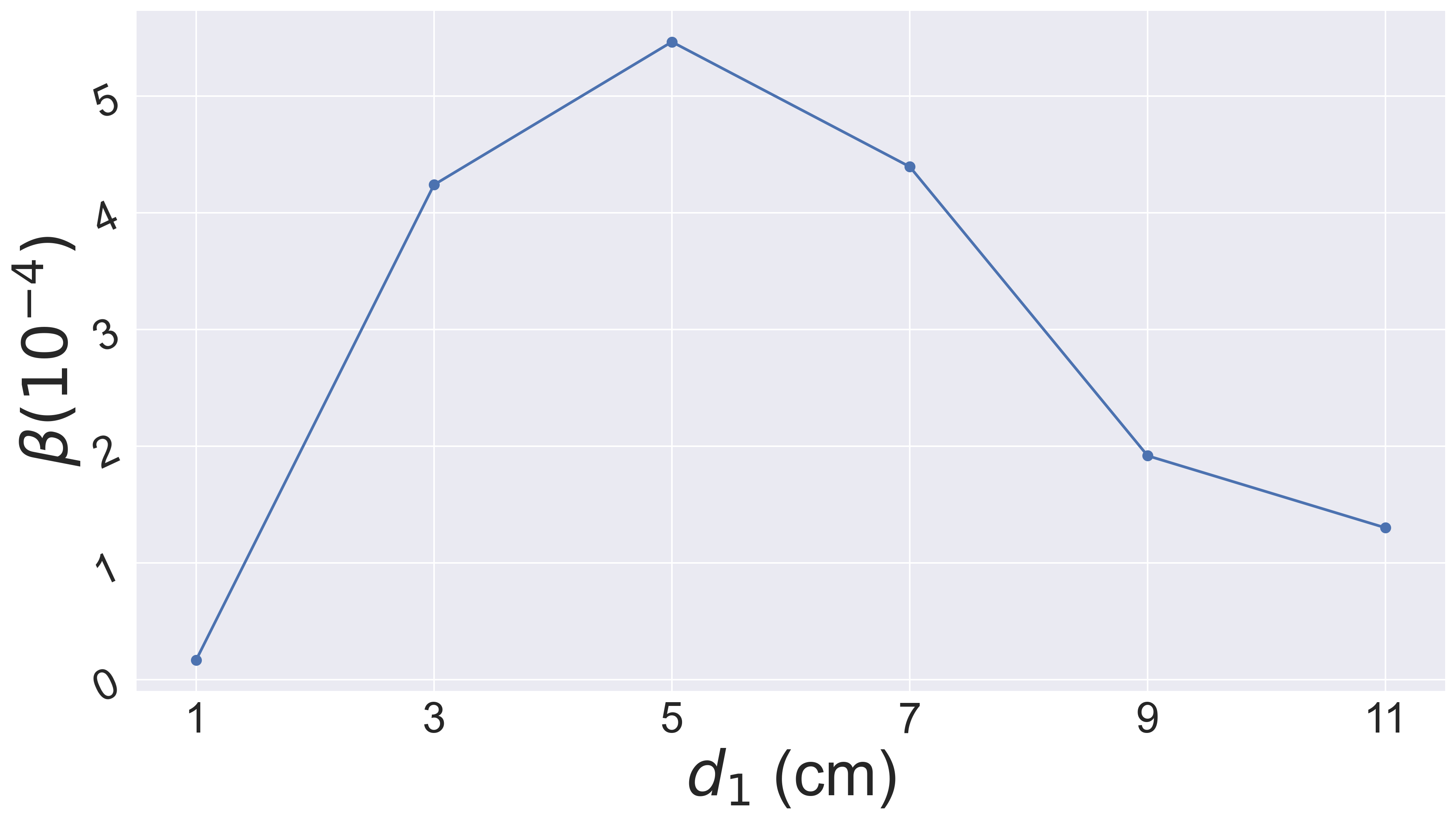}
        \caption{Critical ``temperature" }
        \label{fig:wet_d1_3_b}
    \end{subfigure}
    \hfill 
    \begin{subfigure}[b]{0.45\textwidth}
        \centering
        \includegraphics[width=\textwidth]{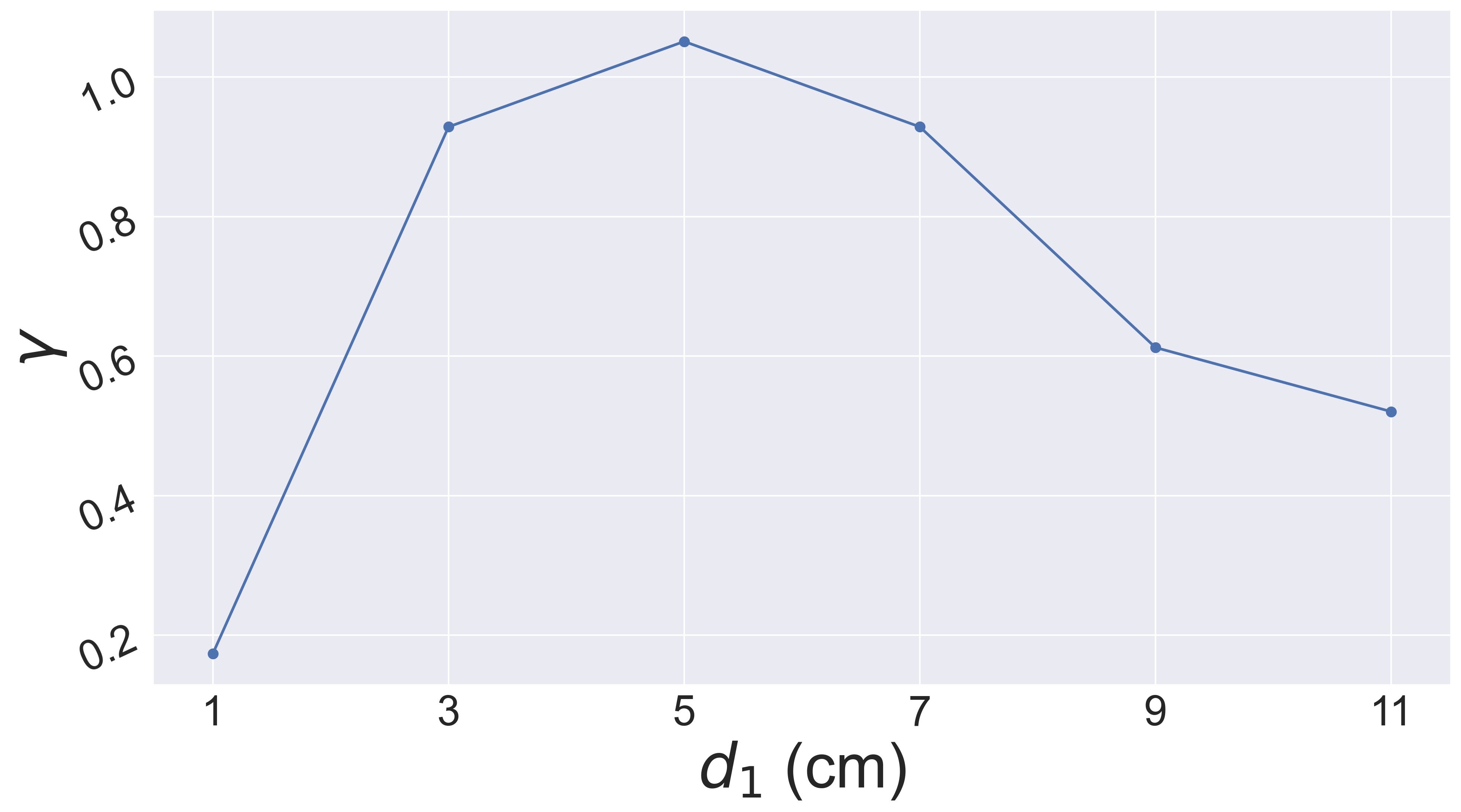}
        \caption{Critical acceleration}
        \label{fig:wet_d1_3_a}
    \end{subfigure}
    \caption{Critical "temperature"  and acceleration, $d_2$=9cm with 3 rods}
    \label{fig:combined_figures 3 d1}
\end{figure}

The ``temperature" $\beta$ can be calculated from equation \eqref{eq:beta}. In figure \ref{fig:critical_vibration_energy 3 d2} and \ref{fig:combined_figures 3 d1}, the critical ``temperature" $\beta$ and acceleration $\gamma$ of the 3 rods structure are plotted as $d_1$ and $d_2$ changes. The maximum $\beta$ ($\gamma$) this structure can withstand is about $5.5 \times 10^{-4}$ (1.0) when $d_1$ and $d_2$ are about 5cm and11cm respectively. The critical $\beta$ and $\gamma$ becomes smaller as $d_1$ or $d_2$ become greater or smaller.

\begin{figure}[H]
    \centering
    \begin{subfigure}[b]{0.45\textwidth}
        \centering
        \includegraphics[width=\textwidth]{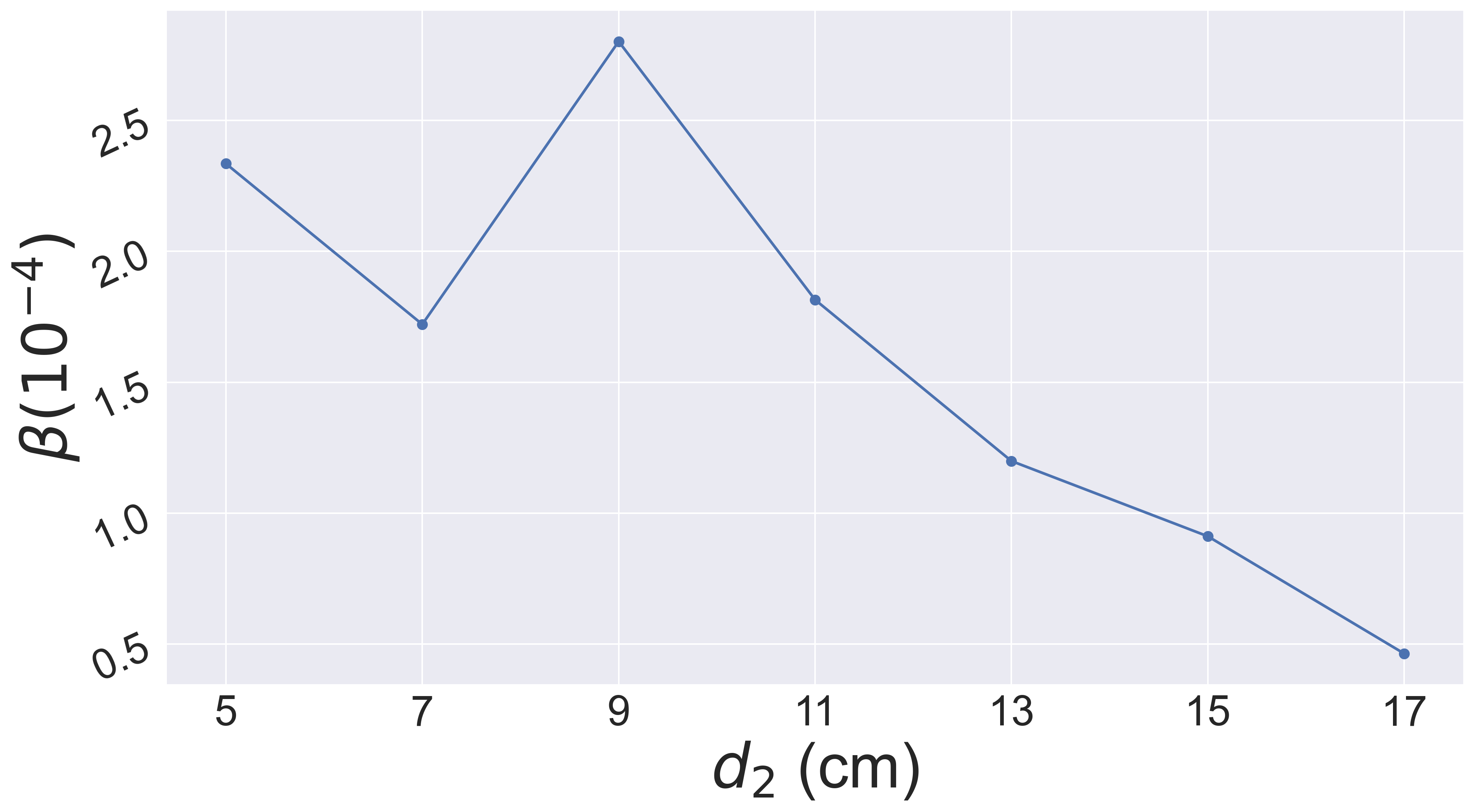}
        \caption{Critical ``temperature“}
        \label{fig:wet_d2_4_b}
    \end{subfigure}
    \hspace{0.05\textwidth} 
    \begin{subfigure}[b]{0.45\textwidth}
        \centering
        \includegraphics[width=\textwidth]{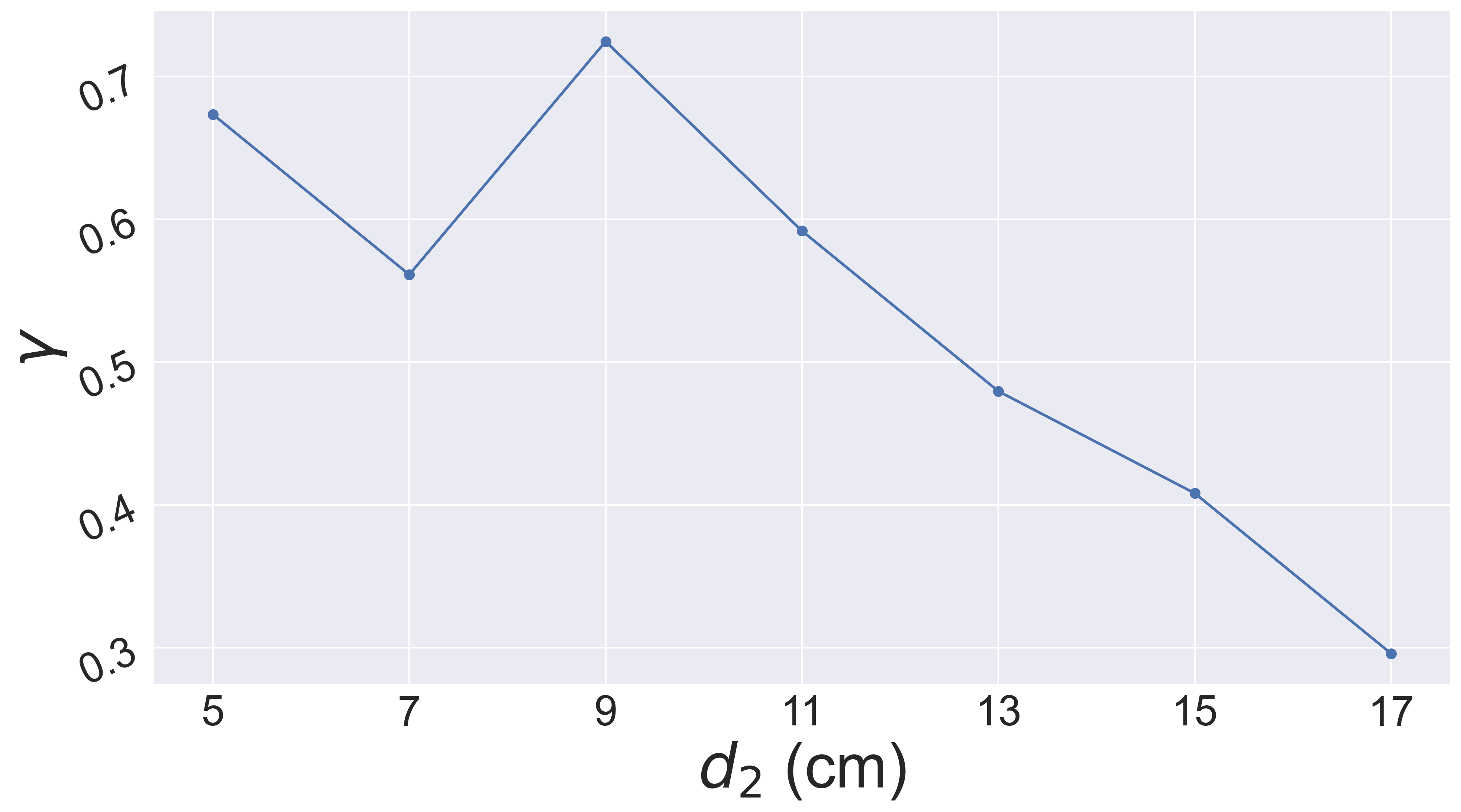}
        \caption{Critical acceleration}
        \label{fig:wet_d2_4_a}
    \end{subfigure}
    \caption{Critical ``temperature"  and acceleration, $d_1$=10cm with 4 rods}
    \label{fig:critical_vibration_energy 4 d2}
\end{figure}

\begin{figure}[H]
    \centering
    \begin{subfigure}[b]{0.45\textwidth}
        \centering
        \includegraphics[width=\textwidth]{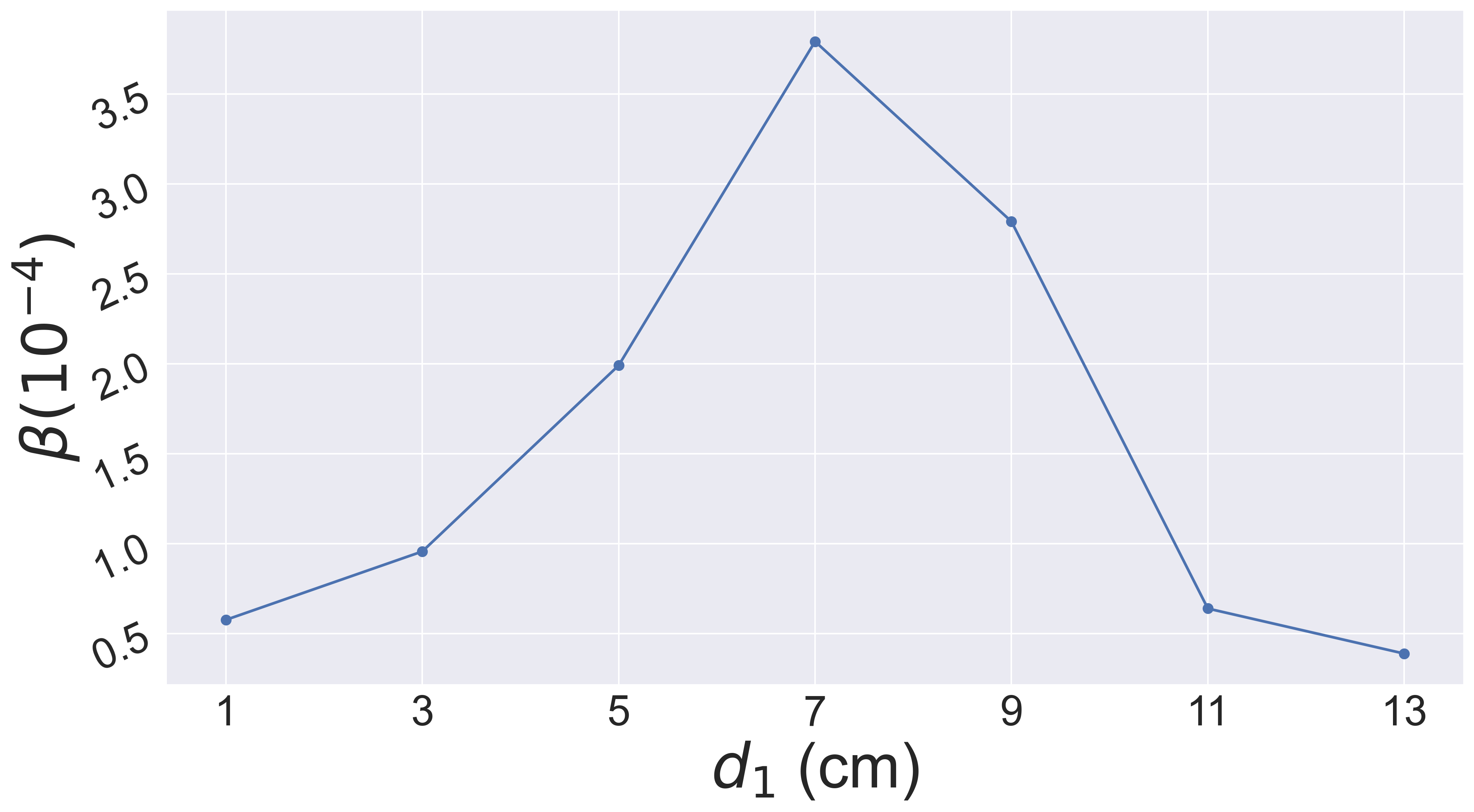}
        \caption{Critical ``temperature"}
        \label{fig:wet_d1_4_b}
    \end{subfigure}
    \hspace{0.05\textwidth} 
    \begin{subfigure}[b]{0.45\textwidth}
        \centering
        \includegraphics[width=\textwidth]{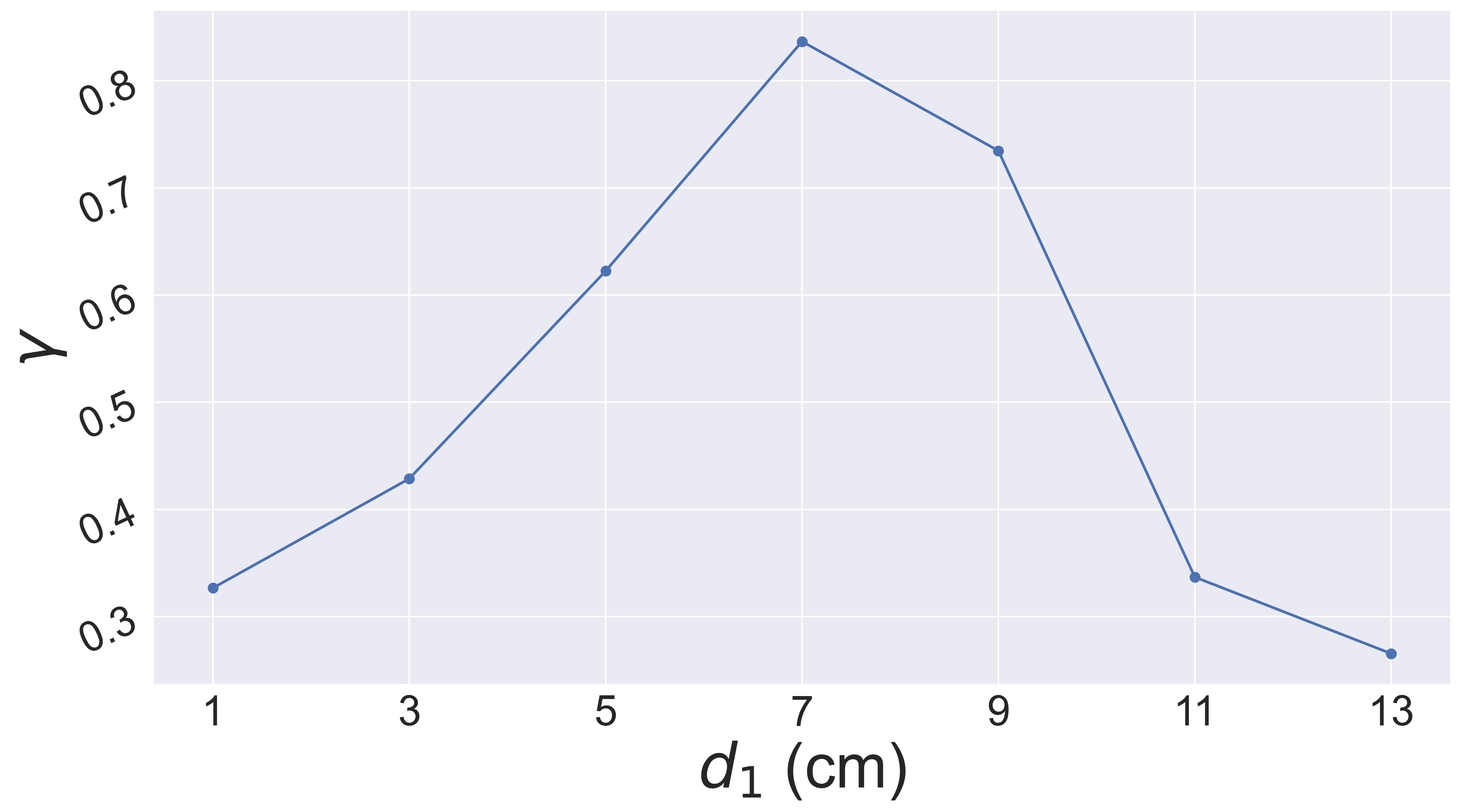}
        \caption{Critical acceleration}
        \label{fig:wet_d1_4_a}
    \end{subfigure}
    \caption{Critical ``temperature"  and acceleration, $d_2$=9cm with 4 rods}
    \label{fig:critical_vibration_energy 4 d1}
\end{figure}

In figure \ref{fig:wet_d2_4_b} and figure \ref{fig:wet_d2_4_a}, the maximum $\beta$ ($\gamma$) for 4 rods structure is about $3 \times 10^{-4}$m (0.7). The overall trends for 3 rods and 4 rods are similar: both have peaks at certain $d_1$ and $d_2$ except that the trends for 4 rod structure as $d_2$ changes is more complicated when $d_2$ is small. The origin of such uncertainty might be that the four rod structure has more degree of freedom and involves more meta-stability. The optimal configuration of $d_1$ and $d_2$  probably results from a favorable position on the potential energy surface. 


\subsection{Vibration Experiment 2}

Rather than observing the sliding critical conditions of the structures, now we investigate the status of the structure in more detail. This time, we use a wet four-rod structure. 

In this experiment, the vibration frequency of the test bench is also set as 18.4Hz. We first put the vibrometer onto the test bench to measure the amplitude and $\gamma$. Then we fix $d_1$ at 7cm, and vary $d_2$ from 7cm to 17cm, increasing 2cm each time. For each $d_2$, we record the initial height of this structure without vibration. If the structure does not move under vibration, record the original height as its steady state. If the structure slides and stops, we record its static height. Else, if the structure falls apart, height is recorded as 0cm. \newline \newline { \large \textbf{Result and Analysis}}

\begin{figure}[H]
    \centering
    \includegraphics[width=0.8\textwidth]{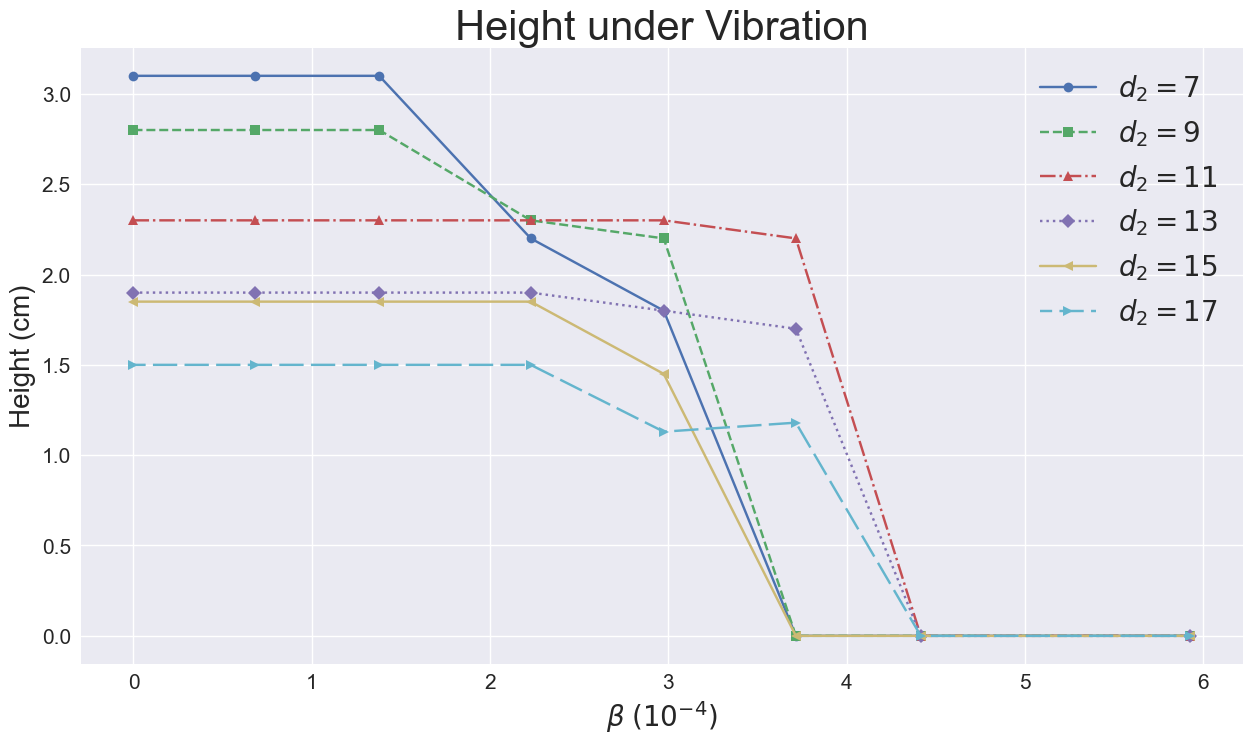}
    \caption{Height change with "temperature" for varying $d_2$}
    \label{Height Change with Vibration for Structures of Varying $d_2$}
\end{figure}
 In figure \ref{Height Change with Vibration for Structures of Varying $d_2$}, starting from an initial height, the height decreases as greater vibration amplitude is applied and eventually collapses.

\begin{figure}[H]
    \centering
    \includegraphics[width=0.8\textwidth]{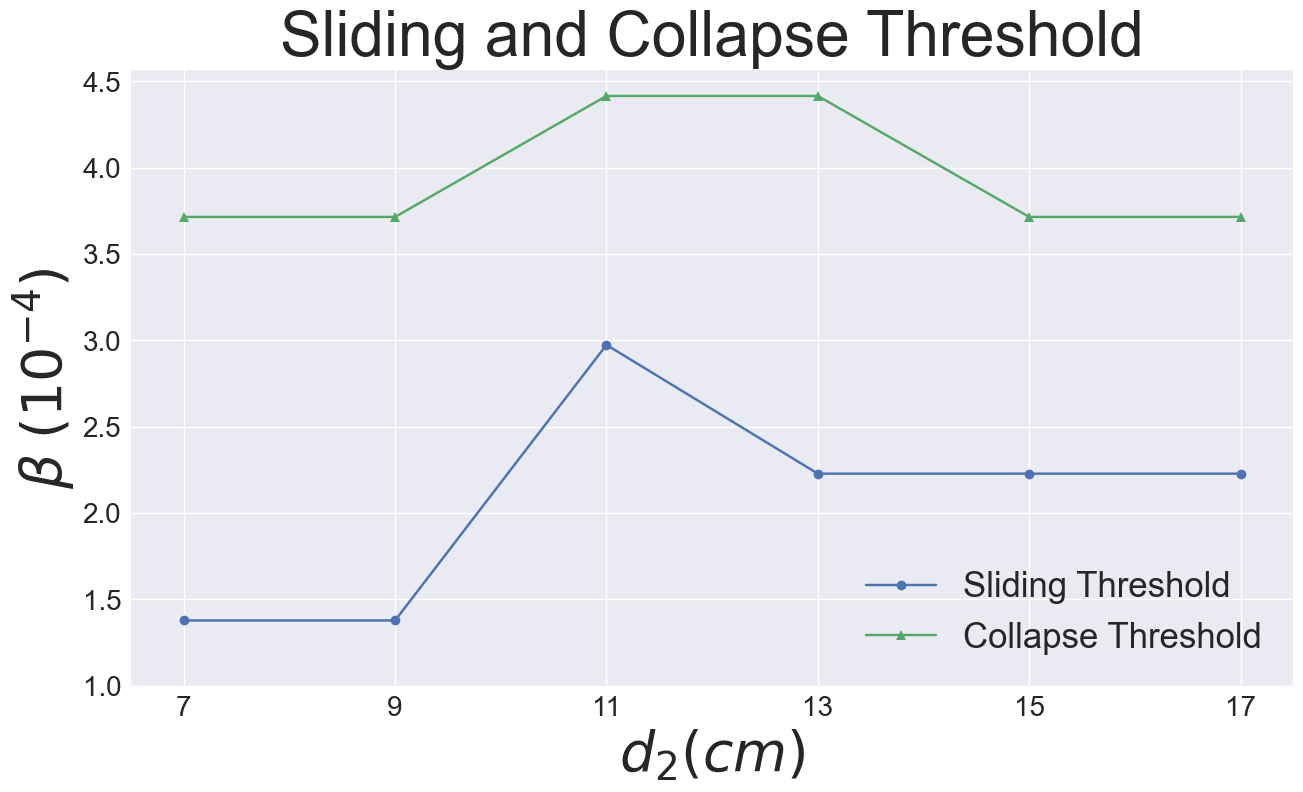}
    \caption{Critical "temperature" of structures for varying d2}
    \label{Temperatures for Structures of Varying d_2}
\end{figure}

From figure \ref{Height Change with Vibration for Structures of Varying $d_2$}, both critical values of $\beta_1$ that the structure starts to slide and $\beta_2$ that it falls apart are plotted in figure \ref{Temperatures for Structures of Varying d_2}. For this structure, the  critical "temperature" for sliding $\beta_1$ has a maximum about $3.5 \times10^{-4}$ when $d_2$ is about 11cm. The  critical "temperature" for collapsing $\beta_2$ has a maximum about $4.4 \times10^{-4}$ when $d_2$ is about 11-13cm.

\subsection{Stability of Bird Nest Structure under Vibration and Applied Weight}
  The experiment sought to comprehend the structure's dynamics under different "temperature" and applied weights.

\textbf{Procedure}
 
The equipment is the same as in Section 4.1. We used a wet four rods structure with $d_2$ = 7cm. Set the frequency of the vibration at 54.6Hz 
at each level of amplitude. Apply different weights, and record the final stable height (a stable state where no further movement is present). If the structure collapse, we record the height as 0. Put the vibrometer onto the test bench to measure the different levels of amplitude and $\gamma$ under different weights, since weights would affect the amplitudes.

\textbf{Result and Analysis}

\begin{figure}[H]
    \centering
    \includegraphics[width=1\textwidth]{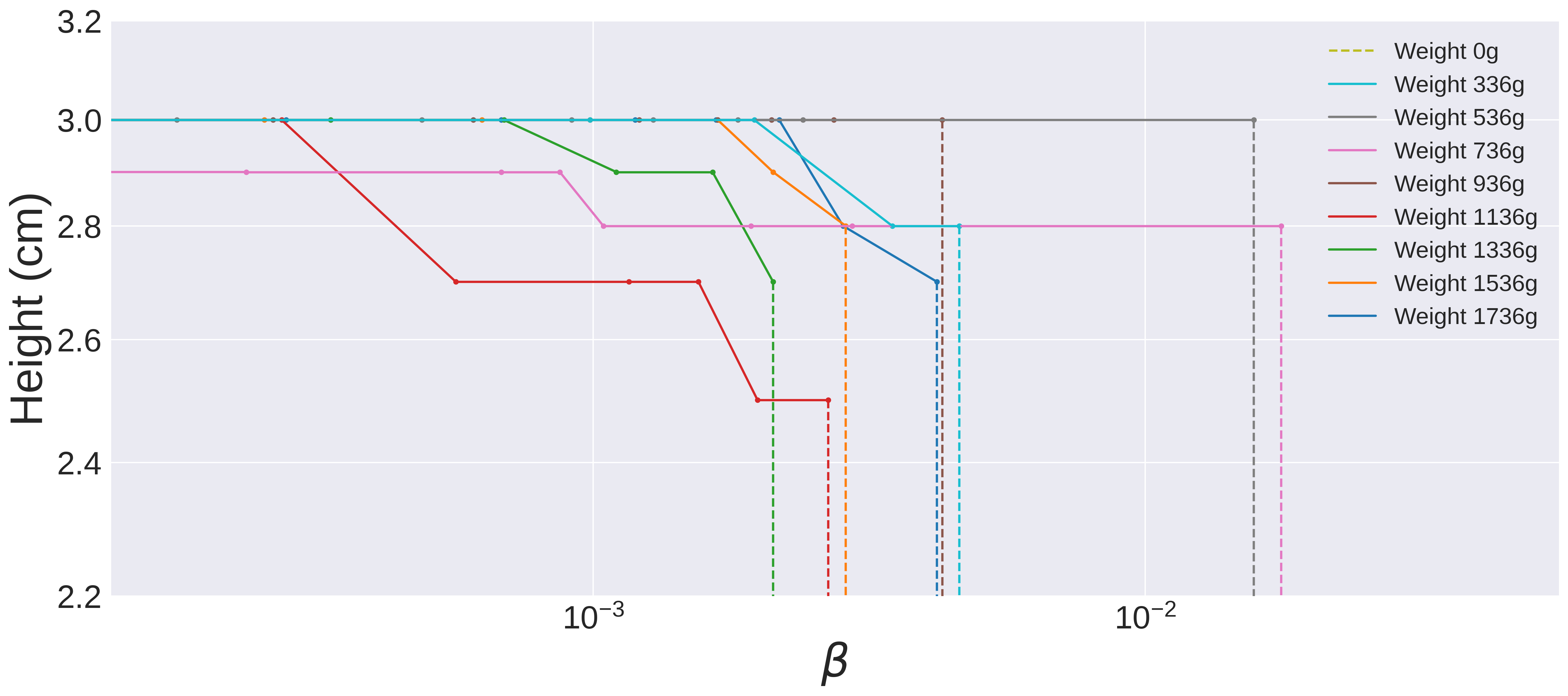}
    \caption{Height change with"temperature" under varying weight (logarithm scaling)}
    \label{Height Change with Vibration under Varying Weight(scaling applied, with phase change indication}
\end{figure}

 
 In figure \ref{Height Change with Vibration under Varying Weight(scaling applied, with phase change indication}, a logarithm scaling is applied. In most cases, the height of the structure declines with increasing "temperature" and finally collapses. The collapse of the structure is indicated with dashed lines in the figure where the next height jumps to 0 suddenly and phase transition from "solid" to "liquid" states happens.

\textbf{Phase Transition Boundary}

\begin{figure}[H]
    \centering
    \begin{subfigure}[b]{0.45\textwidth}
        \centering
        \includegraphics[width=\textwidth]{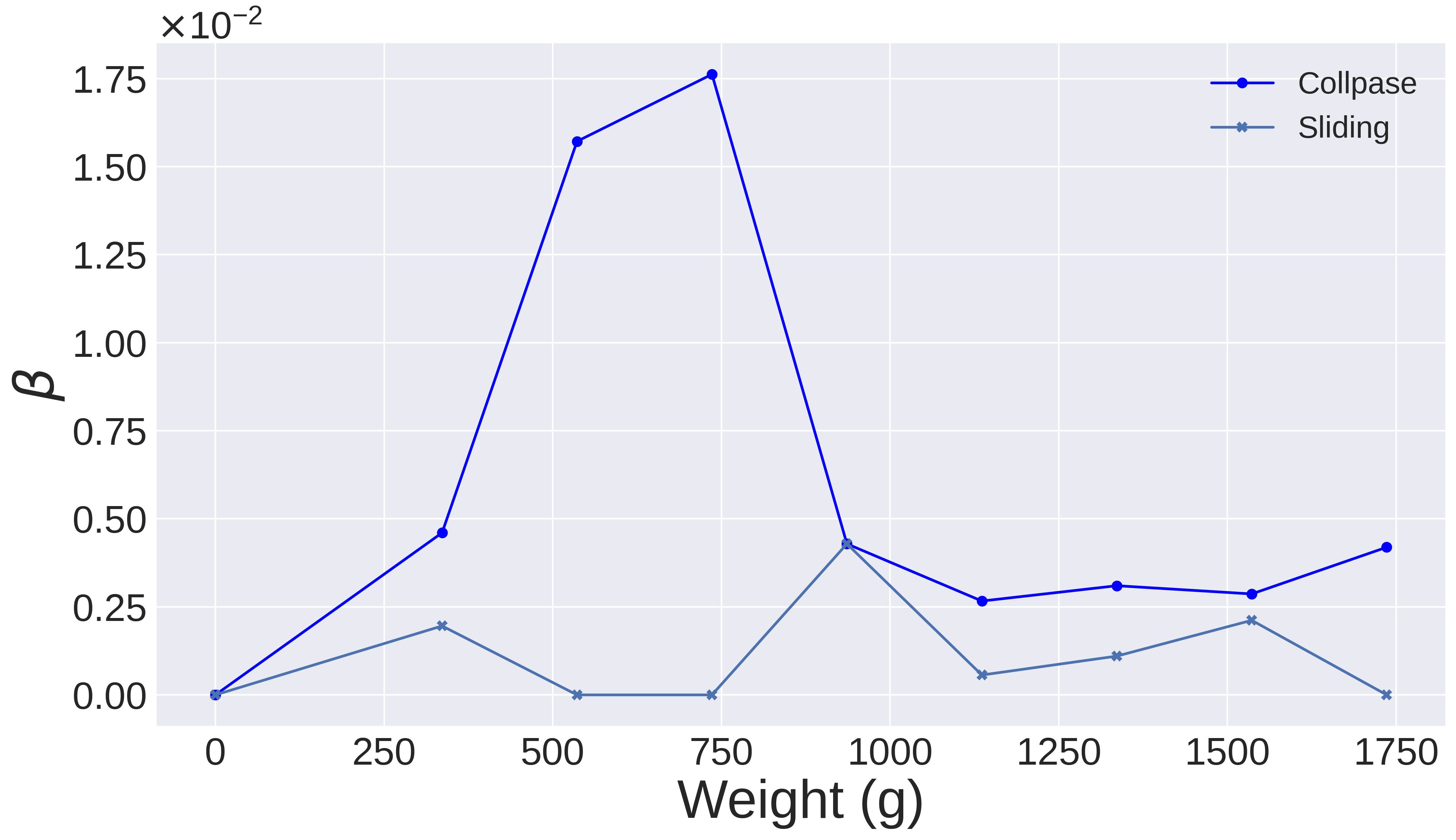}
        \caption{$\beta$ vs weights}
    \end{subfigure}
    \hfill
    \begin{subfigure}[b]{0.45\textwidth}
        \centering
        \includegraphics[width=\textwidth]{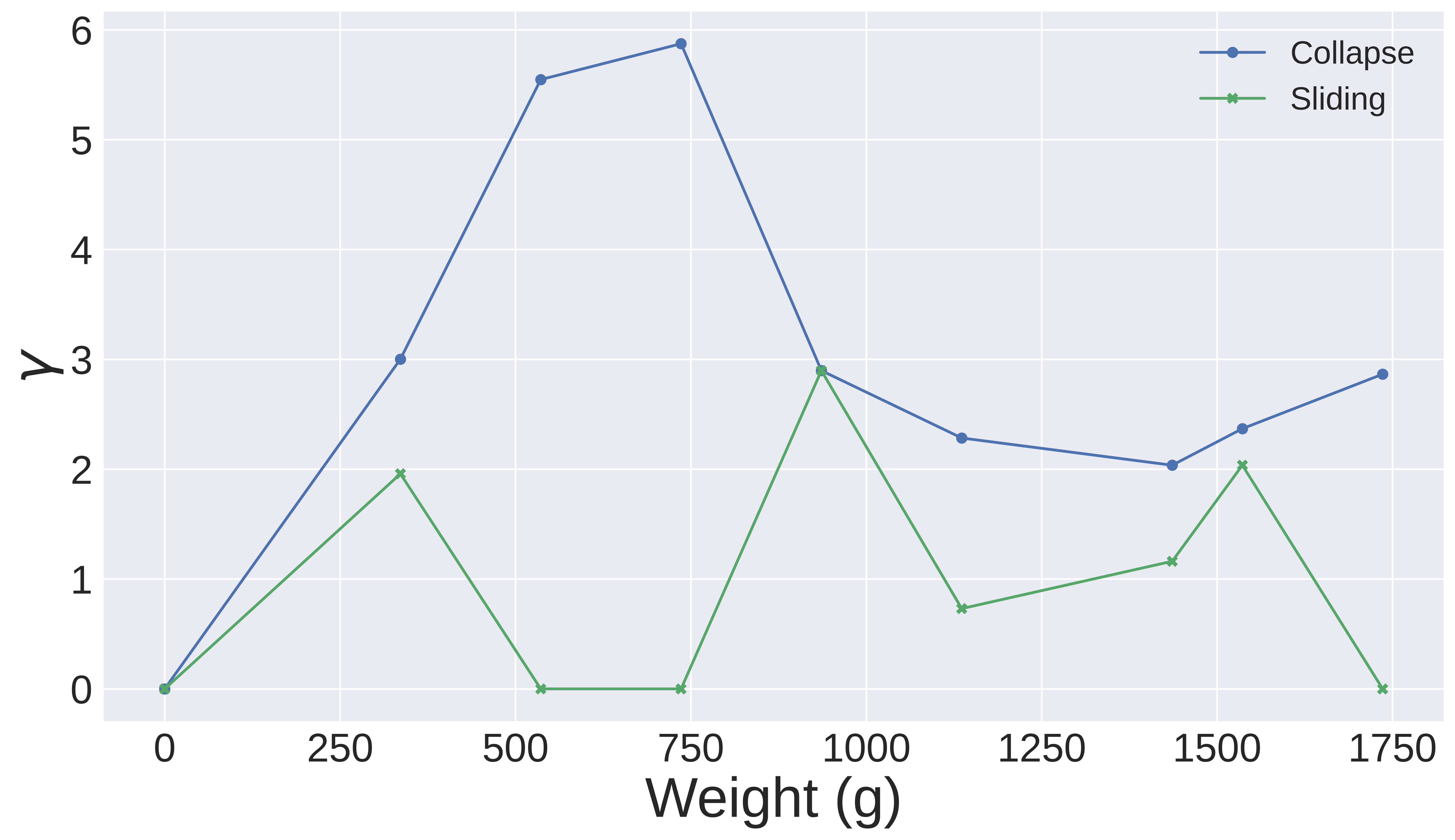}
        \caption{ $\gamma$ vs weights}
        \label{fig:sub2}
    \end{subfigure}
    \caption{Critical "temperature" and acceleration vs Weights}
    \label{fig:Phase Transition Points}
\end{figure}

 Figure \ref{fig:Phase Transition Points} shows that the critical $\beta$ and $\gamma$ (the upper curve is when the structure collapse to 0, and the lower curve is when the structure starts to slide) changes with the applied weight. On the upper curve, for weights in the range of 500-700g, the structure maintains a solid state and can endure temperature up to $\beta \sim 10^{-2}$ and $\gamma \sim 6$, which are two or one orders of magnitude greater, respectively, than those observed in the absence of pressure. It can be seen that appropriate "pressure" can significantly enhance the stability of the structure. $\beta$ and $\gamma$ have minimum at about 1250g weight. It is probably a resonance behavior. On the lower curve, however, there are no clear trend for both  $\beta$ and  $\gamma$ when more weights are applied. Again it verifies the high degree of uncertainty in the meta-stability structure. 

\newpage
\section{Conclusion}
 This study discovers a self-supporting bird nest structure constructed by merely packing a few bamboo rods, with neither container nor fastener. We analyze the stability of the structure under different conditions. For free rods structures (no applied weight or vibration), several conclusions can be derived theoretically. 
 When the coefficient of friction with the ground $\mu'$ is sufficient, the stability of the structure can be determined by the condition for $\mu$, which only depends on $d_1$ and $d_2$. A theoretical phase boundary between the stable and unstable states has been obtained at fixed $\mu$ with varying $d_1$ and $d_2$. The theoretical phase boundary matched the experiment results well.
 
 We also test the stability of the structure when weights are applied. The threshold $\mu'$ under weights would increase by a factor of $\frac{2(M/n+m/2)}{M/n+m}$ in theory. This factor increases quickly at first when applied mass $M$ is small and approaches $1$ as $M$ gets bigger. The rods in structures would slide down (height decrease, $d_2$ increase) to satisfy the condition for $\mu'$. The condition for $\mu'$ also included the parameter $\alpha$, the interior angle of a regular n-gon. A bigger $\alpha$ would yield a smaller threshold $\mu'$, so structures with more rods are more stable. In conclusion, static structures with greater $d_2$ and a number of rods (five>four>three) are more stable. We also used potential energy analysis to explain the stability of the structures. These theoretical predictions are all verified in the experiment. We also find that wet rod structures are more stable than dry ones. The structures can support up to 100 times of it's own weight. 
 
 In the vibration test, the three and four rods structure survives a vibration level up to 
  $\beta \sim 10^{-4}$ and $\gamma \sim 1$. 
 There exists an optimal pairing of $d_1$ and $d_2$ that yields the maximum structural stability. The critical $\beta$ depends on the applied weights. $\beta$ and $\gamma$ can increase by two and one order of magnitude respectively under proper weight. 
 
 Yet, this study is still an oversimplification compared to bird nests in real life. Unlike the structures in which each rod takes the same weights, rods in real bird nests each have different lengths and diameters. The real nests are not necessarily symmetric. Randomness also exists in the process of bird nest construction. These are points to be investigated in further studies. 
\newpage
\printbibliography

@article{song2016bending,
  title={Study on Measuring Young's Modulus of Cylindrical Specimens by Bending Method},
  author={Lianpeng Song and Yu Sun and Yupeng Liu and Li Zhou},
  journal={Physics Experiment},
  number={6},
  pages={22--26},
  year={2016}
}

@article{yashrajbhosale_2022_micromechanical,
  author = {Yashraj Bhosale and Nicholas Weiner and A. Butler and Seung Hyun Kim and Mattia Gazzola and Hunter King},
  title = {Micromechanical Origin of Plasticity and Hysteresis in Nestlike Packings},
  doi = {10.1103/physrevlett.128.198003},
  journal = {Physical Review Letters},
  volume = {128},
  year = {2022}
}

@article{weiner_2020_mechanics,
  author = {N. Weiner and Y. Bhosale and M. Gazzola and H. King},
  title = {Mechanics of Randomly Packed Filaments—The “Bird Nest” as Meta-Material},
  doi = {10.1063/1.5132809},
  journal = {Journal of Applied Physics},
  volume = {127},
  pages = {050902},
  year = {2020}
}

@misc{roberts_2020_why,
  author = {Siobhan Roberts},
  title = {Why Birds Are the World’s Best Engineers},
  url = {https://www.nytimes.com/2020/03/17/science/why-birds-are-the-worlds-best-engineers.html},
  year = {2020}
}

@article{liu2010jamming,
  author={Andrea J. Liu and Sidney R. Nagel},
  title={The Jamming Transition and the Marginally Jammed Solid},
  journal={Annual Review of Condensed Matter Physics},
  volume={1},
  number={1},
  pages={347--369},
  year={2010},
  doi={10.1146/annurev-conmatphys-070909-104045}
}

@misc{icebomb2020,
  title={Ice Bomb - Ice Stick Bomb, Challenge Awaits!},
  author={Popular Science Matters},
  year={2020},
  url={www.163.com/dy/article/FL01GAAB05323H3L.html}
}
\addcontentsline{toc}{section}{References}
\end{document}